\documentclass[apj]{emulateapj}
\def\apjl{ApJL }
\def\aj{AJ }
\def\apj{ApJ }
\def\pasp{PASP }

\def\apjs{ApJS }
\def\araa{ARAA }
\def\aap{A\&A }
\def\nat{Nature }
\usepackage{threeparttable}
\usepackage[maxfloats=100]{morefloats}	
\usepackage{amsmath}

\usepackage[colorlinks=true, linkcolor=red, linkbordercolor={1 0 0}, anchorcolor=magenta, citecolor=blue, citebordercolor={0 1 0}, filecolor=magenta, filebordercolor={0 .5 .5}, menucolor=magenta, menubordercolor={1 0 0}, runcolor=magenta, urlcolor=magenta, urlbordercolor={0 1 1}, frenchlinks=false]{hyperref}
		
\usepackage{multirow}
\usepackage[table]{xcolor}
\usepackage{fancybox}
\usepackage{natbib}
\usepackage{makecell}

\newcommand{\sfg}{\mathrm{S^{4}G}}
\newcommand{\galex}{\textit{GALEX}}
\slugcomment{Submitted to the Astrophysical Journal Supplement (ApJS)}
\shorttitle{The $\galex$\//$\sfg$ radial profiles catalog}\shortauthors{Bouquin et al.}
\begin{document}
\title{The $\galex$\//$\sfg$ surface brightness and color profiles catalog - I. Surface photometry and color gradients of galaxies.}
\author{Alexandre Y. K. Bouquin\altaffilmark{1}, Armando Gil de Paz\altaffilmark{1}, Juan Carlos Mu\~noz-Mateos\altaffilmark{2}, Samuel Boissier\altaffilmark{3}, \\
Kartik Sheth \altaffilmark{4},   Dennis Zaritsky\altaffilmark{5}, Reynier F. Peletier\altaffilmark{6}, Johan H. Knapen\altaffilmark{7,8}, 
Jes\'{u}s Gallego\altaffilmark{1}},
\affil{\altaffilmark{1}Departamento de Astrof\'isica y CC. de la Atm\'osfera, Universidad Complutense de Madrid, E-28040 Madrid, Spain}
\affil{\altaffilmark{2}European Southern Observatory, Casilla 19001, Santiago 19, Chile}
\affil{\altaffilmark{3}Aix Marseille Univ., CNRS, LAM, Laboratoire d'Astrophysique de Marseille, Marseille, France}
\affil{\altaffilmark{4}NASA Headquarters, Washington, DC 20546-0001, USA}
\affil{\altaffilmark{5}Steward Observatory, University of Arizona, 933 North Cherry Avenue, Tucson, AZ 85719, USA}
\affil{\altaffilmark{6}Kapteyn Astronomical Institute, Postbus 800, 9700 AV Groningen, the Netherlands}
\affil{\altaffilmark{7}Instituto de Astrof\'isica de Canarias, V\'ia L\'actea, S/N, 38205 La Laguna, Tenerife, Spain}
\affil{\altaffilmark{8}Departamento de Astrof\'isica, Universidad de La Laguna, E-38205 La Laguna, Tenerife, Spain}
\begin{abstract}
We present new, spatially resolved, surface photometry in FUV and NUV from images obtained by the \textit{Galaxy Evolution Explorer} ($\galex$), 
and IRAC1 (3.6\,$\mu$m) photometry from the \textit{Spitzer Survey of Stellar Structure in Galaxies} ($\sfg$) \citep{Sheth2010}. 
We analyze the radial surface brightness profiles $\mu_{FUV}$, $\mu_{NUV}$, and $\mu_{[3.6]}$, 
as well as the radial profiles of (FUV\,$-$\,NUV), (NUV\,$-$\,[3.6]), and (FUV\,$-$\,[3.6]) colors in 1931 nearby galaxies (z $<$ 0.01). 
The analysis of the 3.6\,$\mu$m surface brightness profiles also allows us to separate the bulge and disk components in a quasi-automatic way, and to compare
their light and color distribution with those predicted by the chemo-spectrophotometric models for the evolution of galaxy disks of \citet{BP00}. 
The exponential disk component is best isolated by setting an inner radial cutoff and an upper surface brightness limit in stellar mass surface density.
The best-fitting models to the measured scale length and central surface brightness values yield distributions of spin and circular velocity within a factor of two to those obtained via direct kinematic measurements.
We find that at a surface brightness fainter than $\mu_{[3.6]}=20.89$\,mag\,arcsec$^{-2}$, or below $3\times 10^{8}$\,$M_{\odot}$\,kpc$^{-2}$ in stellar mass surface density,
the average specific star formation rate for star forming and quiescent galaxies remains relatively flat with radius.
However, a large fraction of $\galex$ Green Valley galaxies \citep[defined in][]{Bouquin2015} shows a radial decrease in specific star formation rate. 
This behavior suggests that an outside-in damping mechanism, possibly related to environmental effects, 
could be testimony of an early evolution of galaxies from the blue sequence of star forming galaxies towards the red sequence of quiescent galaxies.
\end{abstract}

\section{Introduction}
Observing the ultraviolet (UV) part of the electromagnetic spectrum is
a direct way to determine the current star formation
rate in nearby galaxies.  The far-ultraviolet (FUV)
($\lambda_{\mathrm{eff}}$\,=\,1516 \AA) band and near-ultraviolet (NUV)
($\lambda_{\mathrm{eff}}$\,=\,2267 \AA) band luminosities are tracers of the most
recent star formation in galaxies, up to about 100 million years, because they are mainly
produced by short-lived O and B stars, and are directly related to the
current star formation rate (SFR) of galaxies \citep{Kennicutt1998}.
Consequently, the FUV observations of nearby galaxies by the \textit{Galaxy Evolution Explorer} ($\galex$) space telescope \citep[][]{Martin2005} allow us to obtain the amount of stars formed in nearby disk galaxies and dwarfs.
In the last two decades, rest-frame UV
observations have also been used to analyze the evolution of the SFR
throughout the history of the Universe \citep[see the review by][]{Madau2014}. 
However, a detailed analysis of the spatial distribution of the SFR, starting from local galaxies, is needed,
if we want to understand the origin and mechanisms involved 
in the evolution of the SFR in general and 
in the observed decay in the SFR since z$\sim$1.

In spite of the rather quick evolution sinze z=1, many galaxies have kept forming stars until now, some of them
vigorously at all galactocentric distances 
\citep[the so-called extended UV-disk galaxies constitute a prime example in that regard;][]{gdp2005, gdp2007B, Thilker2005, Thilker2007a}.
However, many others (especially massive ones but not exclusively) have had 
their star formation quenched or, at least, damped, 
in the sense that their star formation substantially decreased (and not in the sense that gas has been exhausted), 
at different epochs and at different galactocentric distances.
Our ultimate goal is to address the study of these objects using multiwavelength surface photometry combined for an unprecendented large sample of galaxies in the local Universe.
The sensitivity of the UV emission to even small amounts of star formation
allows us to identify objects that are going through
a transition phase and to determine whether this transition occurs
at all radii at the same time or in an outside-in or inside-out fashion. 
However, in order to relate the current SFR with that having occurred in the past,
the distribution of the UV emission must be
compared with that of the galaxy's stellar mass all the way to the
very faint outskirts of galaxies. 
Deep rest-frame near-infrared imaging data are key in that regard, 
such as those provided by IRAC onboard the \textit{Spitzer} satellite
in the case of nearby galaxies
and soon by the \textit{James Webb Space Telescope}(\textit{JWST}) at intermediate-to-high redshifts.
These observations allow us to probe the radial variations of the SFR in relation to the stellar mass surface density.
Spatially resolved radial color profiles are a powerful
diagnostic tool to gain insight into the relative number of young to old stars. 
However, most of the results obtained to date have 
focused on the global properties of galaxies, even in nearby galaxies.
There are noticeable exceptions such as the works of \citet{MunozMateos2011} and \citet{Pezzulli2015}, but usually for relatively small samples (75 and 35 nearby spiral galaxies respectively in these examples).

Studies of the integrated (NUV\,$-$\,$r$) vs $r$ color-magnitude
diagram for nearby galaxies have revealed a clear bimodal distribution
\citep[e.g.][]{Wyder2007,Martin2007}: quiescent, early-type galaxies (ETGs) 
are seen to form a ``red sequence", whereas actively
star-forming late-type galaxies are seen to form a ``blue
sequence". This has been seen both in the field galaxy population and in
nearby clusters such as Virgo \citep{Boselli2005}.
A recent study of the so-called ``Green Valley Galaxies'' (GVG) using the Sloan Digital Sky Survey (SDSS) data, 
and defined in the ($u$\,$-$\,$r$) color-mass diagram by \citet{Schawinski2014}
shows that GVGs span a wide range of colors and masses.
As pointed out by \citet{Schawinski2014}, using UV-optical bands helps constrain the star formation quenching timescale.
We have shown in \citet{Bouquin2015} that using the (FUV\,$-$\,NUV) vs (NUV\,$-$\,3.6\,$\mu$m) color-color diagram constrains the star formation quenching timescale to be less than 1 Gyr.

Integrated color-color diagrams have been widely used in the past to investigate integrated properties of galaxies.
For example, the (FUV\,$-$\,NUV) versus (NUV\,$-$\,$K$) \citep{gdp2007B}
or the (FUV\,$-$\,NUV) versus (NUV\,$-$\,[3.6]) color-color diagrams \citep{Bouquin2015} can separate well the star-forming galaxies from quiescent galaxies.
\citet{Bouquin2015} have shown that the combination of UV and IR reveals a better
sequential distribution than the ``classical" optical-IR color-color diagrams, especially for star-forming (Blue Clouds) systems.
These color-color diagrams separate nearby galaxies into a very narrow sequence of star-forming galaxies populated mostly by late-type galaxies, 
which we dubbed the $\galex$ Blue Sequence (GBS), and a broader sequence, the $\galex$ Red Sequence (GRS),
where quiescent galaxies such as early-type galaxies are distributed.

The above studies utilise global properties of galaxies, which do not assess the distribution of star formation \textit{within} galaxies.
It is of crucial importance that we understand how star formation is happening \textit{within} nearby galaxies,
where the active zones are, and, based on that information, determine
what mechanism(s) are in effect for activating or suppressing star formation,
in order to compare star formation of galaxies at high redshift.
Looking at the spatially resolved radial profiles is of utmost importance 
as it can give us insight into galaxy disk growth and on how quenching takes places (from inside-out or from outside-in).

Recently, a deep infrared survey of nearby galaxies, the
\textit{Spitzer} Survey of Stellar Structure in Galaxies
\citep[$\sfg$,][]{Sheth2010} has been undertaken using
the Infrared Array Camera (IRAC) onboard the \textit{Spitzer Space
  Telescope}. We used the $\sim$2300 $\sfg$ galaxies as our base sample and complemented
it with the publicly available $\galex$ counterparts (GR6/7) for those
galaxies, and have performed new FUV (1350 - 1750 \AA) and Near-UV (or
NUV) (1750 - 2800 \AA) photometry.  We obtained surface brightness
profiles in FUV and in NUV, as well as
(FUV\,$-$\,NUV) color profiles for
1931 nearby galaxies up to 40\,Mpc. 
These data provide 
both broad wavelength coverage and good physical spatial resolution.
At the median distance of the survey, 23\,Mpc, a $\galex$ PSF of
6$\arcsec$ corresponds to $\sim$700\,pc 
(but varies from 12\,pc to 2737\,pc, 
for ESO245-007 at 0.42\,Mpc, 
to PGC040552 at 94.1\,Mpc
\footnote{One of the sample selection criteria uses the distance inferred from the radial velocity measurements from HI observations, whereas here we use the redshift-independent distance, hence the discrepancy.}
).

This paper follows a classical approach in its structure,
starting with an overview of the criteria used to constrain the initial sample of galaxies (Section~\ref{sec:s4g}, \ref{sec:galex}).
Once the sample is defined, we describe the reduction processes to 
obtain our science-ready products (Section~\ref{sec:analysis}, \ref{sec:uvphotometry})
and the analysis performed (Section~\ref{sec:IRSBprofile}, \ref{sec:colorprofiles}).
Results and the discussion of that analysis are described in 
the section that follows (Section~\ref{sec:results}). 
Then, we also show in Sections~\ref{sec:modeling}, \ref{sec:velandspin} a study on obtaining the circular velocities and spin parameters from the models of \citet{BP00} (BP00) and how they compare to observed values (Section~\ref{sec:colorgradient}).
This is followed by a discussion of the results of this work in Sections~\ref{sec:discussion}, \ref{sec:vcandspin}, \ref{sec:radialdistrib}, \ref{sec:evolutionofGGV}.
Finally, the summary and conclusions are in Section~\ref{sec:summaryandconclusions}.
The derivation of stellar mass surface density from the 3.6\,$\mu$m surface brightness is included in appendix~\ref{appA}, followed by the derivation of the specific star formation rate (sSFR) from the (FUV\,$-$\,[3.6]) color in appendix~\ref{appB}. 

We assume a standard $\Lambda$CDM cosmology, 
with $\textit{H}$$_{0}$\,=\,75\,km\,s$^{-1}$\,Mpc$^{-1}$ and 
all magnitudes throughout this paper are given in the AB system unless stated otherwise.

\section{Sample} \label{sec:sample}
In this section, we briefly describe the criteria used to select the $\sfg$ sample (Section~\ref{sec:s4g}), 
and more in detail the method of retrieval of the cross-matched UV data (Section~\ref{sec:galex}).
However, the reader is referred to \citet{Sheth2010} for more details about the $\sfg$ sample selection.
This study is based uniquely on imaging data.
\subsection{S$^{\textit{4}}$G} \label{sec:s4g}
The \textit{Spitzer Survey of Stellar Structure in Galaxies} ($\sfg$) galaxy sample is 
a deep infrared survey of a (mainly) volume-limited sample of nearby galaxies within $d$\,$<$\,40\,Mpc, 
observed at 3.6\,$\mu$m and 4.5\,$\mu$m with the Infrared Array Camera \citep[IRAC,][]{Fazio2004}
\citep[see][for a full description of the survey]{Sheth2010}. 
Additional selection criteria are: size-limited with $D_{25}$\,$>$\,1$\arcmin$, 
magnitude-limited in $B$-band (Vega)\,$<$\,15.5 mag, 
and above and below the Galactic plane, $\vert b \vert$\,$>$\,30\,$\arcdeg$.
The total sample size is 2352 galaxies. 
A follow-up survey was done to include more ETGs,
but those data are not included in this catalog.

A multiwavelength analysis of the $\sfg$ sample has since been carried out 
as part of the Detailed Anatomy of Galaxies (DAGAL) project, 
and it is now complemented with FUV and NUV data from $\galex$
\citep[see also][for preliminary analyses of the UV-observed sample]{Zaritsky2014, Bouquin2015, Zaritsky2015a}, 
\textit{ugriz} images from SDSS, 
and various other data such as HI data cubes \citep[see][]{Ponomareva2016} or 
H$\alpha$ images \citep[e.g.][]{Knapen2004, Erroz-Ferrer2012}.
Additional analyses and catalogues, such as 
a classical morphological classification \citep{Buta2015},
a bulge/disk decomposition \citep[from $\sfg$ P4 pipeline;][]{Salo2015},
a catalog of morphological features \citep{Herrera-Endoqui2015},
and a stellar mass catalog \citep[P5;][]{Querejeta2015},
have also been produced and are publicly available online\footnote{http://www.astro.rug.nl/$\sim$dagal/}.
Much more detailed analysis of specific subsamples within $\sfg$ are also available elsewhere,  
such as a catalogue of structural parameters from BUDDA decomposition \citep{deSouza2004,Gadotti2008} of 3.6\,$\mu$m images \citep{KimTaehyun2016},
or H$\alpha$ kinematic studies of the inner regions \citep{Erroz-Ferrer2016}.

In this paper, we have used the surface photometry at 3.6\,$\mu$m (IRAC1) measurements from 
the output of pipeline 3 (P3) of the $\sfg$ sample 
\citep[see][for a detailed description of the $\sfg$ P3 treatment]{MunozMateos2015}.
We have collected these data from the IRSA database\footnote{http://irsa.ipac.caltech.edu/data/SPITZER/S4G/},
via their dedicated website.
We only used the 3.6\,$\mu$m surface photometry performed with 
a fixed aperture geometry (filenames of the form *.1fx2a\_noclean\_fin.dat) where 
the center, position angle, and ellipticity are all kept fixed and 
only the aperture radius is increased by radial increments of 
2$\arcsec$ along the semi-major axis. 
Subsequent mentions of $\mu_{[3.6]}$ correspond to 
the aperture-corrected surface brightness 
(columns SB\_COR and its error ESB\_COR, 
as well as the cumulative magnitude TMAG\_COR and 
its error ETMAG\_COR) found in these publicly available data.
Since our $\galex$ photometry is performed every 6$\arcsec$ in 
major-axis radius steps, we only use the data outputs obtained at 
the same step values for the 3.6\,$\mu$m photometry.

Table~\ref{Table:sampletable} shows the first galaxies of our $\galex$/$\sfg$ sample sorted by right ascension,
and lists the FUV and NUV asymptotic magnitudes obtained for our sample along the 
3.6\,$\mu$m asymptotic magnitudes obtained by \citet{MunozMateos2015}.
The complete table, with additional columns such as which $\galex$ tiles were used, 
is publicly available online through VizieR \citep{Ochsenbein2000}. 

\begin{table*}
\begin{center}
\caption{The $\galex$/$\sfg$ sample}
\label{Table:sampletable}
\begin{tabular}{lrrrccccc}
\hline
Name\tablenotemark{a} & RA\tablenotemark{b} & Dec\tablenotemark{c} & T\tablenotemark{d} & distance\tablenotemark{e} & FUV\tablenotemark{f} & NUV\tablenotemark{g} & M$_{3.6}$\tablenotemark{h} & Group ID\tablenotemark{i}\\
\hline
 & deg & deg &  & Mpc & ABmag & ABmag & ABmag & \\
\hline
UGC00017 & 0.929725 & 15.218985 & 9.1 & 13.0$\pm$--- & 16.86$\pm$0.08 & 16.59$\pm$0.02 & 14.880$\pm$0.006 & 1211 \\
ESO409-015 & 1.383640 & -28.099908 & 5.4 & 9.8$\pm$--- & 15.94$\pm$0.01 & 15.86$\pm$0.01 & 15.873$\pm$0.001 & 0 \\
ESO293-034 & 1.583550 & -41.497280 & 6.2 & 18.3$\pm$--- & 14.77$\pm$0.01 & 14.38$\pm$0.01 & 11.612$\pm$0.001 & 0 \\
NGC0007 & 2.087407 & -29.914812 & 4.8 & 21.9$\pm$1.6 & 15.48$\pm$0.01 & 15.18$\pm$0.01 & 14.021$\pm$0.002 & 1096 \\
IC1532 & 2.468434 & -64.372169 & 4.0 & 28.7$\pm$5.3 & 16.74$\pm$0.08 & 16.38$\pm$0.01 & 14.590$\pm$0.004 & 1031 \\
NGC0024 & 2.484438 & -24.964018 & 5.1 & 6.9$\pm$2.8 & 14.11$\pm$0.01 & 13.79$\pm$0.01 & 11.492$\pm$0.001 & 355 \\
ESO293-045 & 2.853125 & -41.398099 & 7.8 & 27.9$\pm$5.5 & 16.27$\pm$0.01 & 16.11$\pm$0.01 & 15.784$\pm$0.007 & 0 \\
UGC00122 & 3.323550 & 17.029280 & 9.6 & 11.6$\pm$0.7 & 16.05$\pm$0.01 & 15.89$\pm$0.01 & 15.815$\pm$0.029 & 0 \\
UGC00132 & 3.503175 & 12.963801 & 7.9 & 22.4$\pm$--- & 17.21$\pm$0.02 & 16.69$\pm$0.05 & 14.585$\pm$0.001 & 0 \\
NGC0059 & 3.854846 & -21.444339 & -2.9 & 4.9$\pm$0.6 & 16.10$\pm$0.01 & 15.35$\pm$0.01 & 12.749$\pm$0.001 & 0 \\
UGC00156 & 4.199970 & 12.350260 & 9.8 & 15.9$\pm$--- & 16.60$\pm$0.07 & 15.73$\pm$0.07 & 14.176$\pm$0.001 & 0 \\
NGC0063 & 4.439552 & 11.450338 & -3.4 & 18.8$\pm$0.2 & 16.81$\pm$0.03 & 15.61$\pm$0.02 & 11.838$\pm$0.001 & 1213 \\
ESO539-007 & 4.701543 & -19.007968 & 8.7 & 25.6$\pm$--- & 16.27$\pm$0.07 & 16.05$\pm$0.03 & 15.256$\pm$0.011 & 0 \\
ESO150-005 & 5.607727 & -53.648004 & 7.8 & 15.2$\pm$2.2 & 15.48$\pm$0.01 & 15.26$\pm$0.01 & 14.083$\pm$0.006 & 0 \\
NGC0100 & 6.011113 & 16.486026 & 5.9 & 16.4$\pm$3.1 & 15.79$\pm$0.04 & 15.32$\pm$0.01 & 13.002$\pm$0.002 & 1214 \\
NGC0115 & 6.692700 & -33.677098 & 3.9 & 30.7$\pm$5.3 & 15.16$\pm$0.01 & 14.91$\pm$0.01 & 13.752$\pm$0.001 & 1097 \\
UGC00260 & 6.762137 & 11.583803 & 5.8 & 32.3$\pm$2.3 & 15.36$\pm$0.01 & 15.04$\pm$0.01 & 12.767$\pm$0.001 & 1188 \\
ESO410-012 & 7.073298 & -27.982521 & 4.6 & 20.6$\pm$--- & 17.44$\pm$0.01 & 17.18$\pm$0.01 & 16.736$\pm$0.006 & 0 \\
UGC00290 & 7.284883 & 15.899069 & 9.5 & 9.0$\pm$0.2 & 17.66$\pm$0.21 & 17.36$\pm$0.08 & 16.412$\pm$0.005 & 0 \\
NGC0131 & 7.410483 & -33.259902 & 3.0 & 18.8$\pm$--- & 16.08$\pm$0.01 & 15.65$\pm$0.01 & 13.036$\pm$0.002 & 0 \\
UGC00313 & 7.858420 & 6.206820 & 4.3 & 27.8$\pm$--- & 16.78$\pm$0.11 & 16.34$\pm$0.04 & 13.976$\pm$0.004 & 0 \\
ESO079-003 & 8.009728 & -64.253213 & 3.1 & 39.0$\pm$4.1 & 16.66$\pm$0.02 & 16.14$\pm$0.03 & 11.604$\pm$0.001 & 0 \\
UGC00320 & 8.128720 & 2.574640 & 6.1 & 40.8$\pm$4.7 & 17.36$\pm$0.01 & 17.04$\pm$0.01 & 15.949$\pm$0.001 & 0 \\
IC1553 & 8.167184 & -25.607556 & 7.0 & 33.4$\pm$1.6 & 16.17$\pm$0.02 & 15.87$\pm$0.01 & 12.970$\pm$0.001 & 1300 \\
ESO410-018 & 8.545903 & -30.774519 & 8.9 & 19.0$\pm$--- & 15.39$\pm$0.01 & 15.21$\pm$0.01 & 14.536$\pm$0.057 & 0 \\
NGC0150 & 8.564448 & -27.803522 & 3.4 & 21.0$\pm$3.3 & 14.19$\pm$0.01 & 13.86$\pm$0.01 & 10.918$\pm$0.001 & 1100 \\
NGC0148 & 8.564559 & -31.785999 & -2.0 & 18.4$\pm$--- & 19.37$\pm$0.66 & 17.79$\pm$0.12 & 11.744$\pm$0.001 & 0 \\
IC1555 & 8.636397 & -30.017818 & 7.0 & 23.1$\pm$2.0 & 15.94$\pm$0.01 & 15.52$\pm$0.01 & 14.438$\pm$0.001 & 1096 \\
NGC0157 & 8.694906 & -8.396344 & 4.0 & 19.5$\pm$5.4 & 13.59$\pm$0.01 & 12.96$\pm$0.01 & 10.066$\pm$0.001 & 1105 \\
IC1558 & 8.946172 & -25.374404 & 9.0 & 13.7$\pm$4.6 & 14.73$\pm$0.01 & 14.43$\pm$0.01 & 13.337$\pm$0.002 & 1100 \\
NGC0178 & 9.784857 & -14.172626 & 8.7 & 18.4$\pm$--- & 14.16$\pm$0.01 & 13.99$\pm$0.01 & 13.193$\pm$0.001 & 0 \\
NGC0210 & 10.145717 & -13.872773 & 3.1 & 21.0$\pm$1.3 & 13.99$\pm$0.08 & 13.76$\pm$0.01 & 10.792$\pm$0.001 & 1102 \\
ESO079-005 & 10.182495 & -63.441987 & 7.0 & 23.5$\pm$2.8 & 15.28$\pm$0.01 & 14.97$\pm$0.01 & 13.866$\pm$0.004 & 1032 \\
NGC0216 & 10.363123 & -21.044899 & -1.9 & 19.1$\pm$--- & 15.54$\pm$0.01 & 15.10$\pm$0.01 & 13.059$\pm$0.004 & 0 \\
PGC002492 & 10.439405 & -16.860757 & 2.0 & 20.7$\pm$--- & 15.74$\pm$0.02 & 15.50$\pm$0.01 & 14.155$\pm$0.006 & 0 \\
IC1574 & 10.765448 & -22.245836 & 9.9 & 4.8$\pm$0.2 & 16.57$\pm$0.01 & 16.09$\pm$0.01 & 14.749$\pm$0.017 & 355 \\
NGC0244 & 11.443430 & -15.596570 & -2.0 & 11.6$\pm$--- & 15.19$\pm$0.01 & 14.94$\pm$0.01 & 13.593$\pm$0.003 & 0 \\
PGC002689 & 11.515689 & -11.506472 & 8.8 & 20.2$\pm$--- & 15.37$\pm$0.03 & 15.21$\pm$0.02 & 14.693$\pm$0.009 & 0 \\
UGC00477 & 11.554634 & 19.489885 & 7.9 & 35.8$\pm$0.4 & 15.86$\pm$0.01 & 15.66$\pm$0.01 & 14.328$\pm$0.002 & 1294 \\
ESO411-013 & 11.776317 & -31.581403 & 9.0 & 23.5$\pm$--- & 17.87$\pm$0.25 & 17.36$\pm$0.03 & 16.037$\pm$0.002 & 0 \\
NGC0247 & 11.785305 & -20.760176 & 6.9 & 3.6$\pm$0.5 & 11.42$\pm$0.02 & 11.12$\pm$0.02 & 9.135$\pm$0.001 & 233 \\
NGC0254 & 11.865155 & -31.421775 & -1.2 & 17.1$\pm$--- & 17.71$\pm$0.15 & 16.39$\pm$0.03 & 11.387$\pm$0.001 & 0 \\
NGC0255 & 11.946929 & -11.468734 & 4.1 & 20.0$\pm$--- & 13.98$\pm$0.01 & 13.75$\pm$0.02 & 12.252$\pm$0.002 & 0 \\
PGC002805 & 11.948177 & -9.899568 & 6.7 & 16.4$\pm$0.5 & 15.61$\pm$0.01 & 15.32$\pm$0.01 & 14.624$\pm$0.005 & 1101 \\
ESO540-031 & 12.457500 & -21.012730 & 9.8 & 3.4$\pm$0.2 & 16.85$\pm$0.01 & 16.59$\pm$0.02 & 16.189$\pm$0.048 & 233 \\
ESO079-007 & 12.517568 & -66.552204 & 4.0 & 25.2$\pm$4.5 & 15.21$\pm$0.04 & 14.94$\pm$0.01 & 13.570$\pm$0.001 & 0 \\
NGC0274 & 12.757695 & -7.056978 & -2.8 & 20.3$\pm$1.5 & 14.52$\pm$0.01 & 14.13$\pm$0.01 & 12.091$\pm$0.001 & 1103 \\
NGC0275 & 12.768555 & -7.065730 & 6.0 & 21.9$\pm$--- & 14.50$\pm$0.01 & 14.15$\pm$0.01 & 12.284$\pm$0.001 & 0 \\
PGC003062 & 13.072022 & -3.966015 & 6.8 & 18.8$\pm$--- & 17.04$\pm$0.20 & 16.59$\pm$0.02 & 15.291$\pm$0.009 & 0 \\
NGC0289 & 13.176101 & -31.205822 & 4.0 & 22.8$\pm$4.1 & 13.32$\pm$0.05 & 13.15$\pm$0.06 & 10.522$\pm$0.003 & 1098 \\
...&&&&&&&&\\
\hline
\end{tabular}
\tablenotetext{1}{same as the $\sfg$ nomenclature}
\tablenotetext{2}{right ascension in degrees and in epoch J2000.0}
\tablenotetext{3}{declination in degrees and in epoch J2000.0}
\tablenotetext{4}{numerical morphological type from RC2}
\tablenotetext{5}{mean, redshift-independent, distance measurements with 1$\sigma$ uncertainty from NED if available; see text for details.}
\tablenotetext{6}{Total FUV apparent magnitude with 1$\sigma$ uncertainty. These uncertainties do not include zero point errors, nor errors associated to the misidentification of background or foreground sources.}
\tablenotetext{7}{Total apparent magnitude with 1$\sigma$ uncertainty. Also see above.}
\tablenotetext{8}{Total 3.6$\mu$m apparent magnitude from IRAC1 photometry and 1$\sigma$ uncertainty \citep{MunozMateos2015}}
\tablenotetext{9}{Groups and clusters flag obtained from the Galaxy On
  Line Database Milano Network \citep[GOLDMine,][]{Gavazzi2003} and
  the Cosmicflows-2 \citep{Tully2013} catalogs. 
  When merging the two catalogs into a single column,
  priority was given to the GOLDMine group ID.
  That is, galaxies with a value of 1, 2, 3, 4, or 9 in the GOLDMine catalog, 
  denoting galaxies in the Virgo Cluster, were kept as such.
  A value of 1 is assigned if the galaxy is in
  the Virgo Cluster at 17Mpc, 2 at 23 Mpc, 3 at 32 Mpc, 4 at 37.5 Mpc,
  and 9 for the ones at various other distances. 
  For galaxies not in GOLDMine, we use the Tully group ID,
  but if the Tully group ID happens to be 1, 2, 3, 4, or 9 (which does not necessarily mean that they are in the Virgo Cluster),
  we append the value with the letter `T' to differentiate them from the GOLDMine group ID.
  Only 11 out of 1931 galaxies have a GOLDMine group ID of 0 but a Tully group ID of 1, 2, 3, 4, or 9, namely: 
  UGC07249 (1T),
  UGC07394 (4T), UGC07522 (4T), NGC4409 (4T), NGC4496A (4T), IC0797
  (1T), IC0800 (1T), PGC042160 (1T), UGC07802 (1T), NGC4666 (9T),
  UGC07982 (2T).
  A group ID of 0 means
  that the galaxy is not in a group in either catalog.}
\tablecomments{Our sample of 1931 galaxies, sorted by right ascension.}
\end{center}
\end{table*}

\begin{figure*}[ht!]
\begin{center}
    	\includegraphics[width=0.4\textwidth]{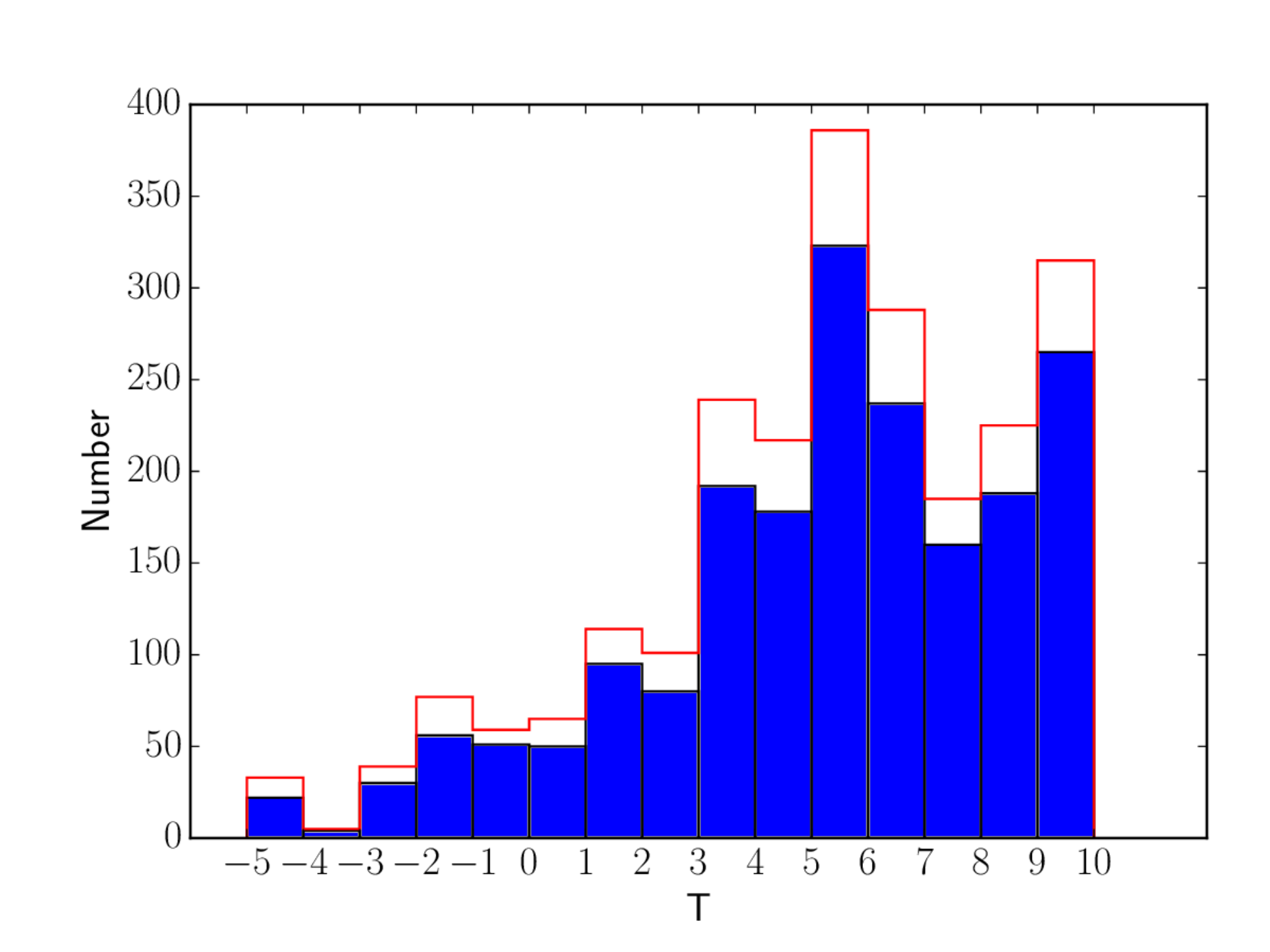}
        	\includegraphics[width=0.4\textwidth]{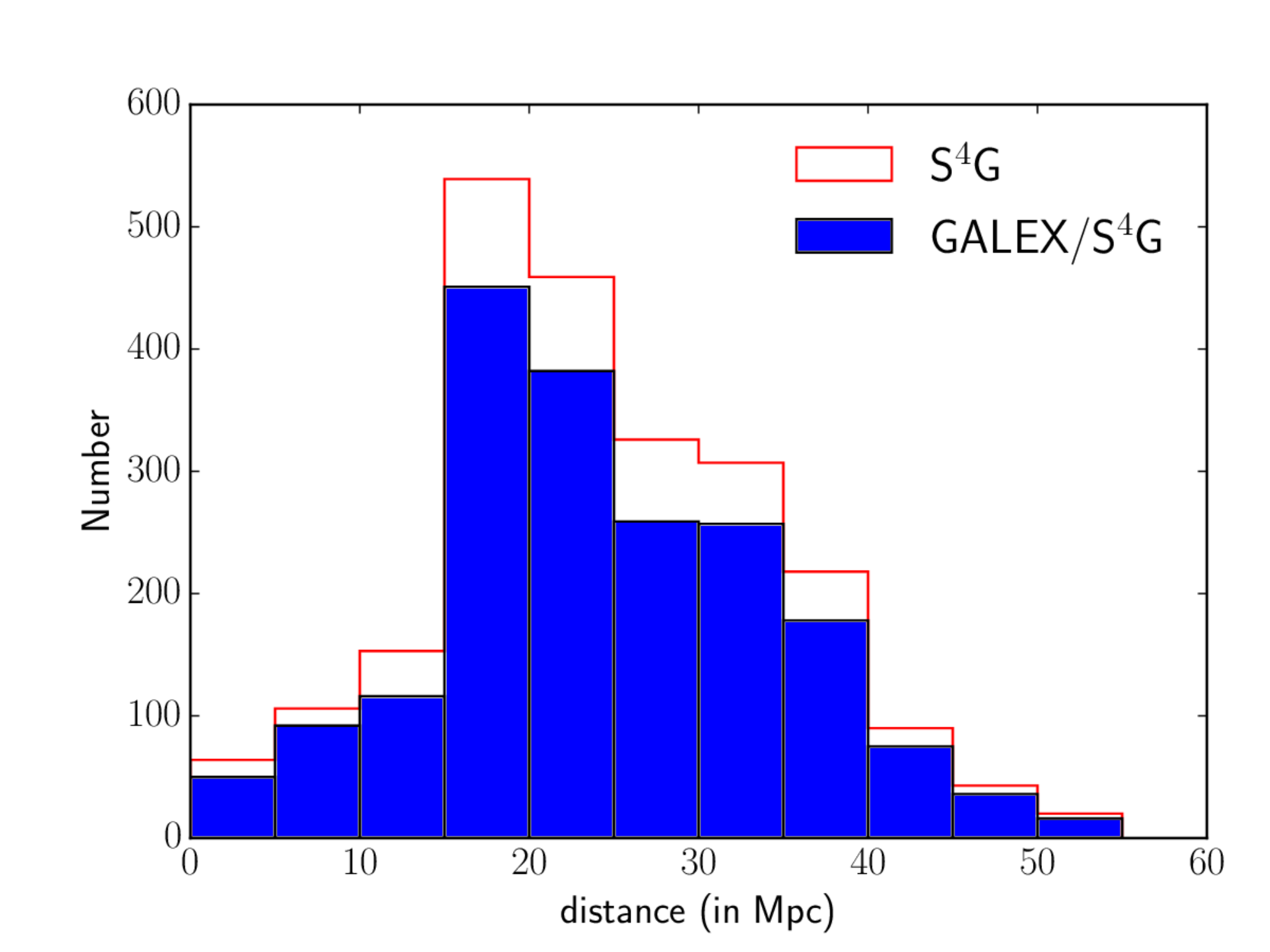}\\
	\includegraphics[width=0.4\textwidth]{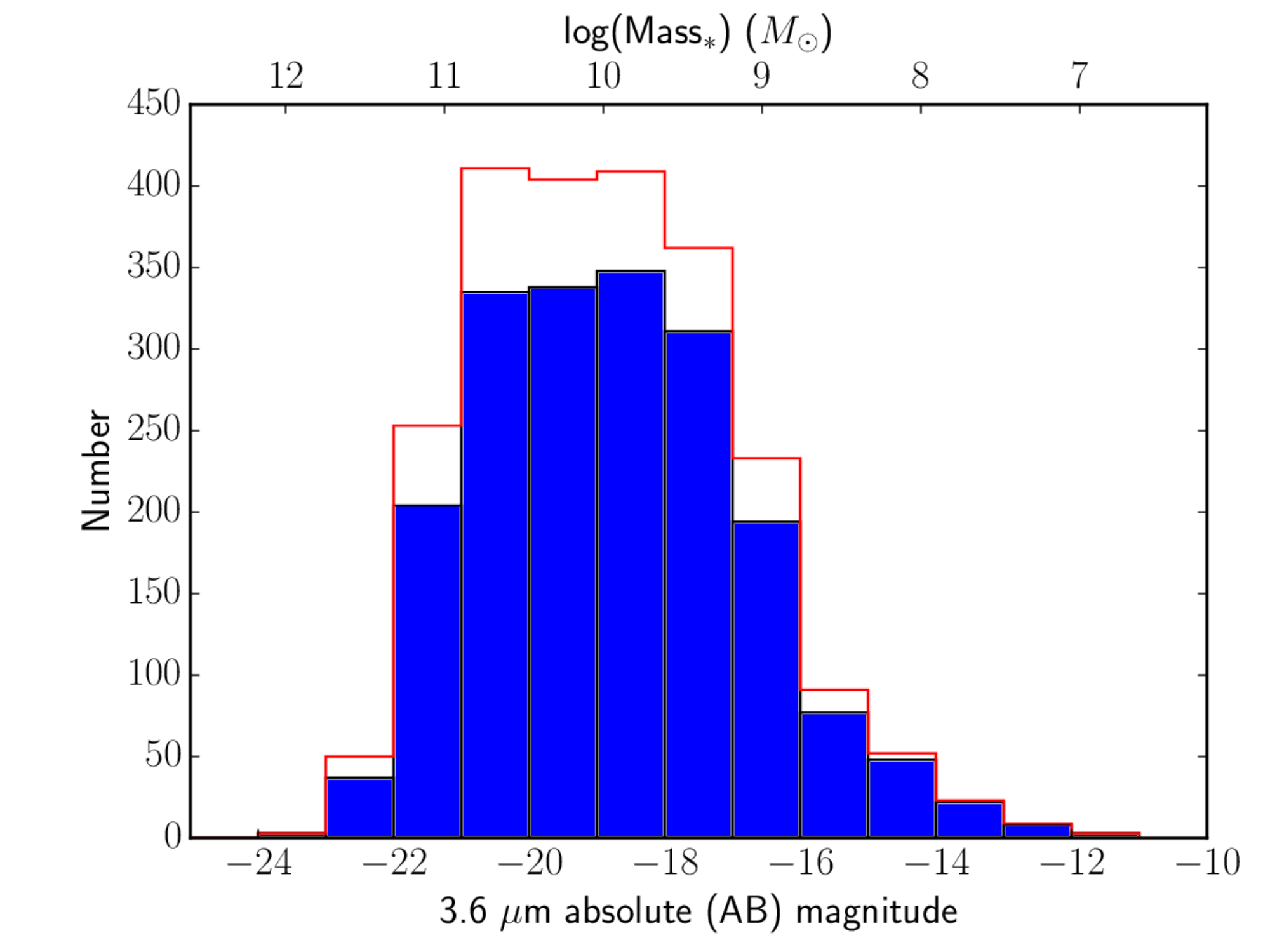}
	\includegraphics[width=0.4\textwidth]{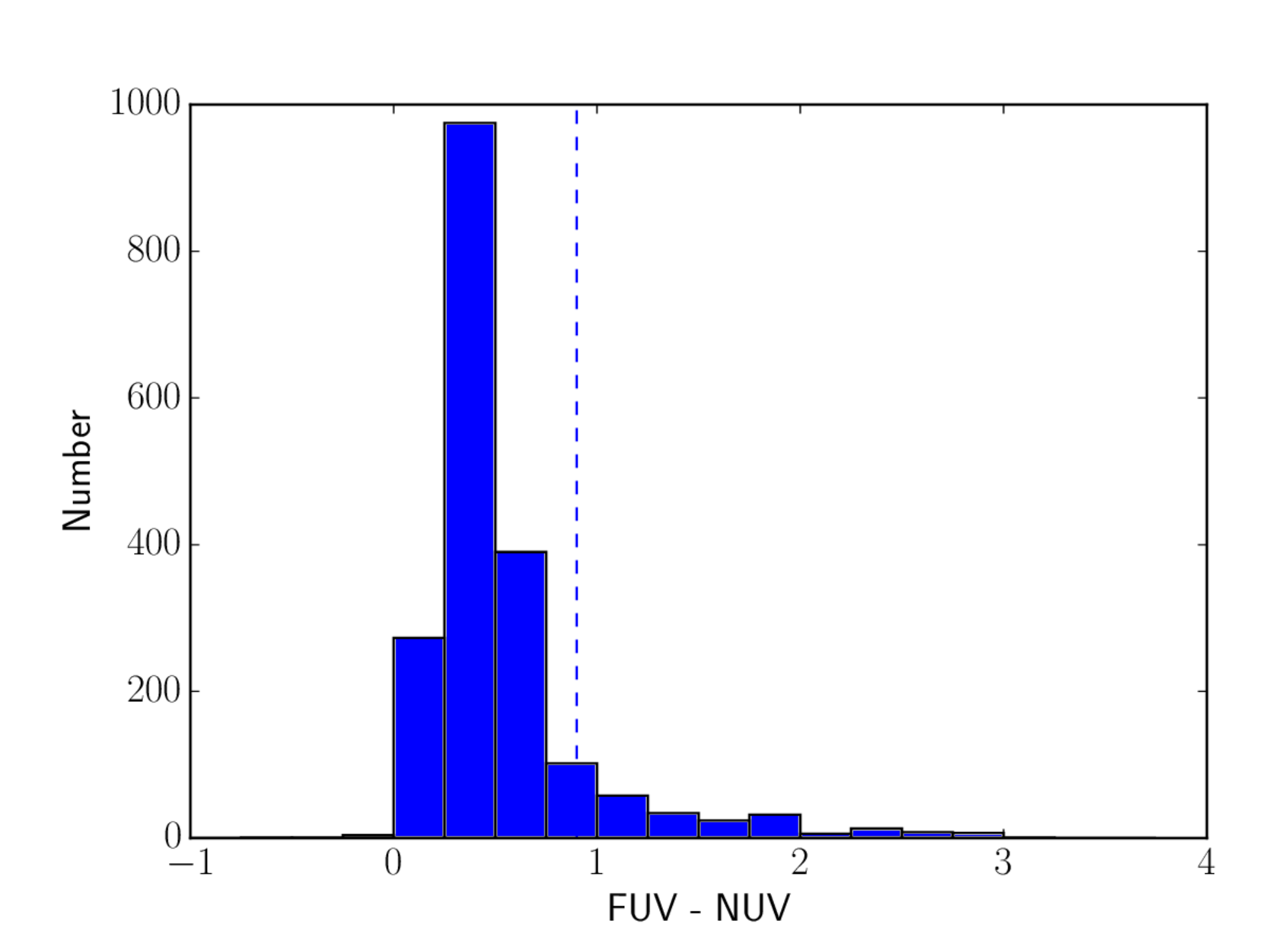}
\caption{
Comparisons of the distributions of the $\sfg$ sample (white bars) with the $\galex$/$\sfg$ subsample (filled bars). 
The distributions of numerical morphological types (T) (binning=1) (top-left), 
distances in Mpc (binning=5Mpc) (top-right), 
3.6$\mu$m absolute AB magnitudes (binning=1mag) (bottom-left), and 
(FUV\,$-$\,NUV) color (binning=0.25 mag) (bottom-right) of both samples are shown. 
The $\sfg$ sample is comprised of 2352 galaxies, and the $\galex$/$\sfg$ subsample of 1931 galaxies. 
The 3.6$\mu$m absolute magnitude is accompanied by the logarithm of the stellar mass in the top x-axis, 
computed using eq.\ref{eq:massfromabsmag} in Appendix~\ref{appA}. 
The vertical line at (FUV\,$-$\,NUV) = 0.9 corresponds to the value used in \citet{gdp2007B} to broadly separate between early and late-type galaxies. \label{fig:histogramsALL}}
\end{center}
\end{figure*}

\subsection{$\galex$ counterparts} \label{sec:galex}
We gathered all available $\galex$ FUV and NUV images and related data
products for 1931 $\sfg$ galaxies that had been observed in at least
one of these two UV bands. Over 200 galaxies do not have $\galex$ data at all.  
We obtained the original $\galex$ data using the $\galex$view\footnote{http://galex.stsci.edu/GalexView/} tool. 
Priority was given to galaxies that have both FUV and NUV images, with the longest FUV exposure time.
If this condition was met for several product tiles (including a very similar exposure time to the NUV in the FUV), we chose the one where the target galaxy was best centered in the field-of-view (FOV).
We collected imaging data from all kinds of surveys such as the All-sky Imaging Survey (AIS), Medium Imaging Survey (MIS), 
Deep Imaging Survey (DIS), Nearby Galaxy Survey (NGS), as well as from Guest Investigator (GIs/GIIs) Programs.

The collected data, once processed, yielded a total of 1931 galaxies
with both FUV and NUV photometry available.  We call this sample,
derived from the $\sfg$ and having FUV, NUV, as well as IRAC1 3.6
$\mu$m photometry, the $\galex$/$\sfg$ sample.  
We compare the $\sfg$ sample and the $\galex$/$\sfg$ sample in Figure~\ref{fig:histogramsALL}.  
The distributions of distances, apparent B-band magnitudes, and morphological types of 
the two samples and the distribution of the integrated
(FUV\,$-$\,NUV) colors of the final
$\galex$/$\sfg$ sample are shown. 
Demographics are shown in
Table~\ref{Table:basictable}. 
Our $\galex$/$\sfg$ sample is
clearly representative of the whole $\sfg$ sample with only minor
differences in the case of the absolute magnitude distribution.
Note that every $\sfg$ galaxy targeteted with $\galex$ was detected and its UV fluxes measured.

We also subdivided the $\galex$/$\sfg$ sample into three other subsamples.
This was done accordingly to the preliminary analysis of the UV-to-IR photometry of \citet{Bouquin2015}, 
where we presented our sample of 1931 galaxies with their asymptotic magnitudes plotted on an 
(FUV\,$-$\,NUV) vs (NUV\,$-$\,[3.6]) color-color diagram.
From this integrated color-color diagram, we were able to select three subsamples of galaxies, 
namely the GBS, the GRS, and the GGV galaxies, and were defined as follows:
\begin{widetext}
\begin{equation}\label{eq:GBS}
\begin{aligned}
\textrm{GBS: }& 0.12 x + 0.16 - 2\sigma_{\mathrm{GBS}} \leq y \leq 0.12 x + 0.16 + 2\sigma_{\mathrm{GBS}}\\
\end{aligned}
\end{equation}
\begin{equation}\label{eq:GRS}
\begin{aligned}
\textrm{GRS: }& -0.23 y+ 5.63 - 1\sigma_{\mathrm{GRS}} \leq x \leq -0.23 y+ 5.63 + 1\sigma_{\mathrm{GRS}}\\
\end{aligned}
\end{equation}
\begin{equation}\label{eq:GGV}
\begin{aligned}
\textrm{GGV: }& y > 0.12 x + 0.16 + 2\sigma_{\mathrm{GBS}} \textrm{ and } x < -0.23 y + 5.63 - 1\sigma_{\mathrm{GRS}}
\end{aligned}
\end{equation}
\end{widetext}
where  $x$ = (NUV\,$-$\,[3.6]), $y$ = (FUV\,$-$\,NUV), $\sigma_{\mathrm{GBS}}$=0.20, and $\sigma_{\mathrm{GRS}}$=0.45.
Both the GBS and GRS are defined to be stripes defined between two parallel lines (Equations~\ref{eq:GBS} and \ref{eq:GRS}).
Note that the GRS equations are expressed in the $yx$-space.
The GGV is the region bluer in (NUV\,$-$\,[3.6]) than the GRS,
but redder in (FUV\,$-$\,NUV) than the GBS (Equation~\ref{eq:GGV}).

The GBS is populated by star-forming galaxies and mostly late-type
galaxies while the GRS is populated by redder systems that lack star
formation (quiescent) and are passively evolving or where only low
levels of residual star formation are present \citep[e.g.,][]{Boselli2005,Yildiz2017}
  and that are, morphologically speaking,
  mostly ETGs. The GGV galaxies are found between the GBS and the GRS
  in this UV-to-IR color-color plane, and they are special in the sense that
  these galaxies can be seen to have decreased star formation
  activity in recent epoch, hence their (FUV\,$-$\,NUV)
  colors are redder than in GBS galaxies and their
  (NUV\,$-$\,[3.6]) colors are bluer than in GRS galaxies.
  However, it should be noted that we do not exclude the possibility
  that this GGV populations could represent GRS galaxies that are
  being rejuvenated thus showing a
  blueing (FUV\,$-$\,NUV) color.
  The important point here is that the quick response of the
  (FUV\,$-$\,NUV) color to even
  small amounts of recent star formation, coupled to
  the tightness of the GBS, allows identifying galaxies that are just
  starting to experience these quenching or rejuvenating events. See
  \citet{Bouquin2015} for further details.

\begin{table}
\caption[Table~caption text]{Galaxy sample demographics}
\label{Table:basictable}
\begin{center}
\begin{tabular}{ l r r l}
\hline
  \multicolumn{2}{c}{Galaxy sample\tablenotemark{a}} &
  \multicolumn{1}{c}{N} &
  \multicolumn{1}{c}{Percentage relative to ()} \\
\hline
  $\sfg$ & & 2352 & 100\% \\
  $\galex$/$\sfg$ & & 1931 & 82.1\% ($\sfg$)\\
\hline
  GBS & & 1753 & 90.8\% ($\galex$/$\sfg$)\\
  GGV & & 70 & 3.6\% --- \\
  GRS & & 79 & 4.1\% --- \\
  Others & & 29 & 1.5\% --- \\
\hline
\multirow{2}{*}{ETGs} & E & 24 & 1.2\% ($\galex$/$\sfg$)\\
& E-S0 & 23 & 1.2\% --- \\
 \hline
\multirow{3}{*}{ETDGs} & S0 & 51 & 2.6\% --- \\
& S0-a & 103 & 5.3\% --- \\
& Sa & 175 & 9.1\% --- \\
\hline
\multirow{5}{*}{LTGs} & Sb & 340 & 17.6\% --- \\
& Sc & 669 & 34.7\% --- \\
& Sd & 168 & 8.7\% --- \\
& Sm & 192 & 9.9\% --- \\
& Irr & 186 & 9.6\% --- \\
\hline
\end{tabular}
\tablenotetext{1}{Name of the samples. GBS = $\galex$ Blue Sequence, GGV = $\galex$ Green Valley, GRS = $\galex$ Red Sequence, ETGs, = Early-Type Galaxies, ETDGs = Early-Type Disk Galaxies, LTGs = Late-Type Galaxies. RC2 morphological types were obtained from HyperLeda.}
\end{center}
\end{table}

\section{ANALYSIS} \label{sec:analysis}\
In this section, we describe our method of analysis of the NIR and UV imaging data acquired by \textit{Spitzer} IRAC1 and $\galex$,
in order to obtain 3.6\,$\mu$m, FUV, and NUV surface photometry.
The acquirement of the 3.6\,$\mu$m surface photometry is not described here as it is already explained in \citet{MunozMateos2015},
and we only focus on the FUV and NUV surface photometry in this article (Section~\ref{sec:uvphotometry}).
We also performed a radial normalization of the 3.6\,$\mu$m radial profiles (Section~\ref{sec:IRSBprofile}).
We also constructed the (FUV\,$-$\,NUV), (FUV\,$-$\,[3.6]), and (NUV\,$-$\,[3.6]) color profiles (Section~\ref{sec:colorprofiles}).

\subsection{UV Surface photometry and asymptotic magnitudes} \label{sec:uvphotometry}
We obtained spatially resolved FUV and NUV surface photometry, as well as
asymptotic magnitudes, for the 1931 galaxies in our $\galex$/$\sfg$
sample. Three types of $\galex$ data products were collected from the database:
\begin{itemize}
\item the intensity maps in FUV (*fd-int.fits) and NUV (*nd-int.fits),
\item the high-resolution relative response maps in FUV (*fd-rrhr.fits) and NUV (*nd-rrhr.fits), and
\item the object masks in both FUV (*fd-objmask.fits) and NUV (*nd-objmask.fits).
\end{itemize}
Once all data were gathered, we proceeded to reduce and analyze our 
$\galex$ UV sample in the same manner as in \citet{gdp2007B}.

First, a sky value was measured from the surroundings of the target
galaxy. This was followed by the preparation of a mask, in two steps.
In the first step of the masking process, we masked automatically unresolved
sources that had 
(FUV\,$-$\,NUV) colors redder than 1 mag,
which masks out most foreground stars.
This was followed by careful visual checks, verifying each and
every single galaxy, and carefully editing the masks one by one, by
manually adding or removing masks, since the automatization could: (a)
falsely detect bulges, fail to select (b) companions, and (c)
foreground blue stars, all for the benefit of preserving very blue
star-forming regions, especially those in the outskirts of disk
galaxies.  We unmasked all affected bulges and tried to include as many
star-forming regions falsely masked, while foreground stars were
masked out as much as possible. In the process we also generated
FUV+NUV RGB images for each galaxy that were used during the manual
masking process in order to have an educated guess on any potential
masking failure encountered. Although great care had been taken
during this masking process, it should be noted that in some cases
(e.g., merging galaxies, galaxies with bright stars nearby, objects at
the edge of the FOV, bad image quality), difficult choices had to be
made.  We acknowledge that in those cases (less than a few percent) the values obtained may differ
from those obtained by other authors (our masks can be provided on
demand). Errors associated to these effects cannot be accounted for and are not included in Table~\ref{Table:sampletable}.

Then, surface brightnesses were measured by averaging over annuli with
the same position angle (PA) and ellipticity ($\epsilon$) as those
used in the analysis of the $\sfg$ sample IRAC data. We used a step
in major-axis radius of 6$\arcsec$ and integrated over a width of
$\pm$3$\arcsec$, also in major-axis radius.  The total uncertainty in
the surface brightness does take into account the contribution of both
local and large-scale background errors \citep{gdp2007B}.

\begin{figure*}
\begin{center}
\ovalbox
{
\includegraphics[trim=2cm -0.7cm 4.5cm 1cm, clip=true, width=0.23\textwidth]{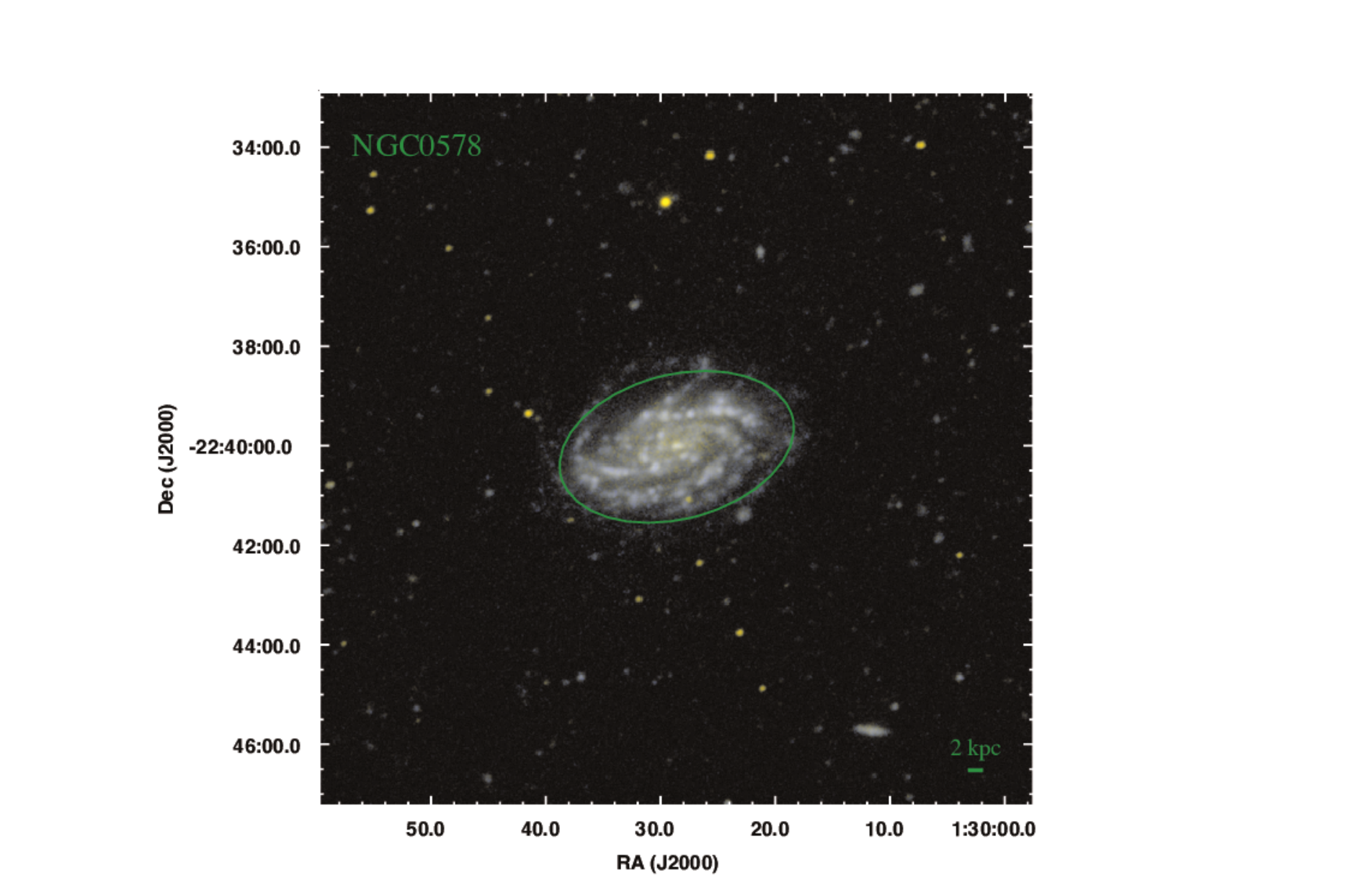}
\includegraphics[trim=1.0cm 2.5cm 0cm 2cm, clip=true, width=0.235\textwidth]{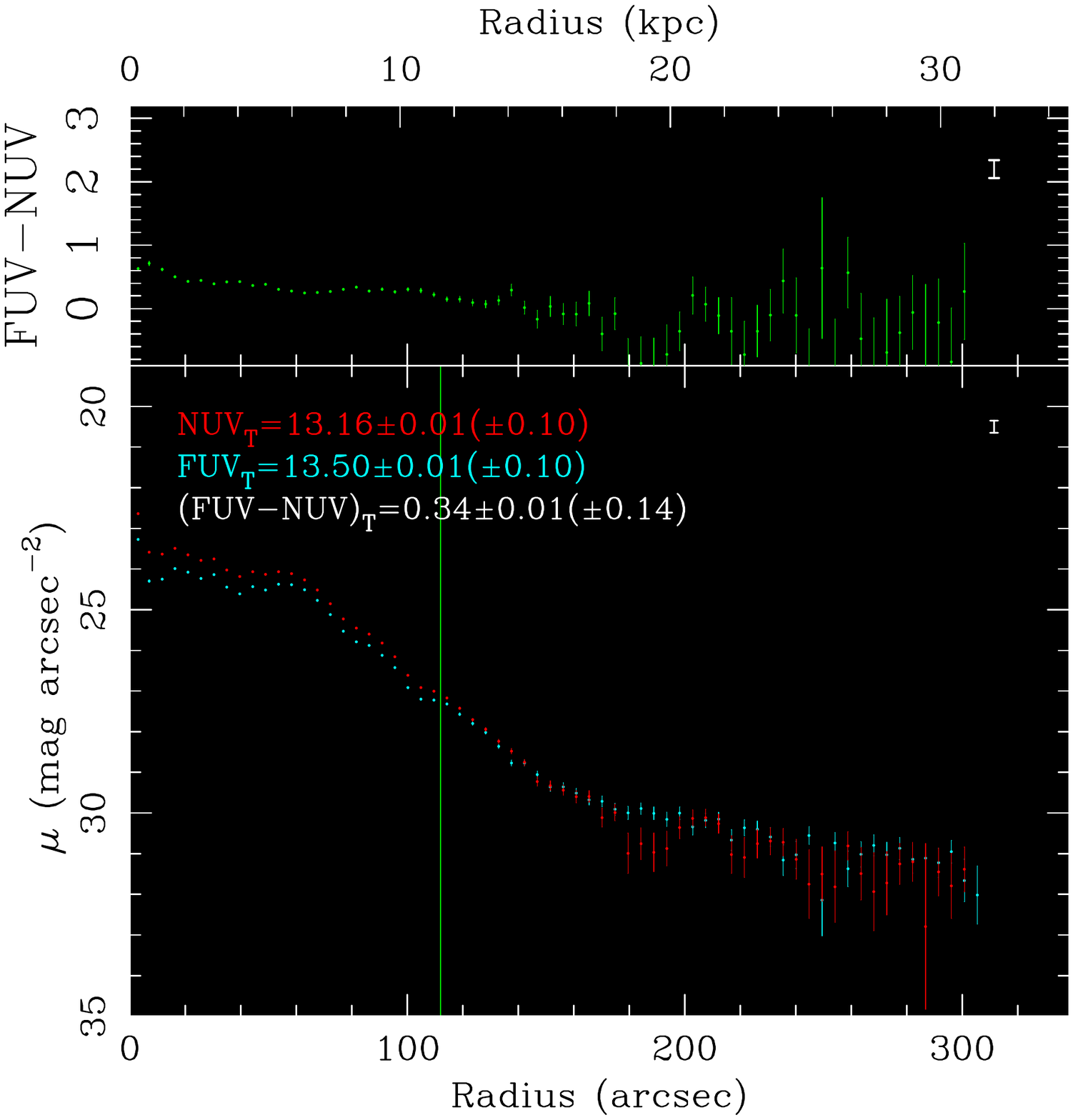}
\includegraphics[trim=2cm -0.7cm 4.5cm 1cm, clip=true, width=0.23\textwidth]{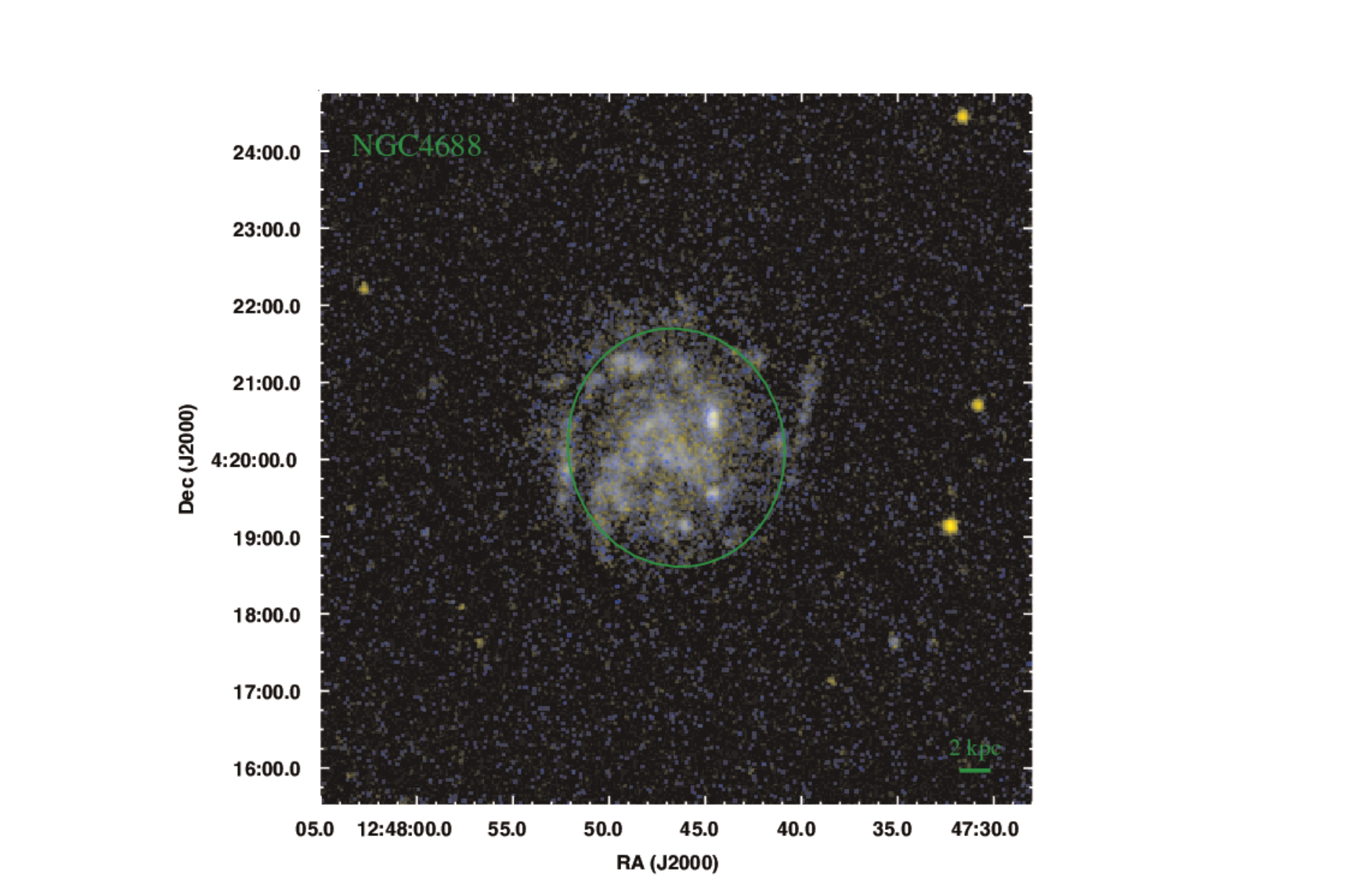}
\includegraphics[trim=1.0cm 2.5cm 0cm 2cm, clip=true, width=0.235\textwidth]{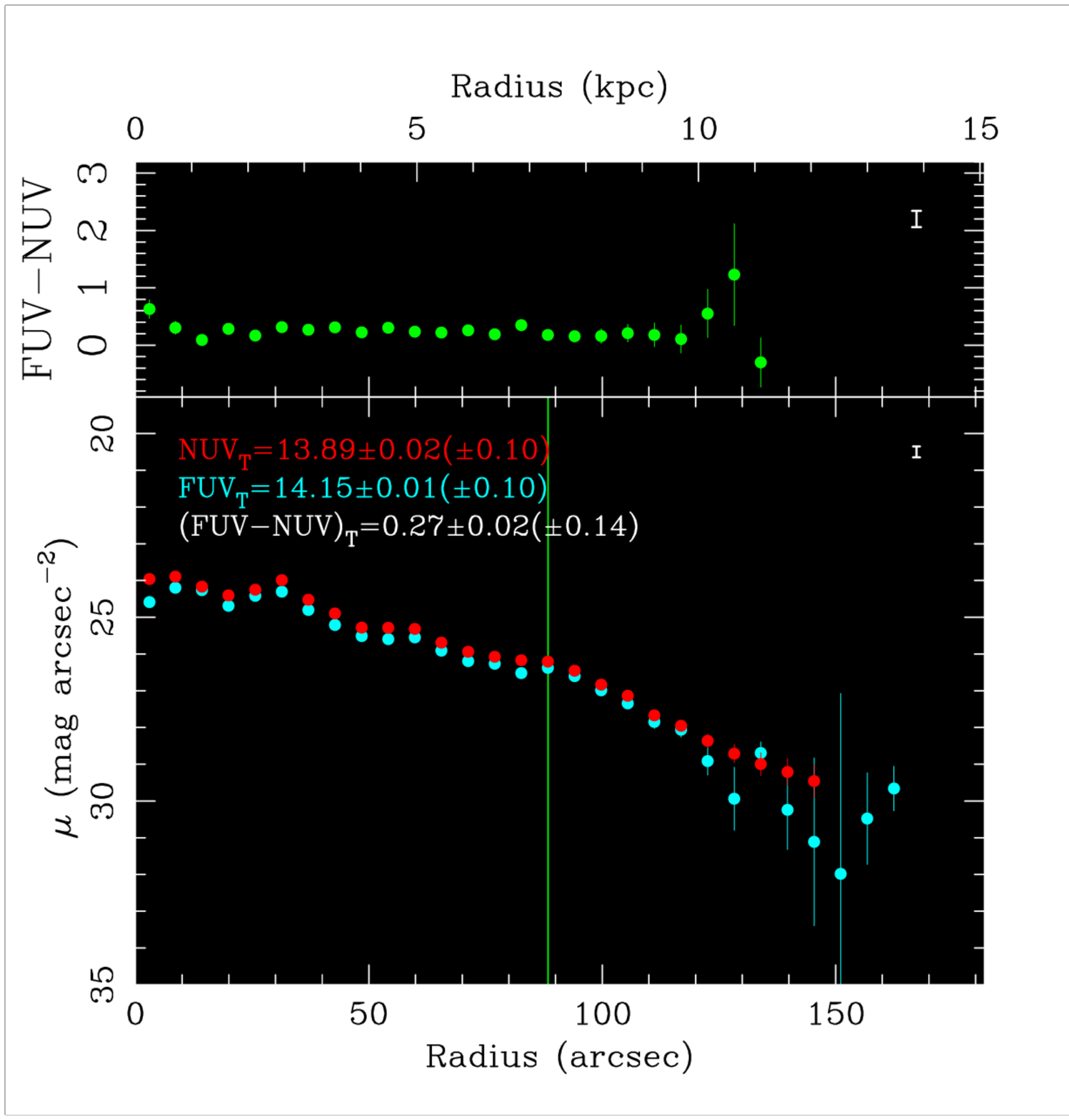}
}
\ovalbox
{
\includegraphics[trim=2cm -0.7cm 4.5cm 1cm, clip=true, width=0.23\textwidth]{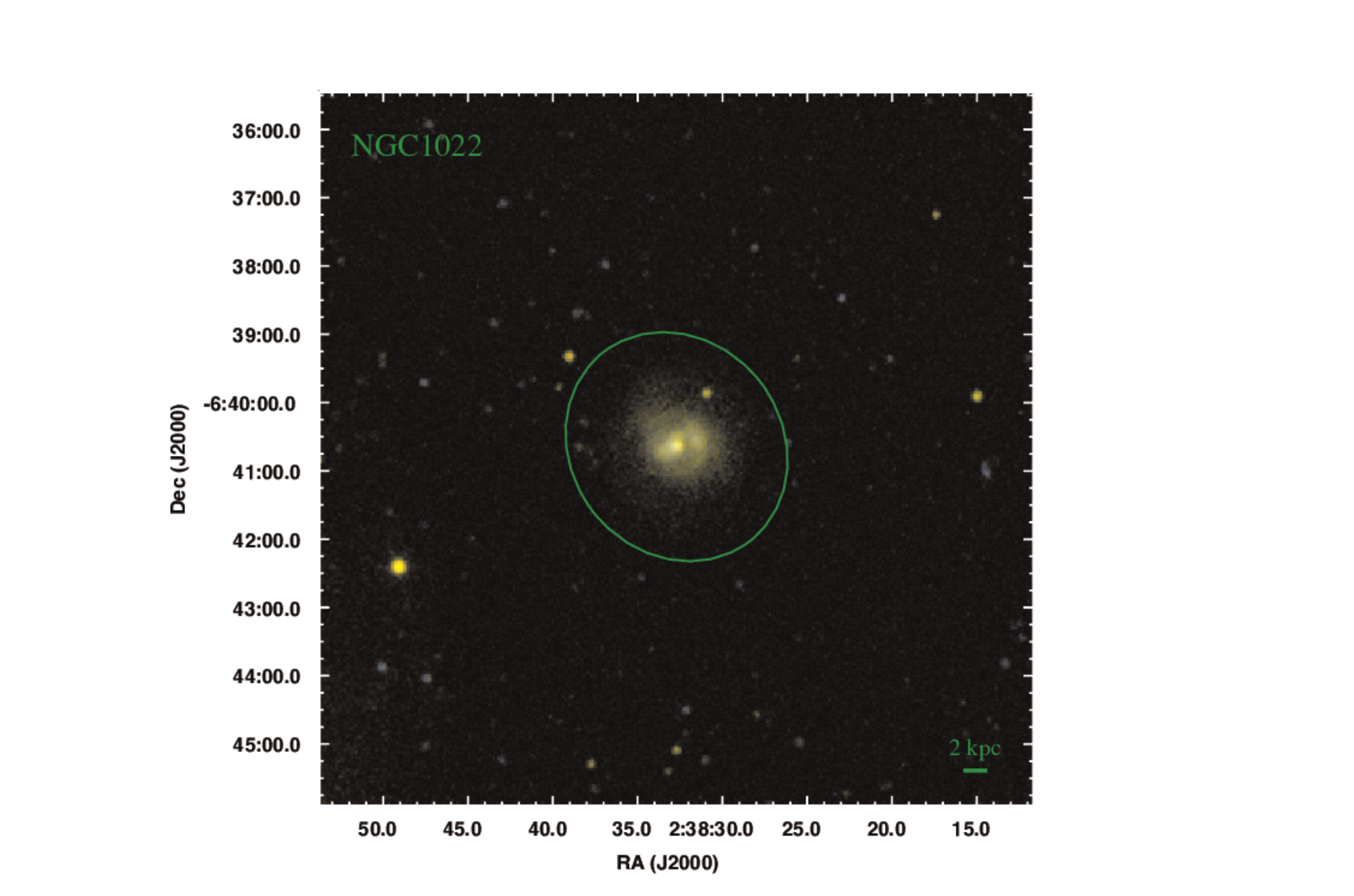}
\includegraphics[trim=1.0cm 2.5cm 0cm 2cm, clip=true, width=0.235\textwidth]{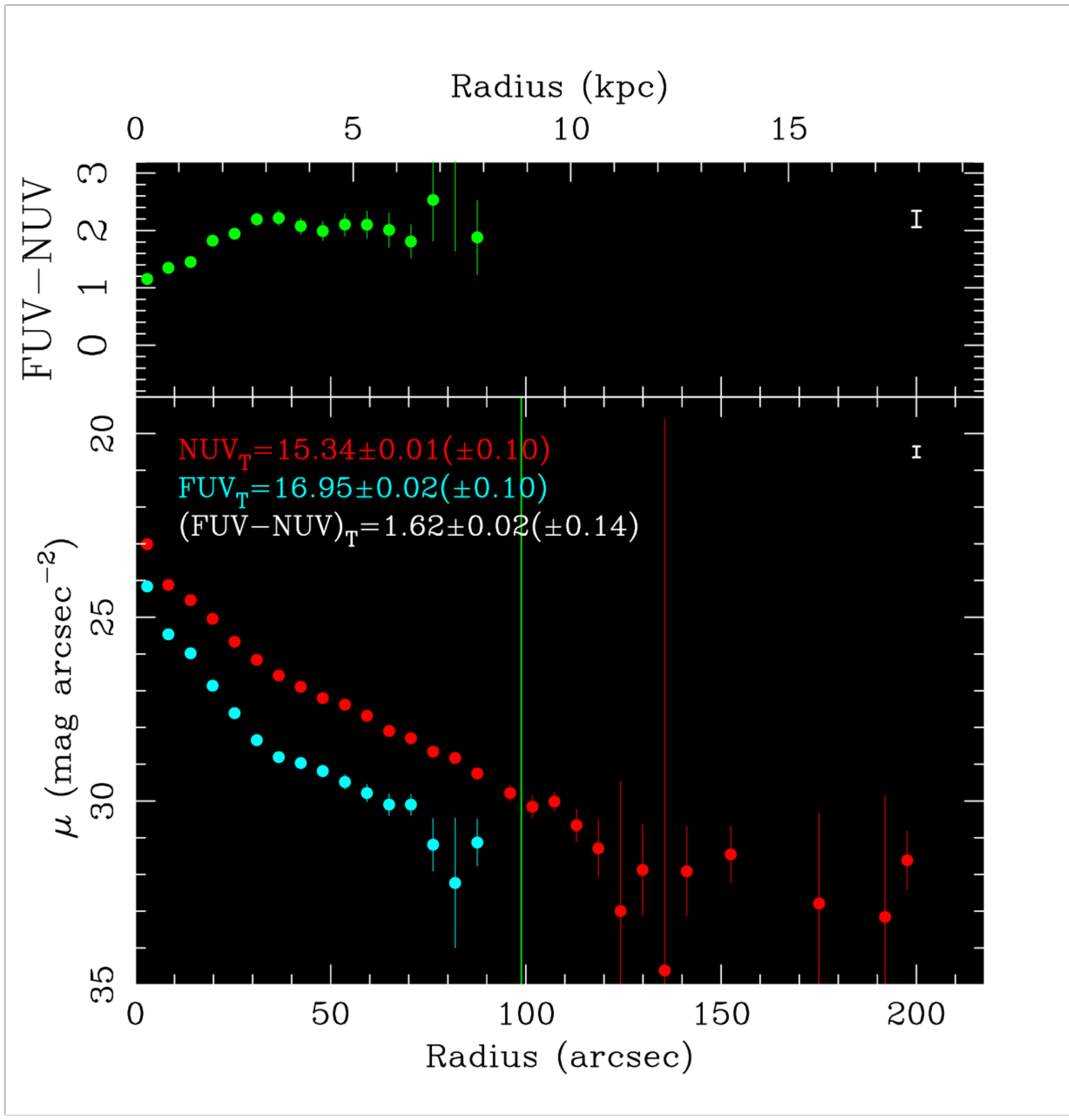}
\includegraphics[trim=2cm -0.7cm 4.5cm 1cm, clip=true, width=0.23\textwidth]{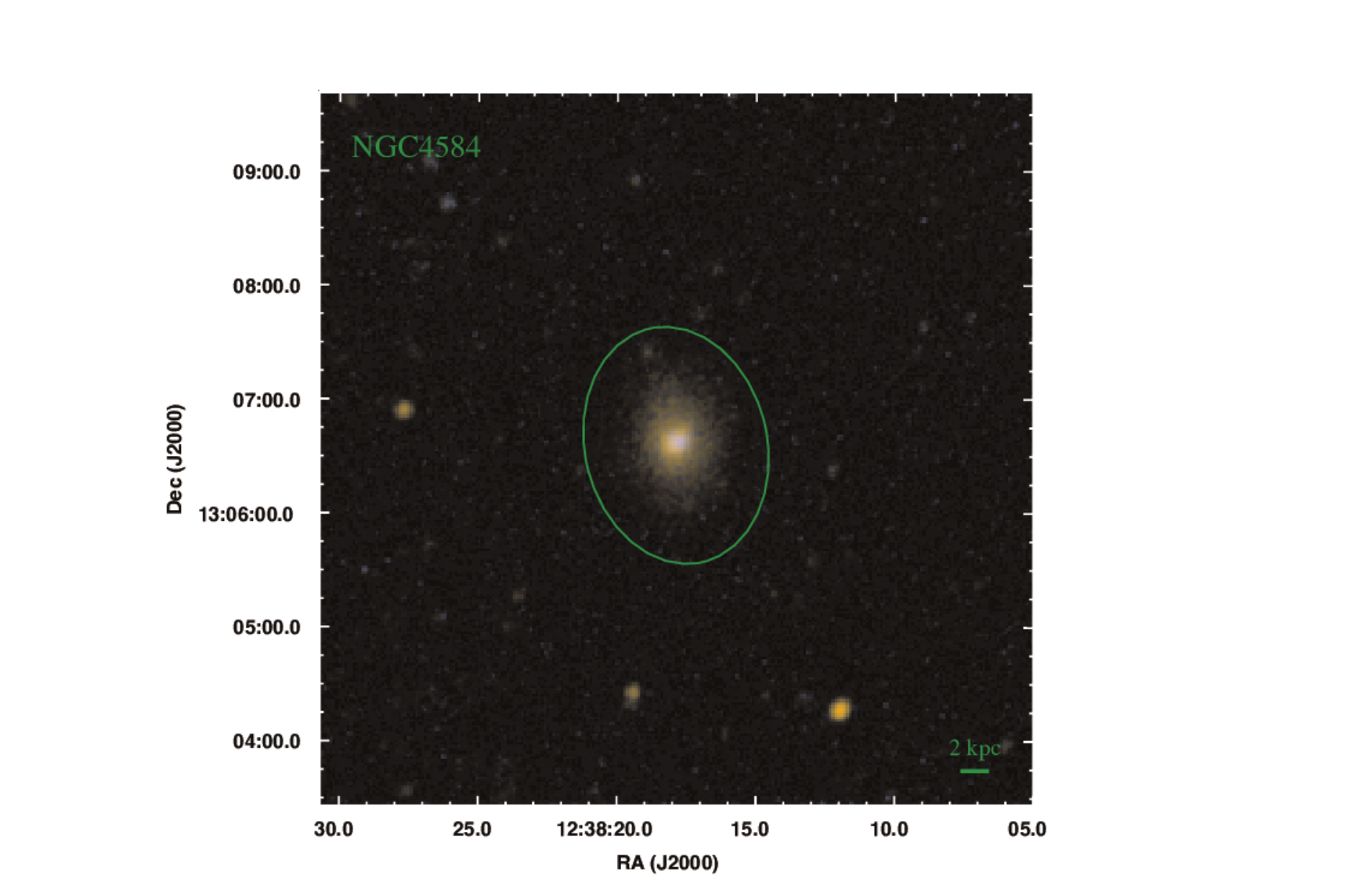}
\includegraphics[trim=1.0cm 2.5cm 0cm 2cm, clip=true, width=0.235\textwidth]{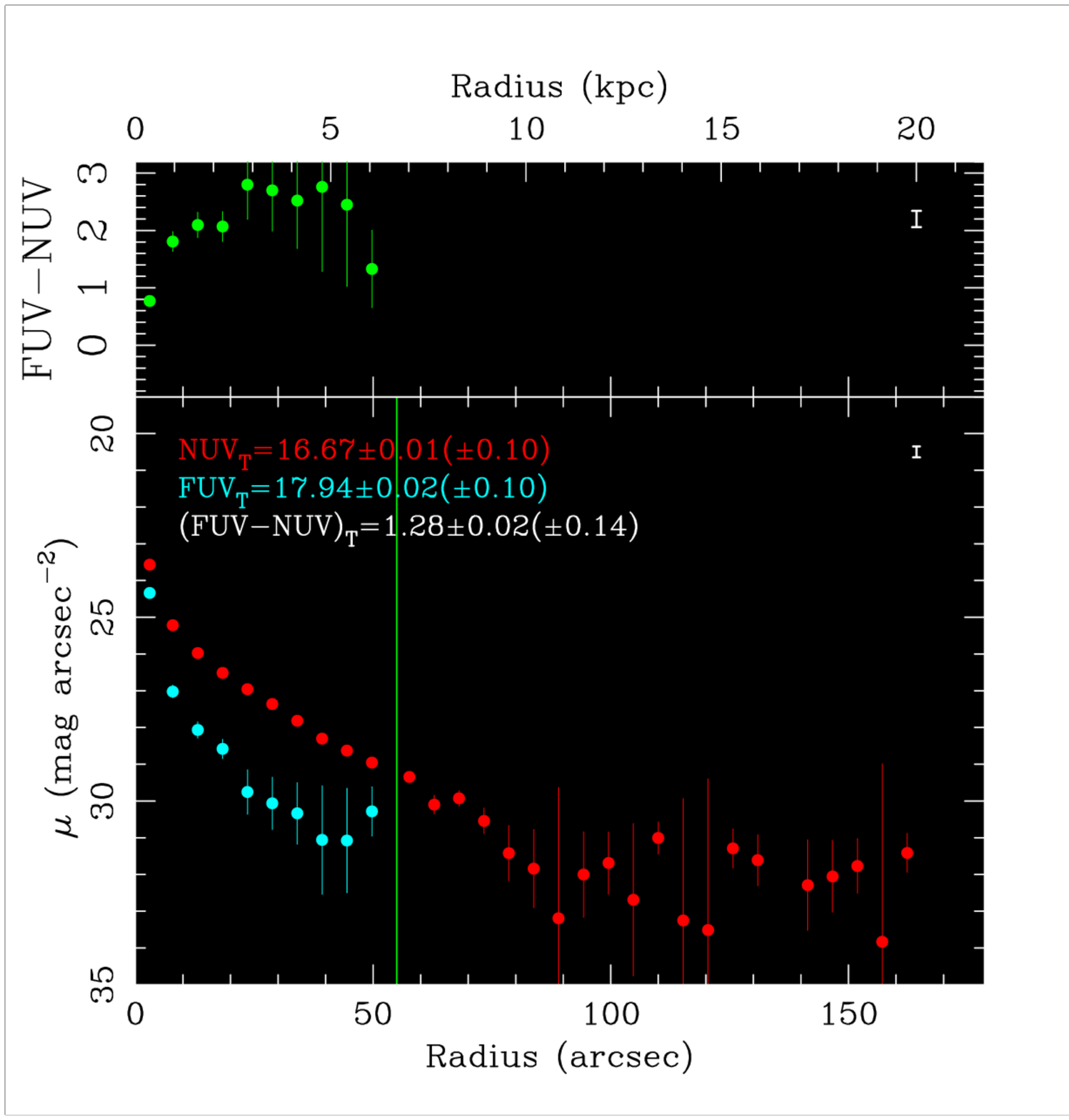}
}
\ovalbox
{
\includegraphics[trim=2cm -0.7cm 4.5cm 1cm, clip=true, width=0.23\textwidth]{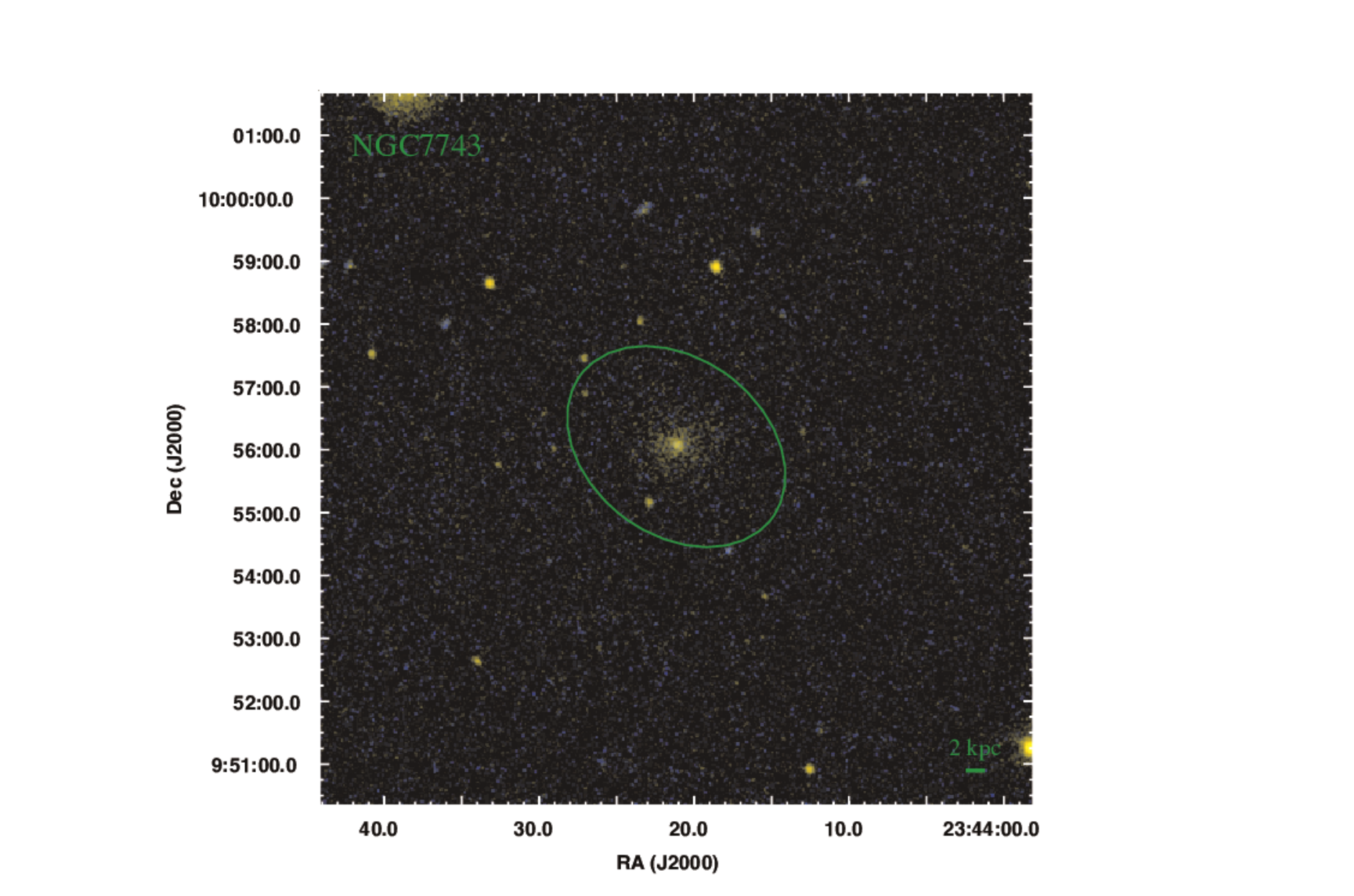}
\includegraphics[trim=1.0cm 2.5cm 0cm 2cm, clip=true, width=0.235\textwidth]{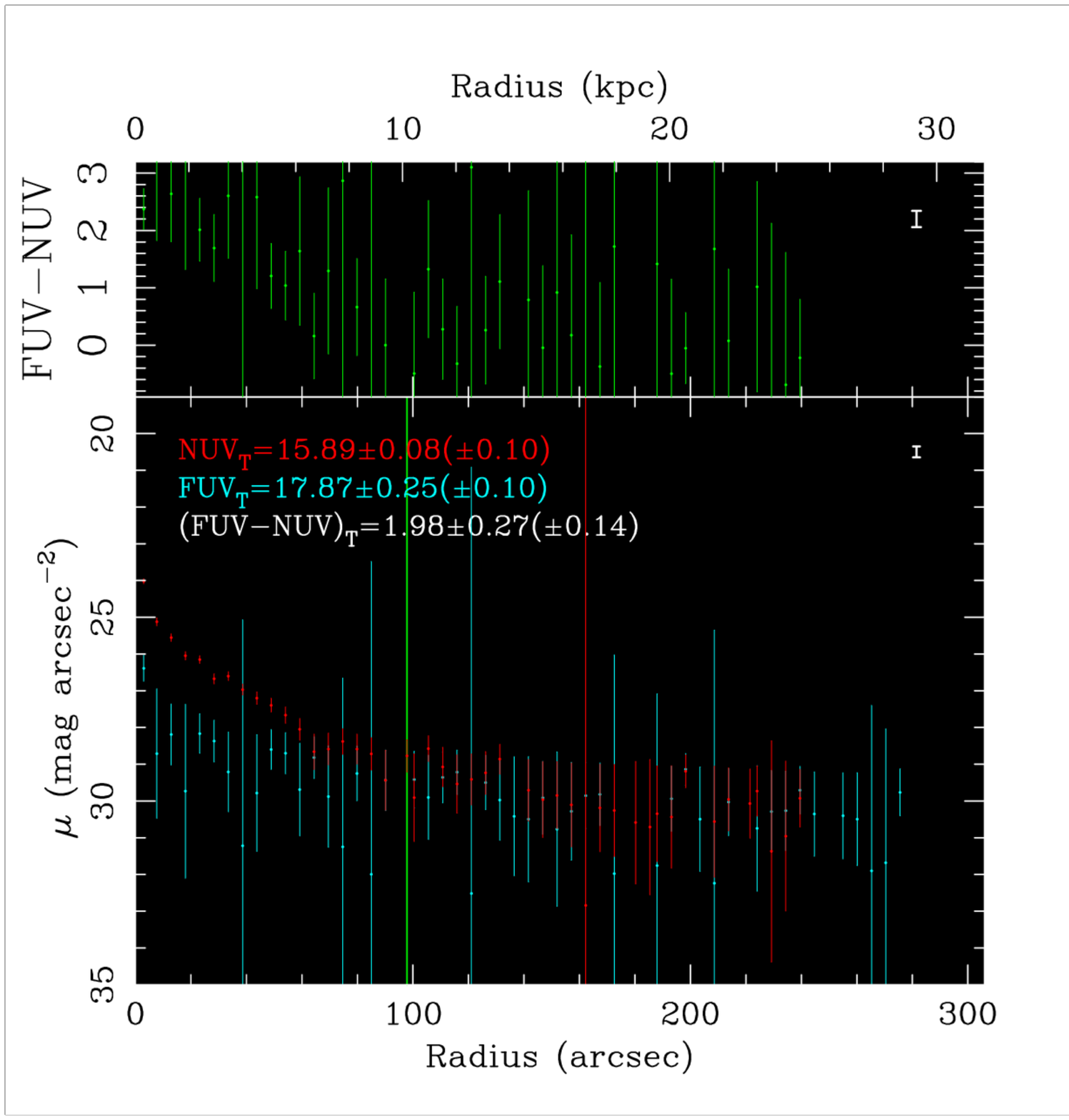}
\includegraphics[trim=2cm -0.7cm 4.5cm 1cm, clip=true, width=0.23\textwidth]{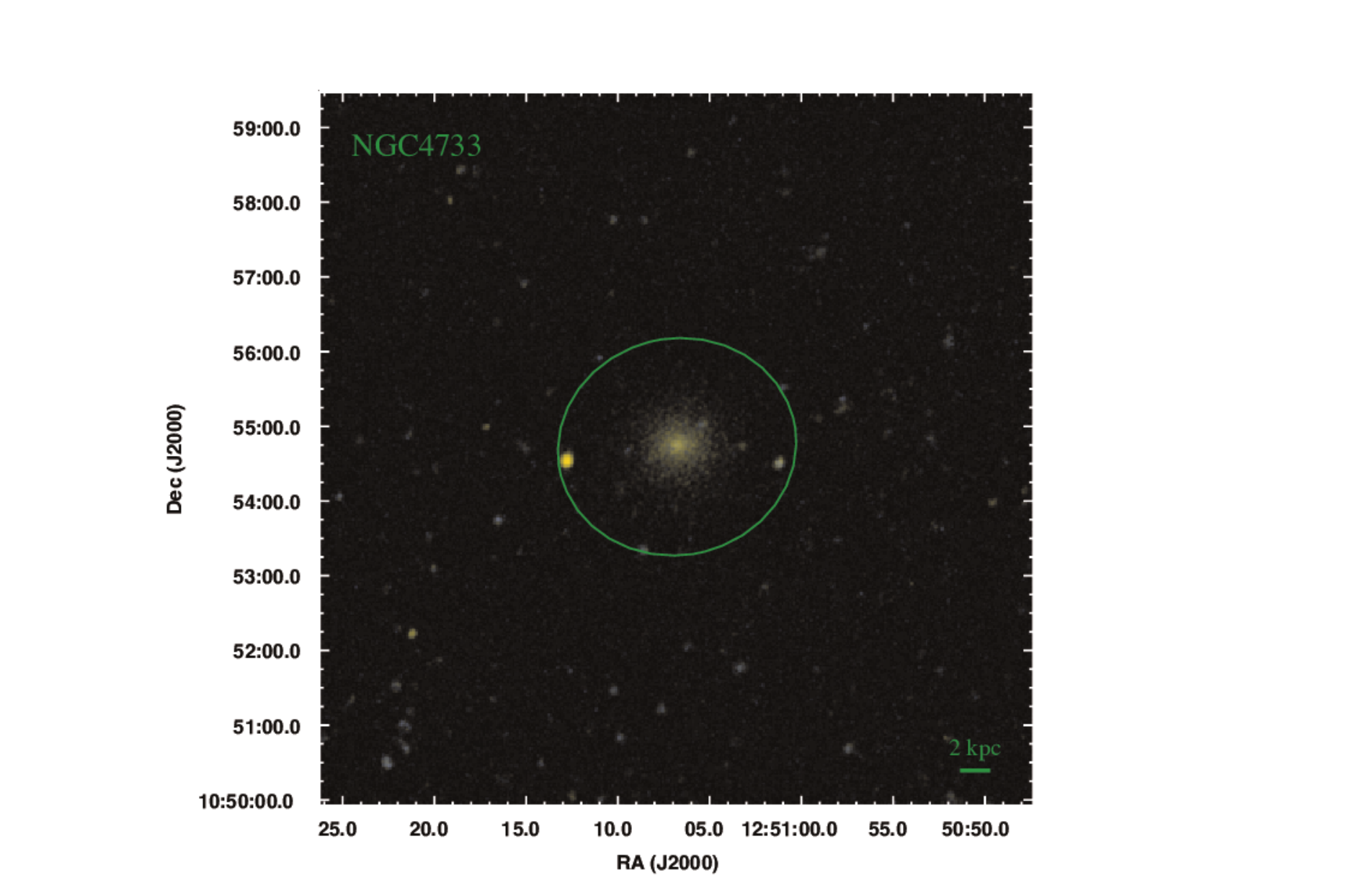}
\includegraphics[trim=1.0cm 2.5cm 0cm 2cm, clip=true, width=0.235\textwidth]{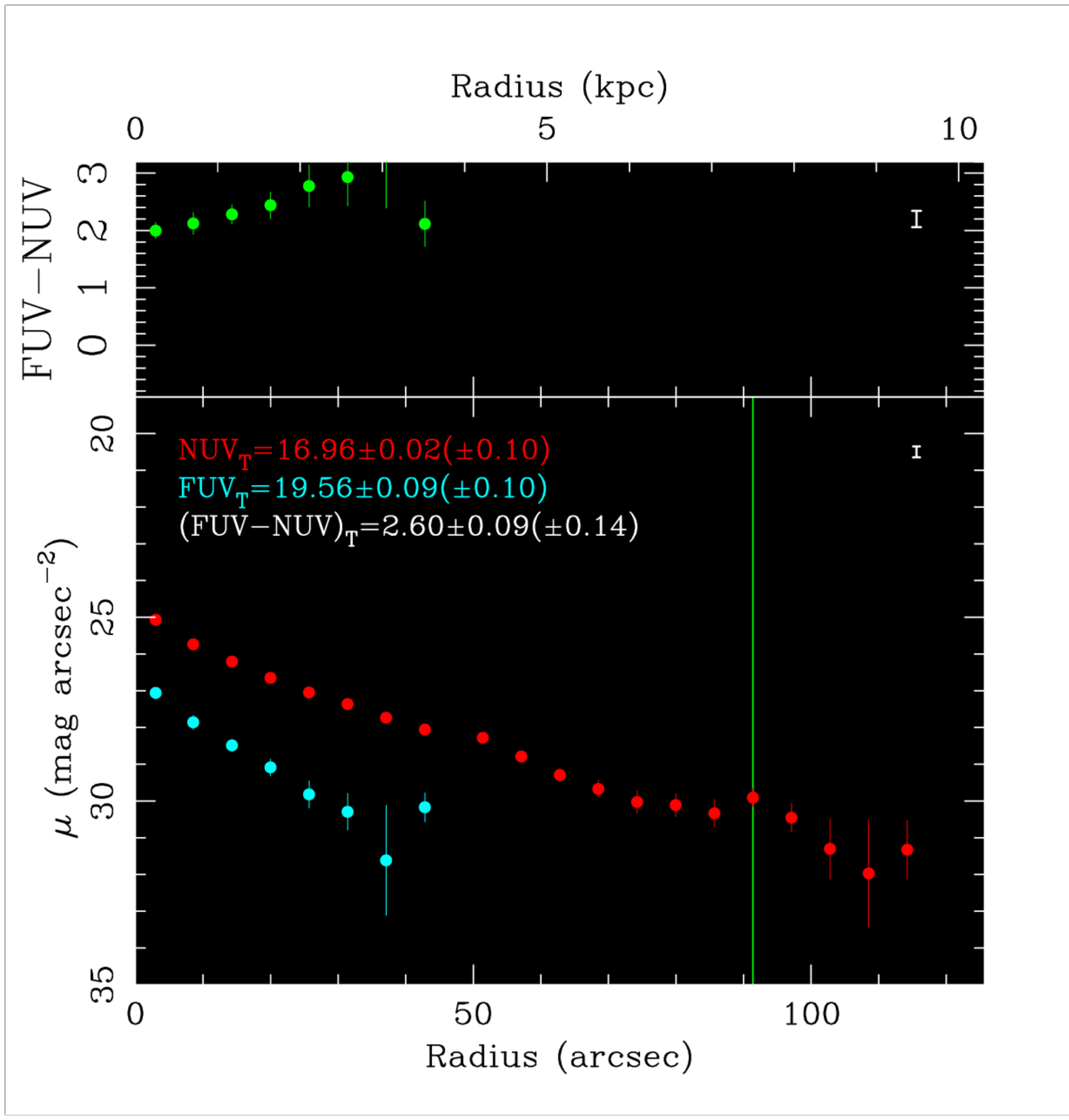}
}\caption{$\galex$ RGB postage stamp images generated from 
FUV and NUV images (left) with 
their respective surface brightness $\mu_{FUV}$, 
$\mu_{NUV}$, and (FUV\,$-$\,NUV) color profiles (right). 
The first row show typical Sc galaxies in the GBS,
the second row show typical Sa galaxies in the GGV,
and the third row show typical E galaxies in the GRS. 
The radial surface brightness profile 
(\textit{red dots} for NUV, \textit{blue dots} for FUV) as well as 
(FUV\,$-$\,NUV) (in mag\,arcsec$^{-2}$) radial color profile (\textit{green dots}), 
are shown. The \textit{green ellipse} in the RGB image corresponds to the isophotal contour D25 at 25 mag arcsec$^{-2}$ in B-band.
A 2\,kiloparsec scale is shown in the bottom-right corner of the image.
\label{fig:profiles}
}
\end{center}
\end{figure*}

In Figure~\ref{fig:profiles}, we show the FUV+NUV RGB postage stamp
images. The resulting products, shown also in this figure, include the surface brightness radial profiles in both FUV and NUV in mag\,arcsec$^{-2}$, 
(FUV\,$-$\,NUV) color profiles in mag\,arcsec$^{-2}$, and asymptotic magnitudes (in mag) for each galaxy.
The obtained values are corrected for
extinction due to the Milky Way. 
This foreground Galactic extinction was obtained following the UV extinction law of \citet{Cardelli1989}, 
assuming a total to selective extinction ratio $R_{V}=A_{V}/E(B-V)=3.1$, 
giving the attenuation values of $A_{\mathrm{FUV}}=7.9 E(B-V)$ and $A_{\mathrm{NUV}}=8.0 E(B-V)$, where the reddening $E(B-V)$ from 
Galactic dust is obtained from the map of \citet{Schlegel1998}.
The surface photometry of the sample is not corrected for internal dust attenuation nor inclination of the host galaxy.
A partial table including FUV and NUV surface photometry for 192 ETGs
was first released by \citet{Zaritsky2015a} and is also available in the VizieR online database \citep{Ochsenbein2000}.

In Table~\ref{Table:sbtable}, examples of the values we obtained are shown.
The graphical rendering of the data is shown in Figure~\ref{fig:SBprofileskpc} and is explained in the next subsection.

\subsection{IR profile radial normalization} \label{sec:IRSBprofile} 
\begin{figure*}
\begin{center}
	\includegraphics[width=0.49\textwidth]{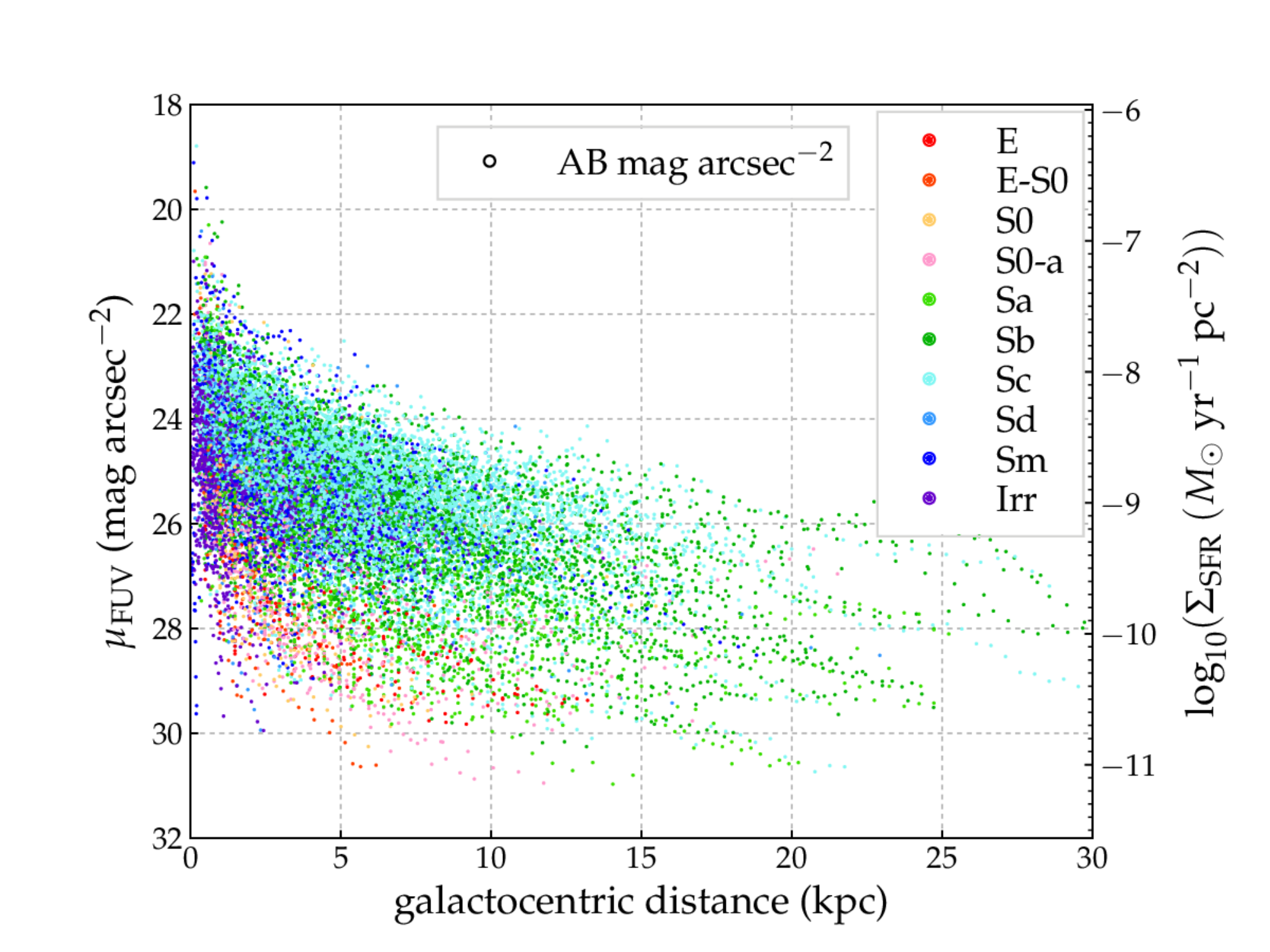}
	\includegraphics[width=0.49\textwidth]{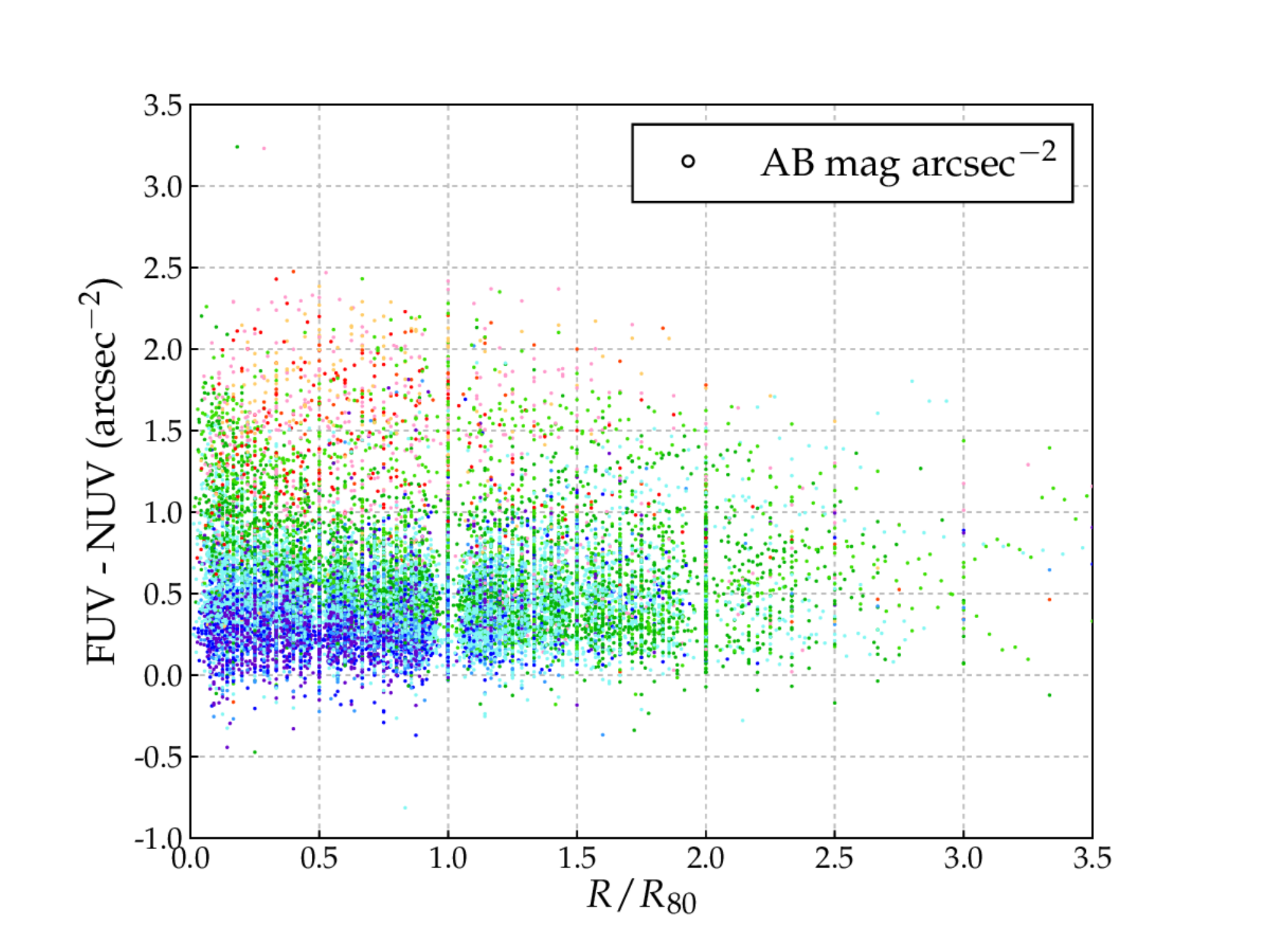}\\
        \includegraphics[width=0.49\textwidth]{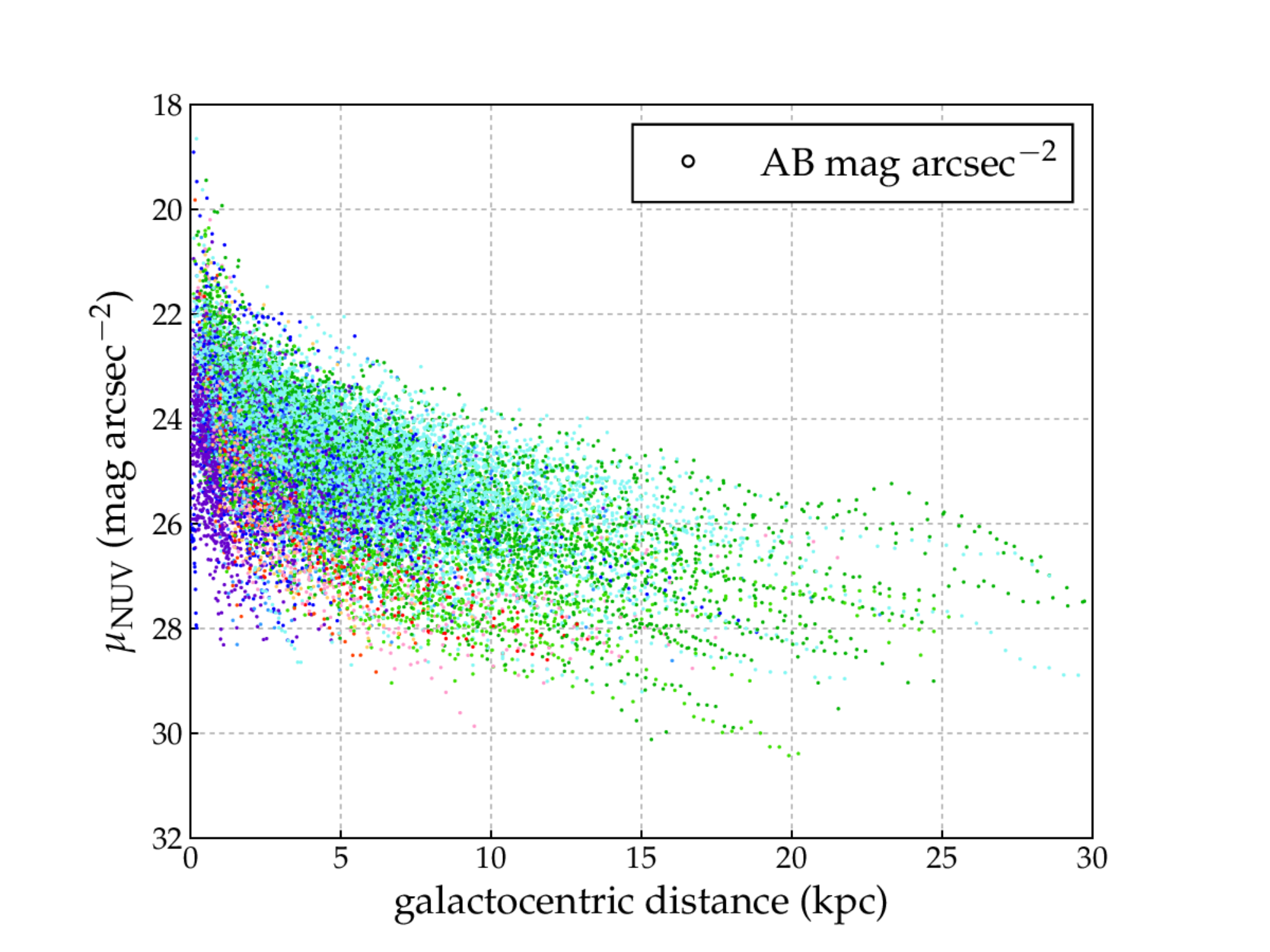}
	\includegraphics[width=0.49\textwidth]{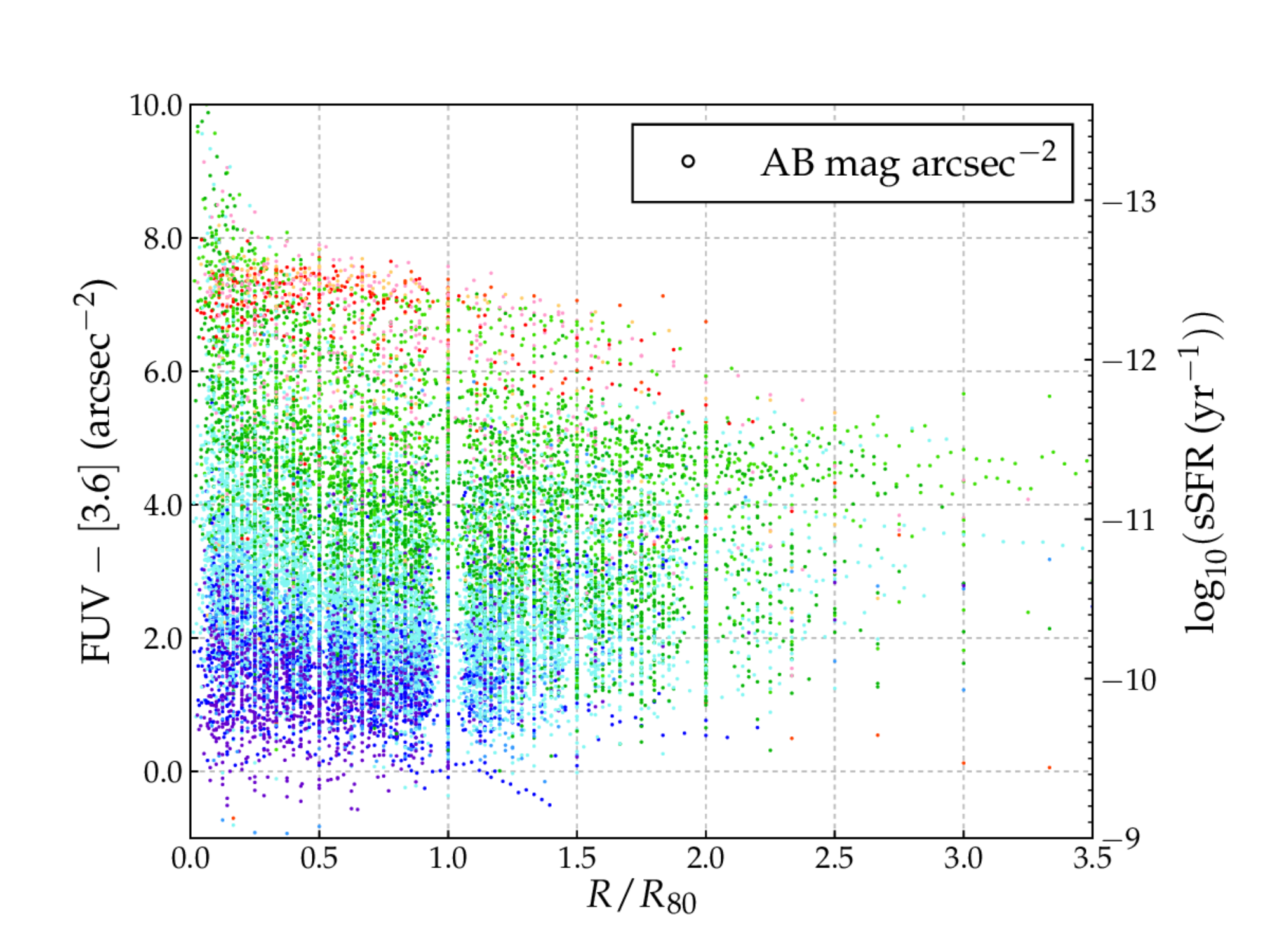}\\
	\includegraphics[width=0.49\textwidth]{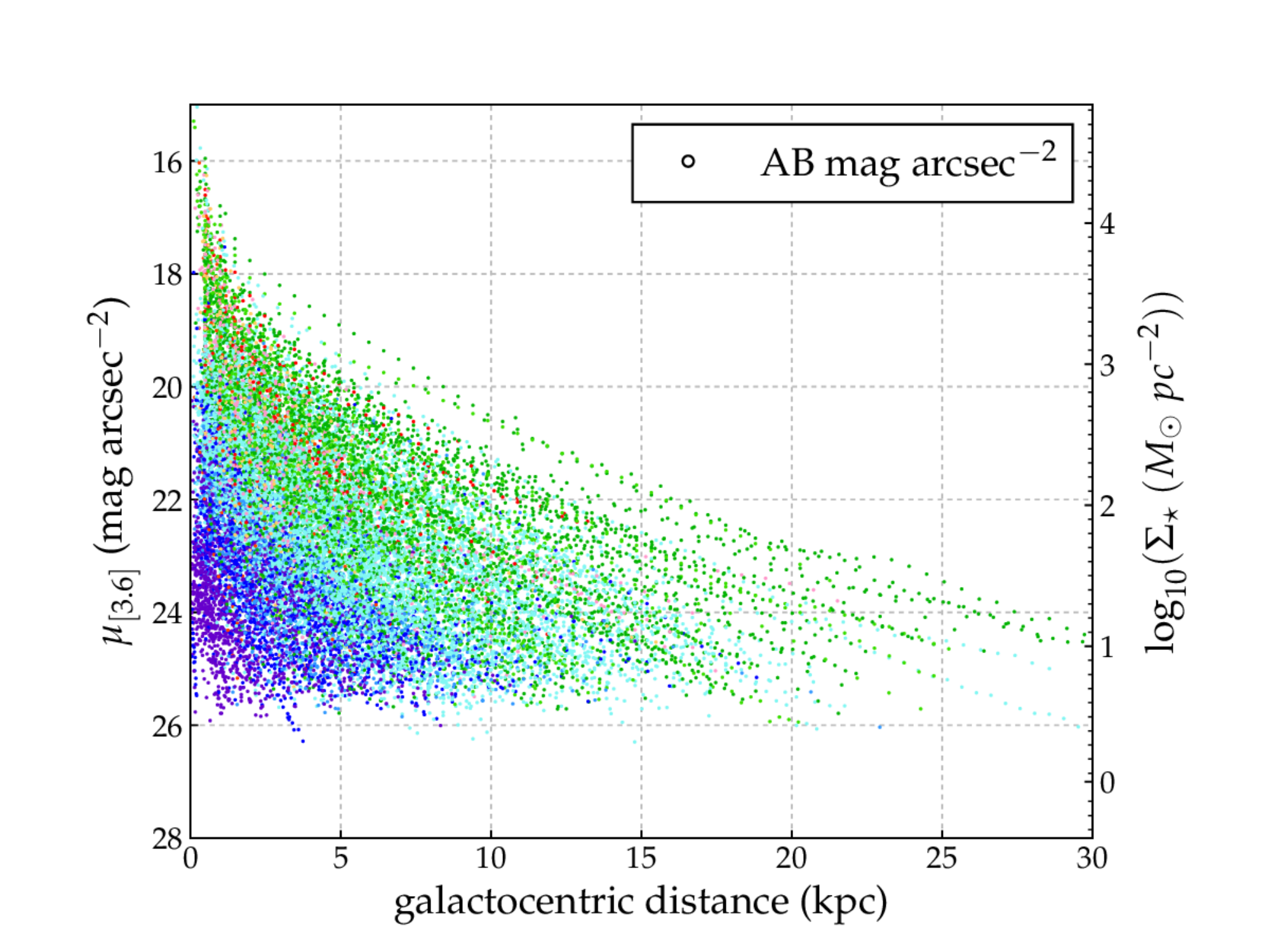}
	\includegraphics[width=0.49\textwidth]{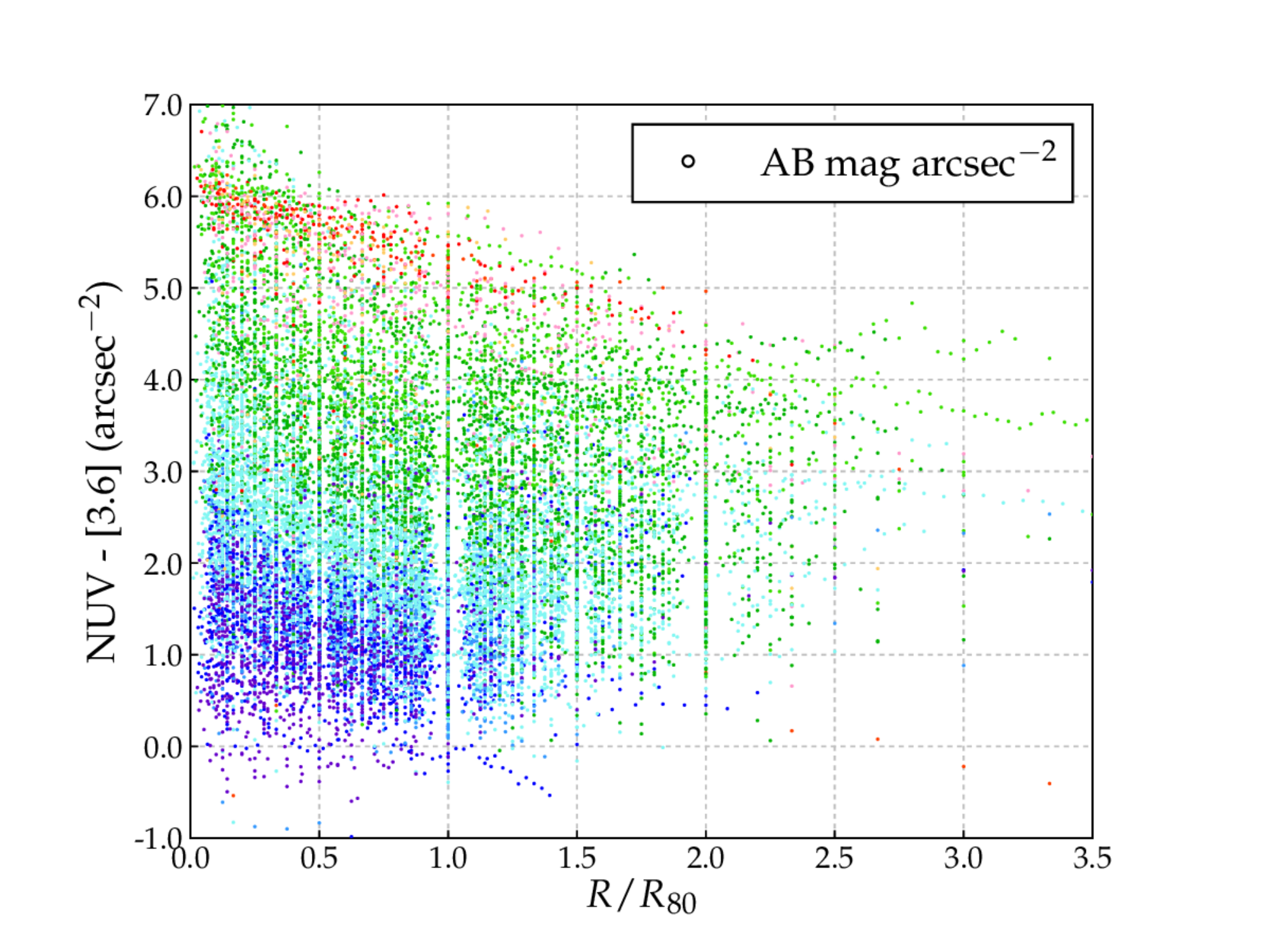}
        \caption{
        \textit{Left column, top to bottom} FUV, NUV, and [3.6] surface brightness versus radius in kiloparsec. \textit{Right column, top to bottom} (FUV\,$-$\,NUV), (FUV\,$-$\,[3.6]), and (NUV\,$-$\,[3.6]) colors vs $R/R80$. Each dot represents a data point. Our entire sample of 1931 galaxies is shown.
        Color-coding is based on the numerical morphological types and is the following:
E is red, E-S0 is orange, S0 is yellow, S0-a is pink, Sa is light-green, Sb is dark-green, Sc is cyan, Sd is light-blue, Sm is dark-blue, and Irr is purple.
The discretization seen in the right-hand plots is due to the fact that the $R80$ values derived from the analysis of our growth curves are obtained from the data point that encompasses a fraction of the light closest to 80\% but it is not interpolated. The figures show that this translates in an error of no more than $\pm$0.1\,$R/R80$.
\label{fig:SBprofileskpc}
        }
\end{center}
\end{figure*}

Figure~\ref{fig:SBprofileskpc} shows the FUV, NUV and 3.6\,$\mu$m surface brightness profiles $\mu_{\mathrm{FUV}}$, $\mu_{\mathrm{NUV}}$, and $\mu_{[3.6]}$
in units of mag arcsec$^{-2}$ plotted against the radius in kiloparsec in one case (left panels), and normalized in units of $R/R80$ in the other (right panels).

$R/R80$ is a distance unit that we devised based on the radius (i.e$.$, semi-major axis of ellipse) that 
encloses 80\% of the total 3.6\,$\mu$m light and that we call $R80$.
The innermost measurement is at 6$\arcsec$ semi-major axis radius, 
and the rest of the measurements radially outward in the disk are represented as small dots for each 6$\arcsec$ step.
The very center is excluded because it could be affected by differences in the PSF
amongst the three bands and by the contribution of an AGN.
The SB measurements are taken up to 3 $\times$ D25, however, for the analysis, we select only measurements having errors less than 0.2\,mag\,arcsec$^{-2}$.
These errors include the total measurement uncertainties, dominated by Poisson noise in the centers and by sky uncertainties in the outskirts, but exclude any systematic zero-point uncertainty.
Color-coding is based on the numerical morphological types and is the following:
E is red, E-S0 is orange, S0 is yellow, S0-a is pink, Sa is light-green, Sb is dark-green, Sc is cyan, Sd is light-blue, Sm is dark-blue, and Irr is purple.
Numerical morphological types were obtained from HyperLeda \citep{Makarov2014} and follow the RC2 classification scheme: 
-5\,$\leq$\,E\,$\leq$\,-3.5, 
-3.5\,$<$\,E-S0\,$\leq$\,-2.5, 
-2.5\,$<$\,S0\,$\leq$\,-1.5, 
-1.5\,$<$\,S0-a\,$\leq$\,0.5, 
0.5\,$<$\,Sa\,$\leq$\,2.5, 
2.5\,$<$\,Sb\,$\leq$\,4.5, 
4.5\,$<$\,Sc\,$\leq$\,7.5, 
7.5\,$<$\,Sd\,$\leq$\,8.5, 
8.5\,$<$\,Sm\,$\leq$\,9.5, and 
9.5\,$<$\,Irr\,$\leq$\,999.
Galaxies with unknown morphological type are assigned the numerical type 999, and are included in the irregular galaxies (Irr) bin,
as these are, in the vast majority of the cases, systems with ill-defined morphology. 

\subsection{Color profiles} \label{sec:colorprofiles}
The right column of Figure~\ref{fig:SBprofileskpc} shows each galaxy's spatially resolved radial
color profiles in 
(FUV\,$-$\,NUV), (FUV\,$-$\,[3.6]) and (NUV\,$-$\,[3.6]) 
as a function of galactocentric distance both in kpc and $R/R80$ units. 
Each plot shows the corresponding color profile distribution for each galaxy, color-coded by morphological type. 
As mentioned above, measurements are taken every 6$\arcsec$ from the center of each galaxy, 
and each profile reaches the galactocentric distance where the error in either FUV, NUV or
3.6\,$\mu$m surface brightness becomes 0.2\,mag\,arcsec$^{-2}$ or larger, 
thus rejecting the data that follow. It should be noted that measurements are available up to 3 $\times$ D25,
but are more dominated by sky uncertainties as we move radially outward.

\begin{figure*}
\begin{center}
	\includegraphics[width=0.49\textwidth]{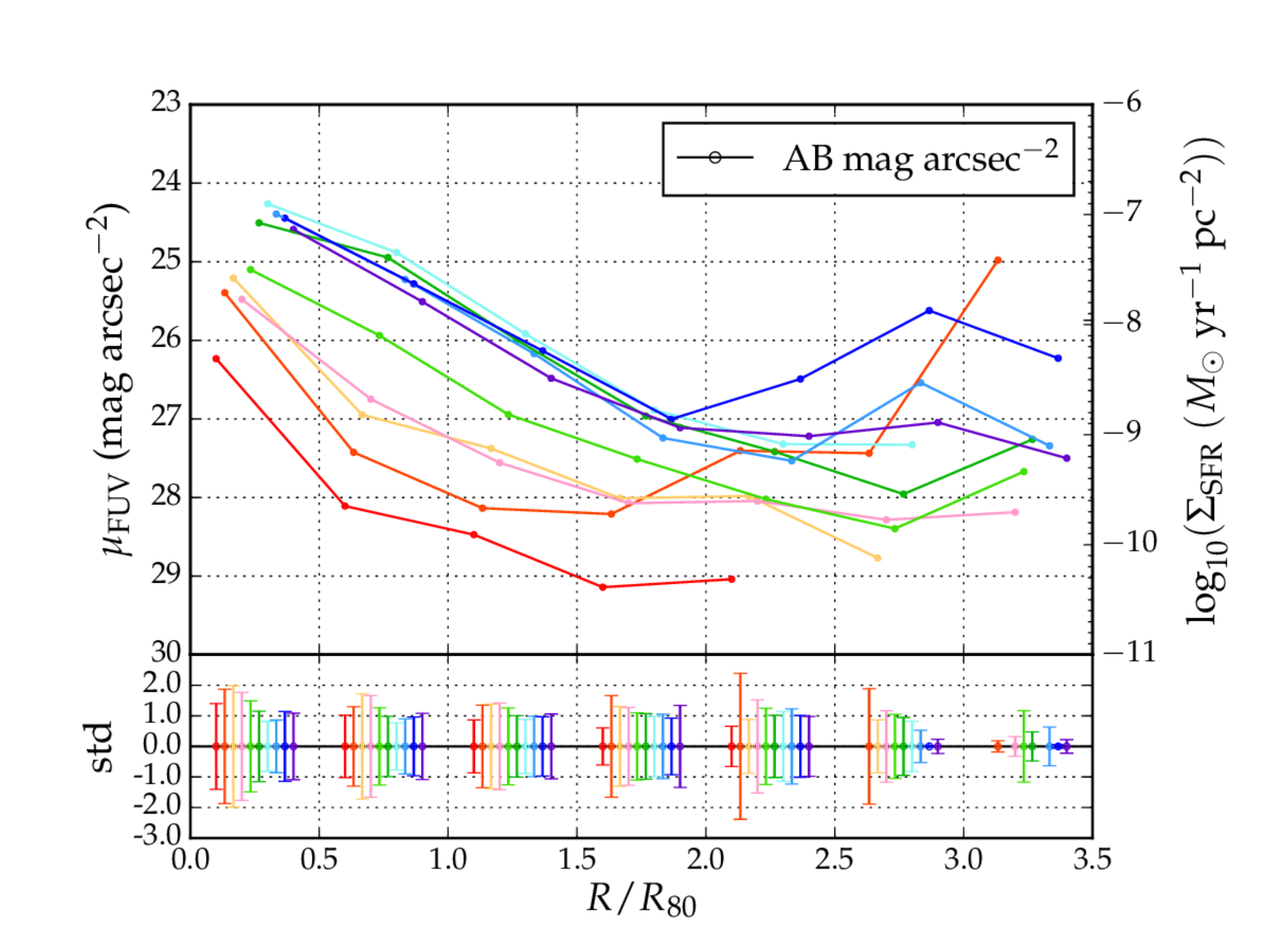}
	\includegraphics[width=0.49\textwidth]{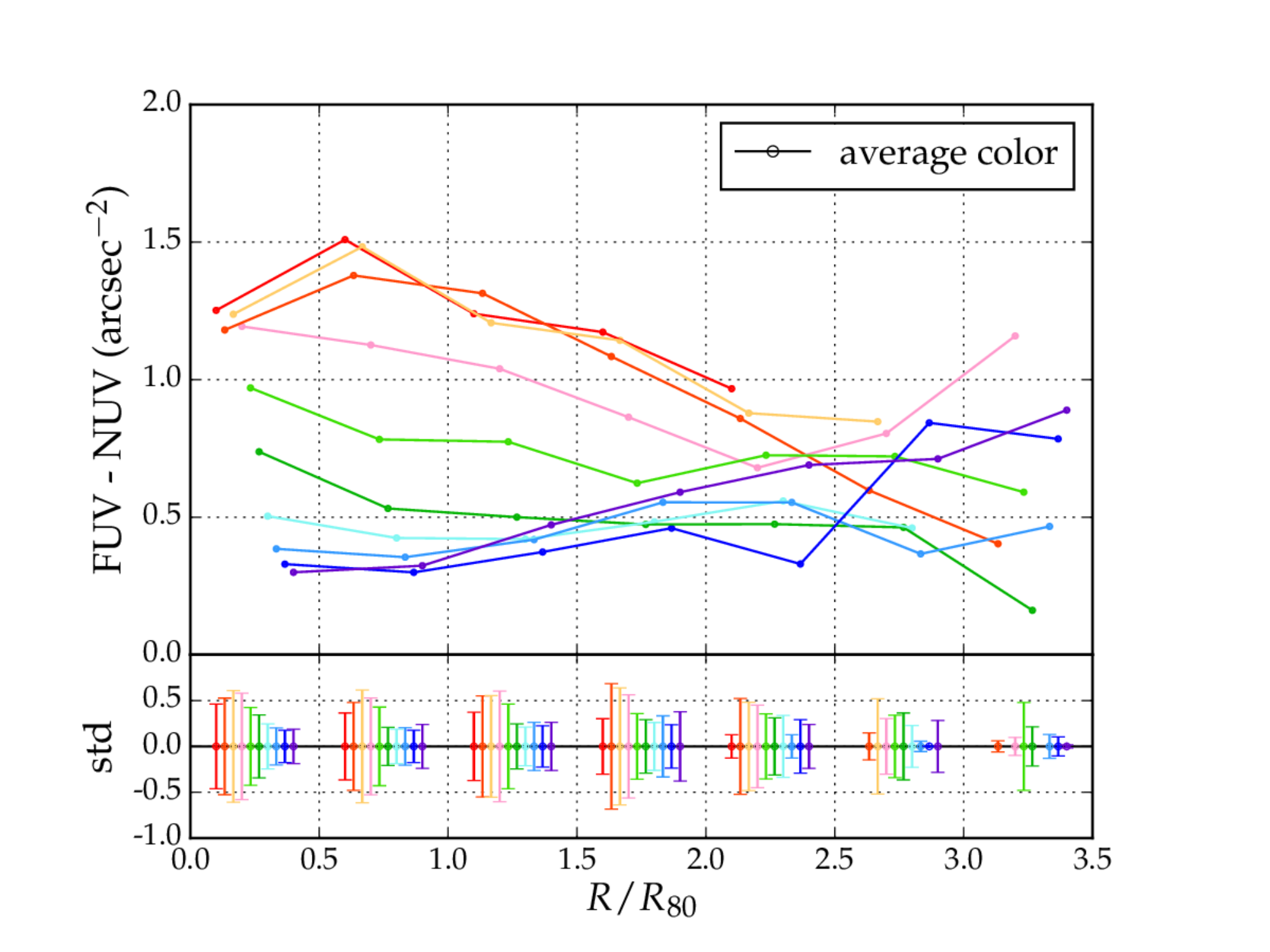}\\
	\includegraphics[width=0.49\textwidth]{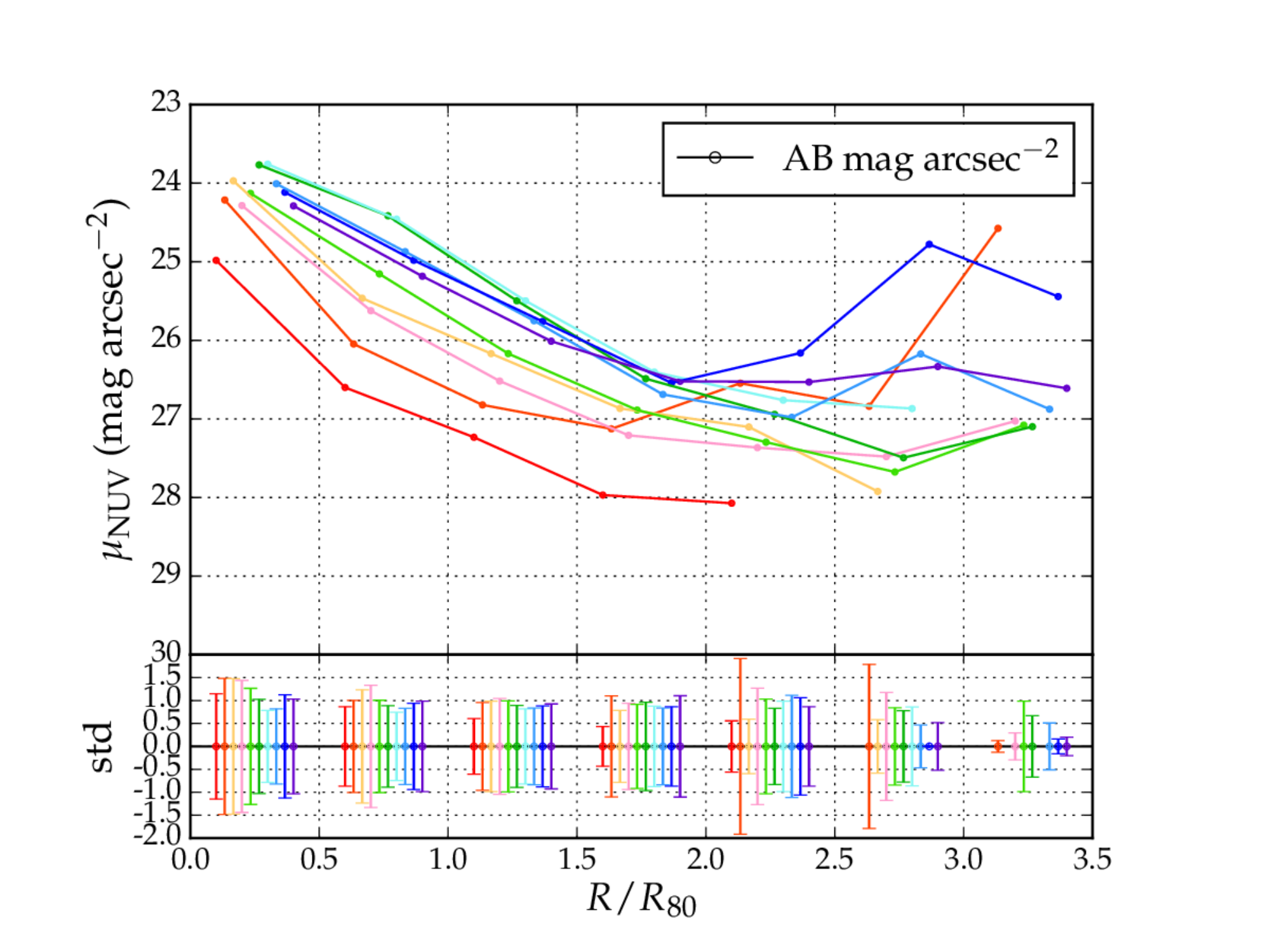}
	\includegraphics[width=0.49\textwidth]{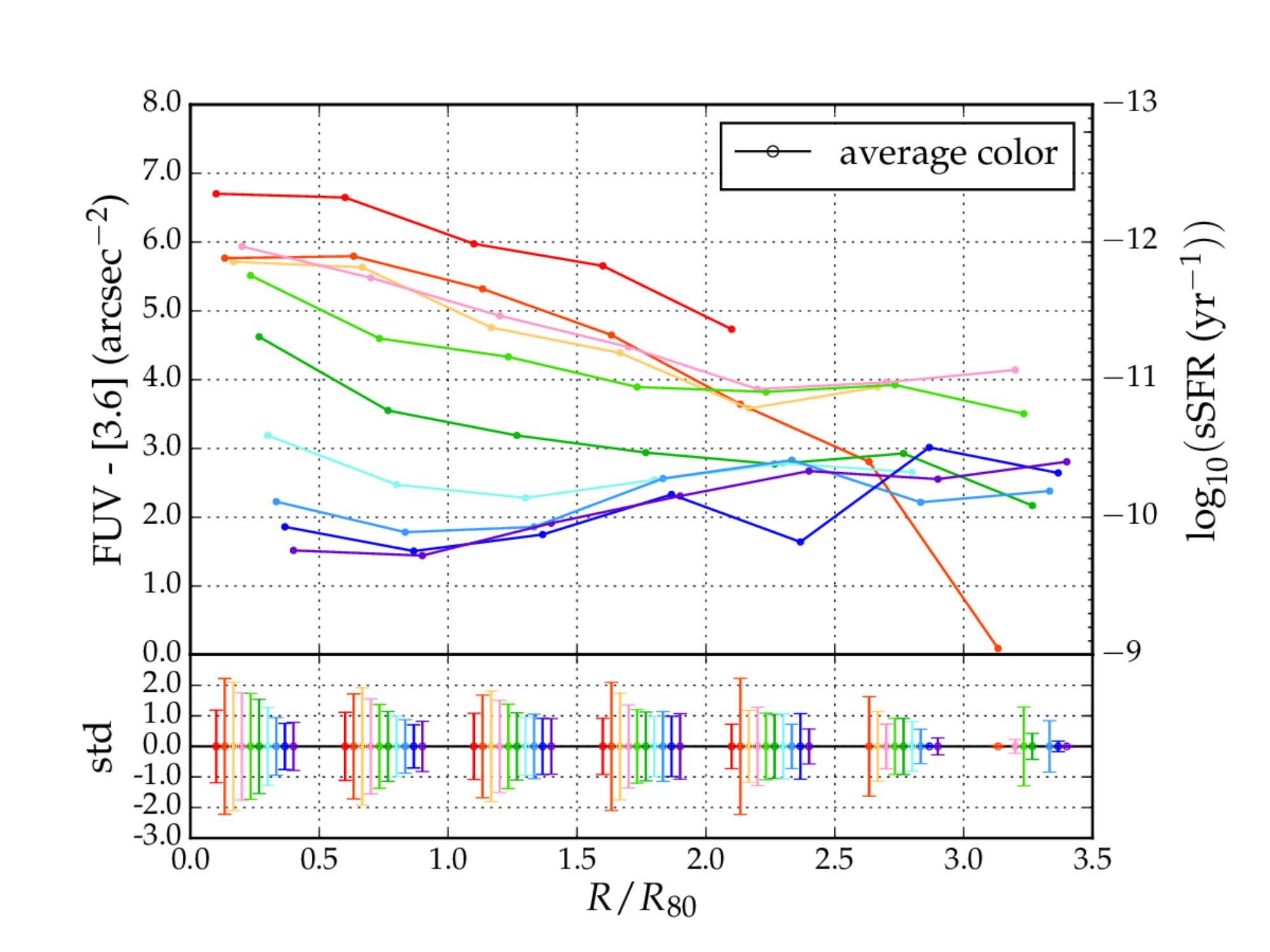}\\
	\includegraphics[width=0.49\textwidth]{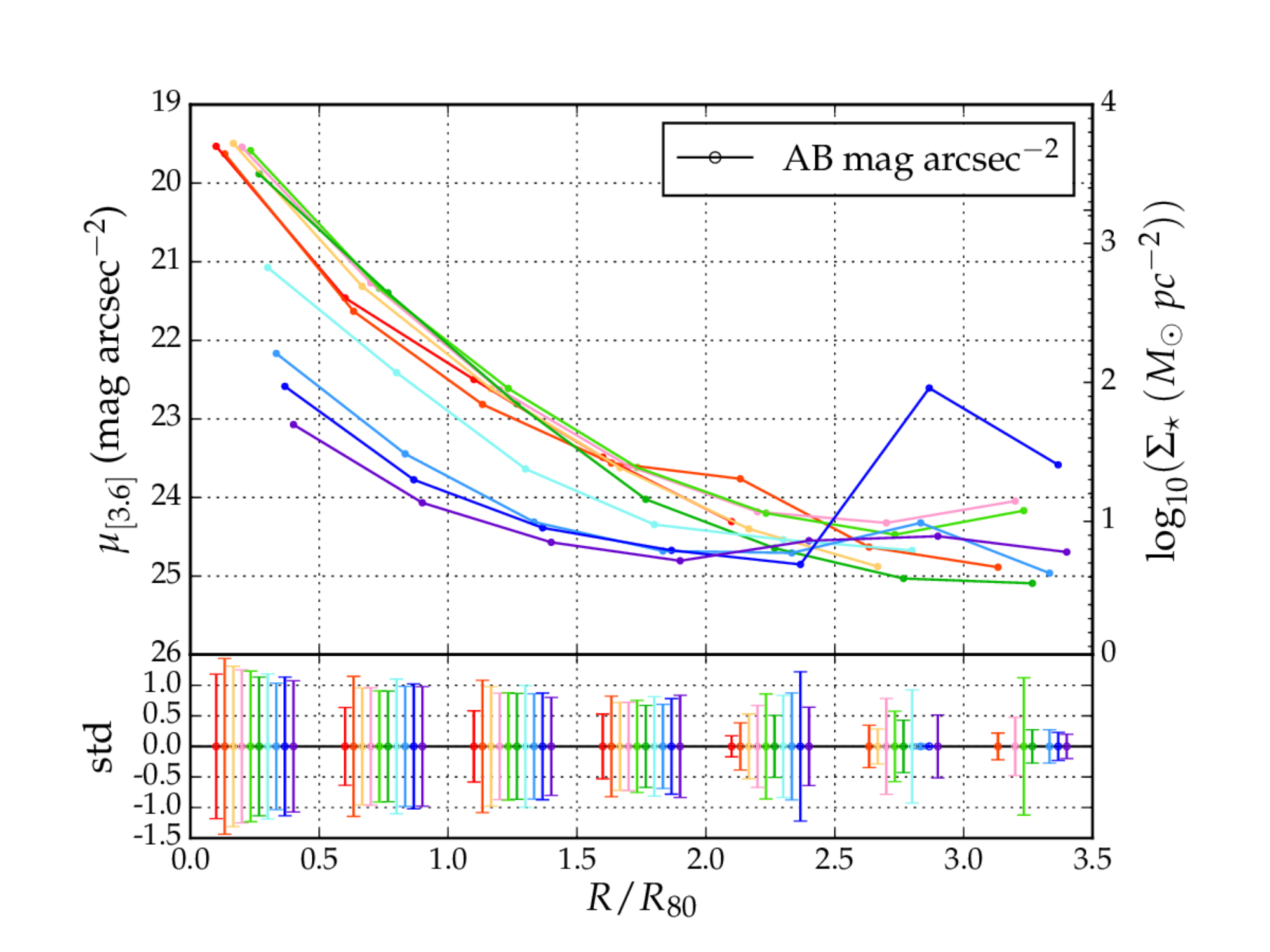}
	\includegraphics[width=0.49\textwidth]{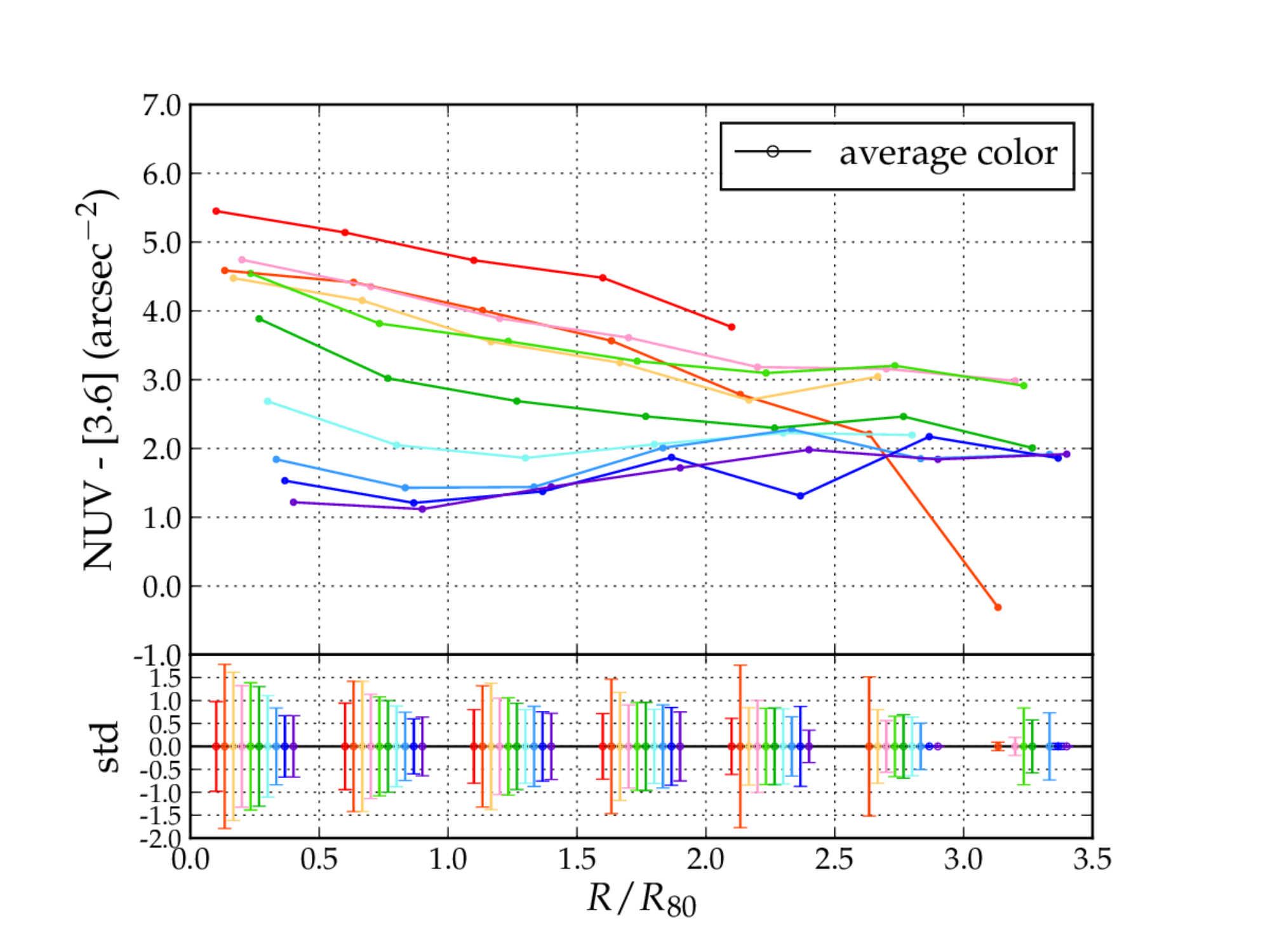}
	\caption{\textit{Left column}: \textit{top panel}: Average surface brightness color-coded per morphological type per $R/R80$ bin of width 0.5. \textit{Bottom panel}: the standard deviation (std) of the scatter from the mean, including the uncertainty, within each bin.
	A translation in x is applied for better visibility. It should be noted that the sample size substantially drops beyond $R/R80 >1.5$ due to the observation limits in each band.
	\textit{Right column}: The same but for color profiles.
	\label{fig:average} 
	}
\end{center}
\end{figure*}

Figure~\ref{fig:average} shows the average surface brightnesses and colors per $R/R80$ bin of width 0.5, as well as the range
of the scatter from the mean value in each bin, and per morphological type.
It should be noted that the range appears to diminish as we move radially outward, 
but this is due to reaching the observation limits in each band.

\begin{figure*}[ht!]
\begin{center}
	\includegraphics[width=0.95\textwidth]{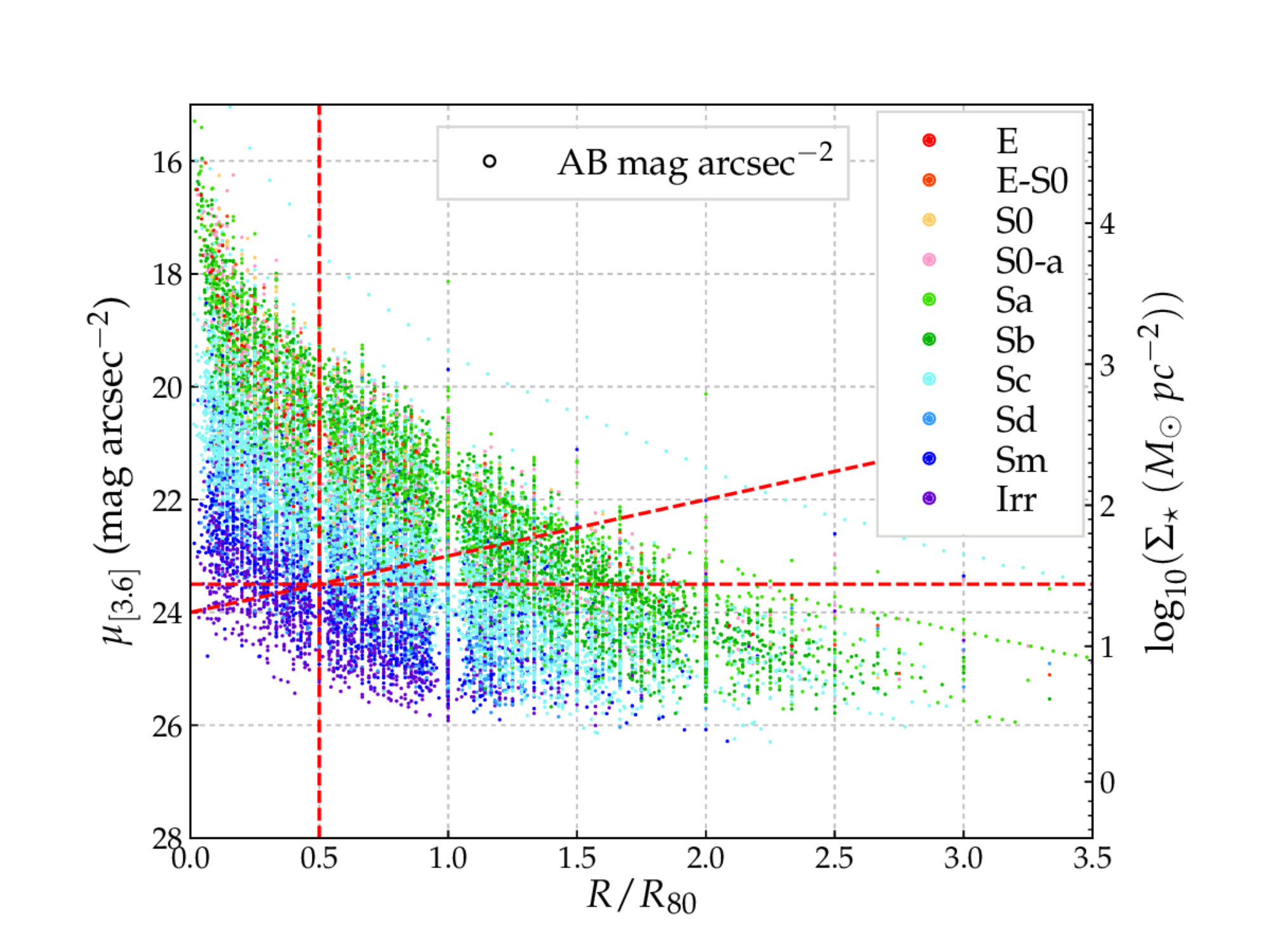}
        \caption{3.6 $\mu$m surface brightness profiles normalized to $R80$. The red dashed lines represent one example of the vertical and horizontal cutoffs, and another example of an oblique cutoff.
\label{fig:mu36SBprofilesR80}
        }
\end{center}
\end{figure*}
\begin{table}
\caption[Table~caption text]{Surface brightness data (examples)}
\label{Table:sbtable}
\begin{center}
\begin{tabular}{ c c c c c }
\hline
  \multicolumn{1}{c}{Name} &
  \multicolumn{1}{c}{r} &
  \multicolumn{1}{c}{$\mu_{FUV}$} &
  \multicolumn{1}{c}{$\mu_{NUV}$} &
  \multicolumn{1}{c}{$\mu_{[3.6]}$} \\
\hline
& ($\arcsec$) & mag/($\arcsec$)$^{2}$ & mag/($\arcsec$)$^{2}$ & mag/($\arcsec$)$^{2}$ \\
\hline
UGC00017 & 6 & 26.14$\pm$0.09 & 25.76$\pm$0.05 & 23.16$\pm$0.03\\
 & 12 & 26.32$\pm$0.10 & 25.86$\pm$0.05 & 23.60$\pm$0.05\\
 & 18 & 25.96$\pm$0.05 & 25.83$\pm$0.03 & 24.01$\pm$0.07\\
 & 24 & 26.47$\pm$0.05 & 26.23$\pm$0.03 & 24.49$\pm$0.11\\
 & 30 & 26.65$\pm$0.05 & 26.38$\pm$0.03 & 24.80$\pm$0.14\\
 & 36 & 26.91$\pm$0.06 & 26.63$\pm$0.03 & 24.99$\pm$0.17\\
 & 42 & 26.66$\pm$0.05 & 26.49$\pm$0.03 & 25.27$\pm$0.21\\
 & ... & ... & ... & ...\\
\hline
ESO409-015 & 6 & 21.92$\pm$0.01 & 21.89$\pm$0.01 & 22.72$\pm$0.02\\
 & 12 & 23.39$\pm$0.02 & 23.23$\pm$0.01 & 23.50$\pm$0.04\\
 & 18 & 24.85$\pm$0.03 & 24.50$\pm$0.02 & 24.24$\pm$0.07\\
 & 24 & 25.95$\pm$0.05 & 25.52$\pm$0.03 & 24.98$\pm$0.14\\
 & 30 & 26.93$\pm$0.07 & 26.22$\pm$0.03 & 25.08$\pm$0.15\\
 & 36 & 27.63$\pm$0.09 & 26.87$\pm$0.05 & 25.62$\pm$0.24\\
 & 42 & 28.04$\pm$0.11 & 27.34$\pm$0.06 & 25.85$\pm$0.29\\
  & ... & ... & ... & ...\\
 \hline
\end{tabular}
\end{center}
\end{table}
%

\section{RESULTS} \label{sec:results}
The FUV and NUV are most sensitive to the presence and amount of (recently born) massive stars and, in particular, the FUV can be directly linked (modulo IMF) to the (observed) SFR, at least for late-type galaxies. 
In our preliminary work \citep{Bouquin2015}, we have seen that the majority of star-forming disk galaxies in our sample are distributed along the GBS, however, there exists some disk galaxies with redder, integrated, (FUV\,$-$\,NUV) color that are located in the GGV.  Spatially resolved color profiles allow us to see which parts of the galaxy are actually forming stars or not. Note that the (FUV\,$-$\,NUV) color is quite reddening-free (but not extinction-free) for MW-like foreground dust and that even if that is not the case, the effect of dust in disks, especially in its outskirts, is smaller than that found between GBS and GRS galaxies \citep{MunozMateos2007}.

In order to study in more detail the disk component of a galaxy, we first need to isolate it by separating it from the bulge component.
However, galaxies come in different shapes and sizes: some galaxies are bulgeless and only have a disk, whereas some others are diskless and only have a massive spheroidal component. We devised a method to isolate the disk component only from the 3.6\,$\mu$m SB profiles, by applying a radial cutoff and a SB cutoff and finding the best linear fit to the outer parts of these NIR profiles (Section~\ref{sec:diskseparation}).
This method allows us, regardless of the morphological type, to isolate the disk component and to obtain its scale-length and central surface brightness from the slope and y-intercept of the linear fit. 
With the spatially resolved photometry, we are able to construct a so-called star-forming main sequence, relating the FUV SB, $\mu_{\mathrm{FUV}}$, to SFR surface density and the 3.6\,$\mu$m SB, $\mu_{[3.6]}$, to surface stellar mass density (Section~\ref{sec:sfms}).
The sSFR can be directly obtained from the (FUV\,$-$\,[3.6]) color (Section~\ref{sec:colorandssfr}).
We also explore the color-color diagrams obtained from these bands (Section~\ref{sec:CCD}).
We show how the disks of GGV galaxies are also different from those of other galaxies (Section~\ref{sec:GGV}).

\subsection{Disk separation using near-IR SB profiles.} \label{sec:diskseparation}
Disks are known to have an exponential profile and are therefore close
to a straight line in a surface brightness (a logarithm) versus
galactocentric radius plot, at least in their inner regions. In the
very outer regions, these single exponential profiles commonly bend
\citep[see][and references therein]{Marino2016}. 
It should be noted in this context, however, that the level of either down- or up-bending in
the surface brightness profiles of galaxy disks is usually minimized
at near-infrared wavelengths \citep[e.g.][]{MunozMateos2011} \citep[see also][for a comparison of these bending profiles at different wavelengths and in stellar mass]{Bakos2008}.

In order to isolate the disk component in a coherent and reproducible way among all our
1884 disk galaxies (S0 and beyond) and
to derive their multiwavelength properties,
we have made use of the 3.6\,$\mu$m
surface brightness profiles of our sample and performed an
error-weighted fit to our data points in $\mu_{[3.6]}$ versus galactocentric radius in kpc.
Prior to this fitting, the surface brightnesses were 
corrected for geometrical inclination effects by adding $- 2.5 \log_{10}(b/a)$ (mag arcsec$^{-2}$), 
where $a$ and $b$ are the semi-major and semi-minor axes in the B-band, to each data point. 
No internal dust attenuation correction is applied.
This has the effect of dimming the surface brightness for inclined systems \citep{Graham2008}. 
See Section~\ref{sec:modeling} on how this inclination correction affects the comparison with the models.
Then, we identified the position
beyond which the profile starts to be best described by an
exponential law at these wavelengths. In order to exclude the bulge
(i.e$.$ either the region where the S\'ersic index is significantly larger
than unity or the steepening associated to a pseudo-bulge) and given that we
have in hand $R80$ measurements (major-axis radius where 80\% of the IR
light is enclosed) for the entire sample 
we remove the inner part of the profile up to some factor of $R80$ to
perform different sets of fits. For this analysis we explored $R/R80$
cutoff factors of 0, 0.25, 0.50, 0.75, 1.00,
and 1.25 and evaluated how far we should go from the galaxy center in
each case to
have good linear fits as given by the corresponding sample-averaged reduced
$\chi^{2}$ values (see below). We combined this inner cutoff in $R/R80$ with cutoffs in surface
brightness magnitude in the range
$\mu_{[3.6]}$=21.5 $\sim$ 24\,mag\,arcsec$^{-2}$, 
so only points fainter than
the corresponding cutoff would be considered for the fit. 

The rationale for using a combination of the two parameters is 
that we should normalize to the size of the objects to 
(1) do a first-order separation between bulges and disks and 
(2) take into account the fact that early-type systems usually have large, 
massive bulges with brighter near-infrared surface brightnesses than the disks of late-type spirals.
Thus, when we cut in surface brightness we exclude larger regions in massive early-type systems 
and only the very central regions of very late-type spirals (see Figure~\ref{fig:SBprofileskpc}, bottom-right plot).
However, we should certainly add a quality-of-fit criterion here to determine the goodness of these criteria.

In order to determine the reduced-$\chi^{2}$ for each fit, the
number of degrees-of-freedom (d.o.f.) is computed as 
the number of data points that remain after applying
the corresponding cutoffs minus the 
number $P$ of free parameters, where $P=2$ in our linear fitting case
\citep[see][for a discussion]{Andrae2010arXiv}.
Average reduced-$\chi^{2}$ are computed for each combination of
cutoffs and the results are shown in Table~\ref{Table:chisqtable}.
\begin{table*}
\caption[Table~caption text]{Average reduced-$\chi^{2}$ of the linear-fit with $\mu_{[3.6]}$ and $R/R80$ cuts}
\label{Table:chisqtable}
\begin{center}
\begin{tabular}{rr||cc|cc|cc|cc|cc|cc}
\multicolumn{2}{c}{\multirow{2}{*}{}} & \multicolumn{12}{c}{$R/R80$ cutoffs}\\
\multicolumn{2}{c||}{} & \multicolumn{2}{c|}{0.00} & \multicolumn{2}{c|}{0.25} & \multicolumn{2}{c|}{0.50} & \multicolumn{2}{c|}{0.75} & \multicolumn{2}{c|}{1.00} & \multicolumn{2}{c}{1.25} \\
\cline{3-14}
 & 
 & $<\chi^{2}>$ & N\tablenotemark{a} & $<\chi^{2}>$ & N & $<\chi^{2}>$ & N & $<\chi^{2}>$ & N & $<\chi^{2}>$ & N & $<\chi^{2}>$ & N \\
\hline
\hline
\multirow{6}{*}{\rotatebox[origin=c]{90}{$\mu_{[3.6]}$ cutoffs}} 
 & 21.5 & 26.20 & (1577) & 20.84 & (1554) & 15.72 & (1451) & 9.68 & (1240) & 4.04 & (794) & 2.87 & (535) \\ 
 & 22 & 10.64 & (1489) & 8.63 & (1474) & 6.97 & (1387) & 5.85 & (1191) & 3.26 & (781) & 2.48 & (530) \\ 
 & 22.5 & 4.89 & (1384) & 4.28 & (1375) & 3.54 & (1298) & 3.11 & (1126) & 2.02 & (756) & 1.62 & (518) \\ 
 & 23 & 2.34 & (1232) & 2.17 & (1228) & 1.81 & (1165) & 1.63 & (1014) & 1.14 & (693) & 0.98 & (482) \\ 
 & 23.5 & 1.37 & (1034) & 1.28 & (1033) & \cellcolor{gray!25}1.12 & \cellcolor{gray!25} (987) & 0.96 & (863) & 0.68 & (591) & 0.56 & (419) \\ 
 & 24 & 0.78 & (755) & 0.77 & (754) & 0.73 & (723) & 0.67 & (630) & 0.40 & (426) & 0.35 & (296) \\ 
 \end{tabular}
 \tablenotetext{1}{N is the number of galaxies remaining after applying the cutoffs and on which the linear-fitting is performed.}
\end{center}
\end{table*}
\begin{table*}
\caption[Table~caption text]{Average reduced-$\chi^{2}$ of the linear-fit in the $\mu_{[3.6]}$ vs $R/R80$ plane with oblique cuts}
\label{Table:chisqtable2}
\begin{center}
\begin{tabular}{rr||cc|cc|cc|cc|cc|cc}
\multicolumn{2}{c}{\multirow{2}{*}{}} & \multicolumn{12}{c}{slope (a) cutoff}\\
\multicolumn{2}{c||}{} & \multicolumn{2}{c|}{-6} & \multicolumn{2}{c|}{-5} & \multicolumn{2}{c|}{-4} & \multicolumn{2}{c|}{-3} & \multicolumn{2}{c|}{-2} & \multicolumn{2}{c}{-1} \\
\cline{3-14}
 & 
 & $<\chi^{2}>$ & N & $<\chi^{2}>$ & N & $<\chi^{2}>$ & N & $<\chi^{2}>$ & N & $<\chi^{2}>$ & N & $<\chi^{2}>$ & N \\
\hline
\hline
\multirow{6}{*}{\rotatebox[origin=c]{90}{\thead[l]{y-intercept (b) \\ cutoff}}} 

 & 20 & 777.55 & (1717) & 649.14 & (1716) & 569.63 & (1713) & 469.68 & (1712) & 393.01 & (1707) & 304.54 & (1699) \\
 & 22 & 205.61 & (1697) & 158.46 & (1691) & 129.09 & (1684) & 88.95 & (1668) & 52.78 & (1644) & 27.10 & (1592) \\
 & 24 & 61.96 & (1633) & 44.50 & (1607) & 28.48 & (1563) & 13.20 & (1528) & 5.58 & (1412) & \cellcolor{gray!25}2.07 & \cellcolor{gray!25}(1233) \\
 & 25 & 33.19 & (1578) & 23.64 & (1540) & 11.19 & (1482) & 6.19 & (1375) & 2.21 & (1202) & 0.78 & (831) \\
 & 26 & 20.87 & (1516) & 10.53 & (1445) & 6.24 & (1339) & 2.50 & (1166) & 0.96 & (845) & 0.44 & (219) \\
 & 28 & 6.20 & (1260) & 3.12 & (1090) & 1.74 & (838) & 0.89 & (471) & 0.79 & (127) & 0.72 & (5) \\
 & 30 & 2.46 & (858) & 1.56 & (596) & 0.87 & (322) & 1.27 & (103) & 1.04 & (8) & \nodata & (0) \\

\end{tabular}
\end{center}
\end{table*}

When doing these fits we excluded elliptical galaxies
($T$\,$\leq$\,$-$3.5) in all cases. 
It should be noted that as we move towards
higher values in both the $R/R80$ and $\mu_{[3.6]}$ cutoffs, 
the number of points used for the linear fit decreases,
and the number of galaxies that can be analyzed becomes smaller.
This is because some galaxy profiles do not reach beyond the cutoffs,
or only one data point is beyond them.
Besides, eventually the reduced-$\chi^{2}$ goes below unity,
telling us that we are overfitting the data.
This is in part due to the effect of correlated errors associated to the uncertainties in the
sky subtraction in the very outer surface brightness measurements. We
find that the best set of $R/R80$ and $\mu_{[3.6]}$ cutoffs, i.e$.$, the
one that yields an average reduced-$\chi^{2}\sim1$ with still a large
number of galaxies, is at $R/R80$=0.5 and $\mu_{[3.6]}$=23.5
mag\,arcsec$^{-2}$, where $<$$\chi^{2}$$>$=1.12 and the number of galaxies
is 987 ($\sim$51\% of the $\galex$/$\sfg$ sample; see Figure~\ref{fig:mu36SBprofilesR80}).

\begin{figure*}
\begin{center}
\includegraphics[width=0.4\textwidth]{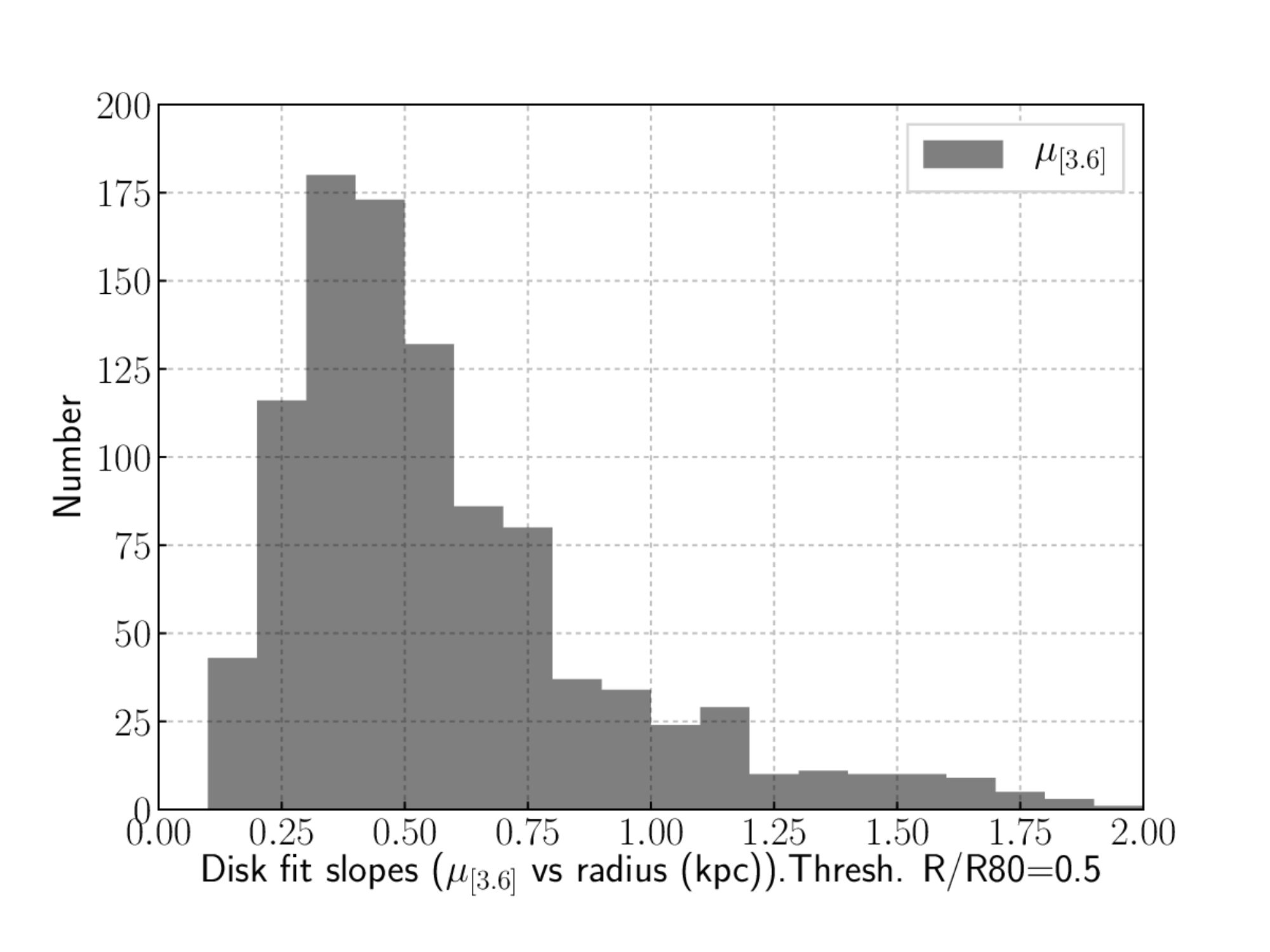}
\includegraphics[width=0.4\textwidth]{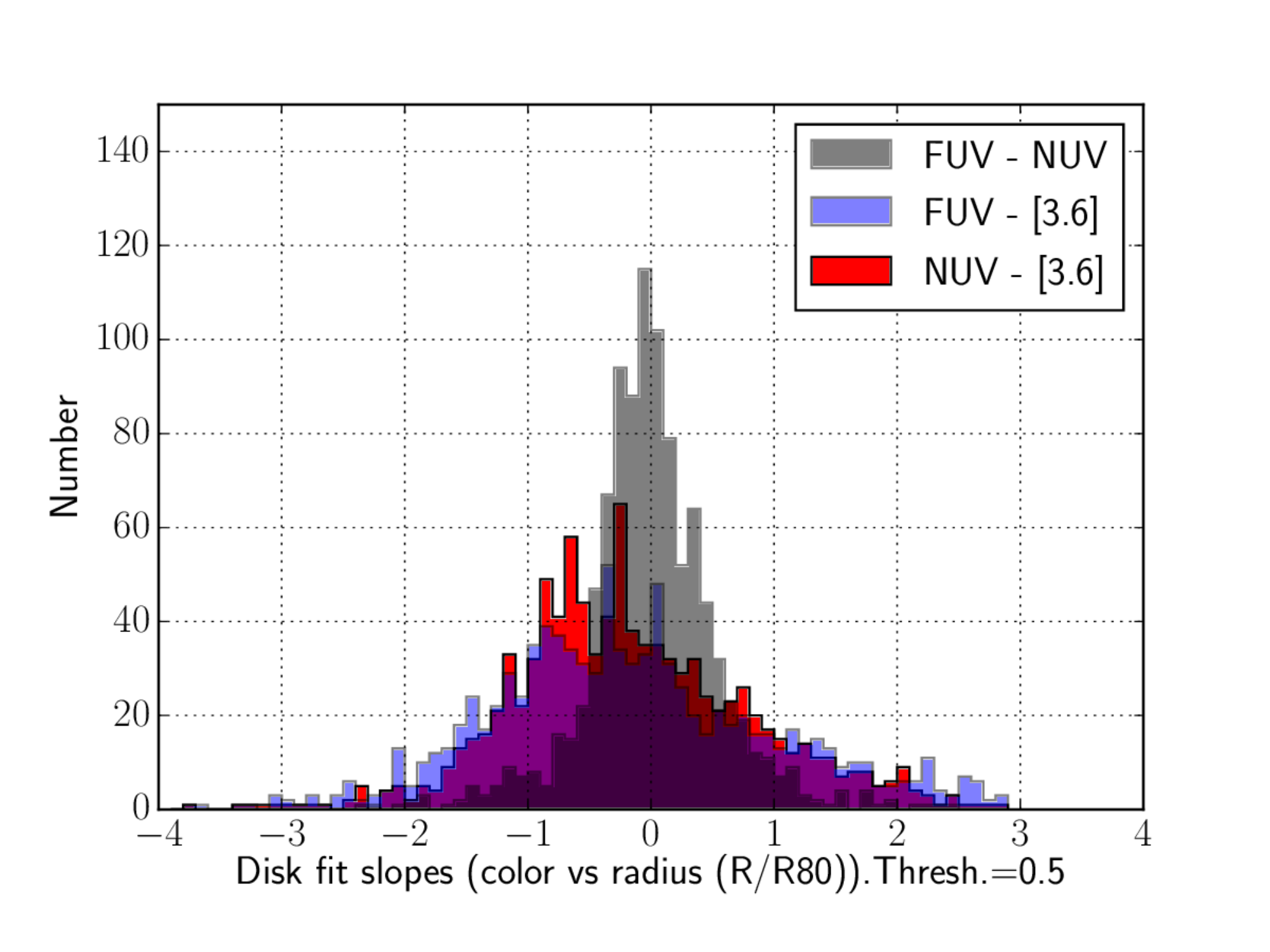}\\
\includegraphics[width=0.4\textwidth]{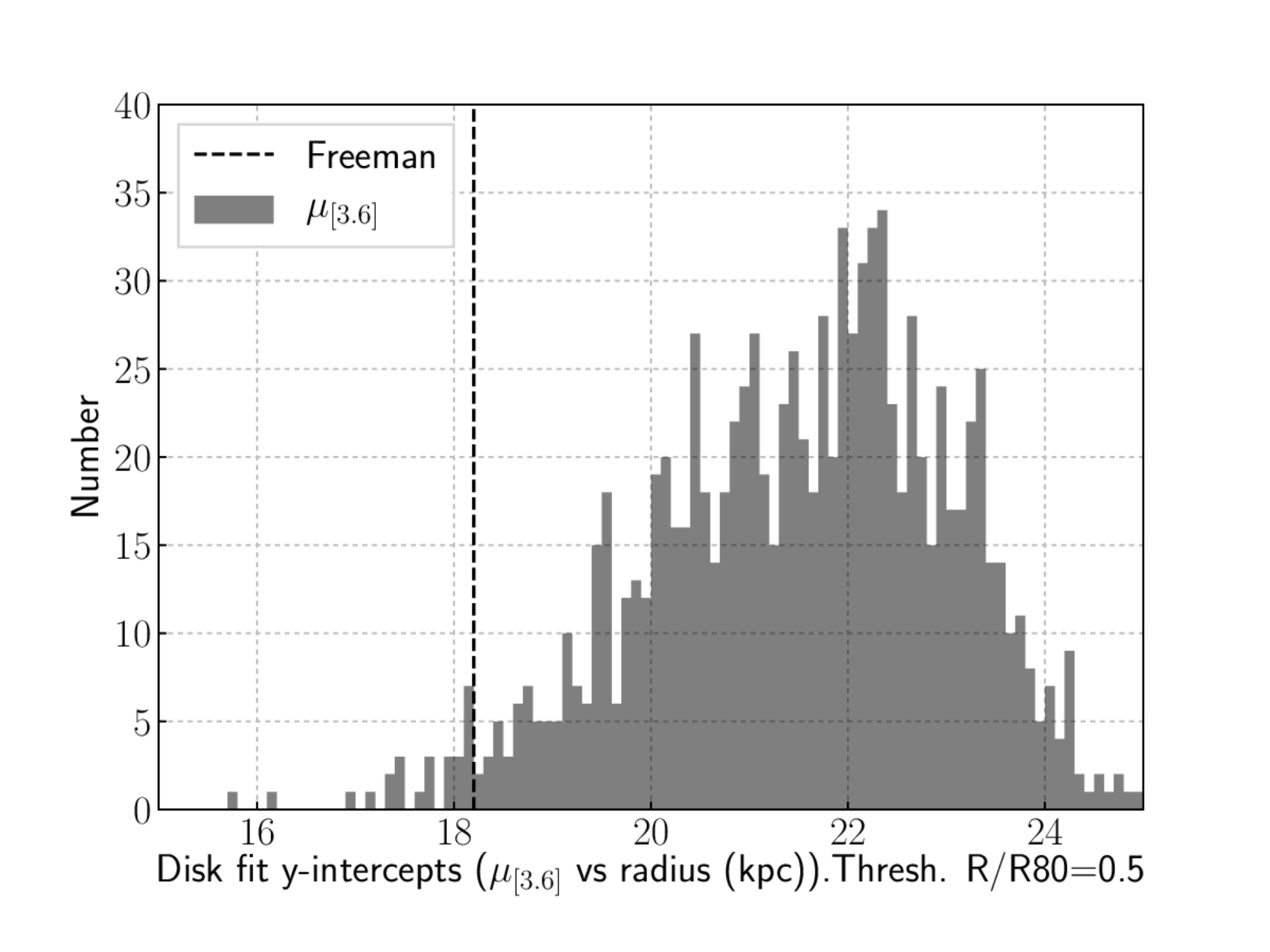}
\includegraphics[width=0.4\textwidth]{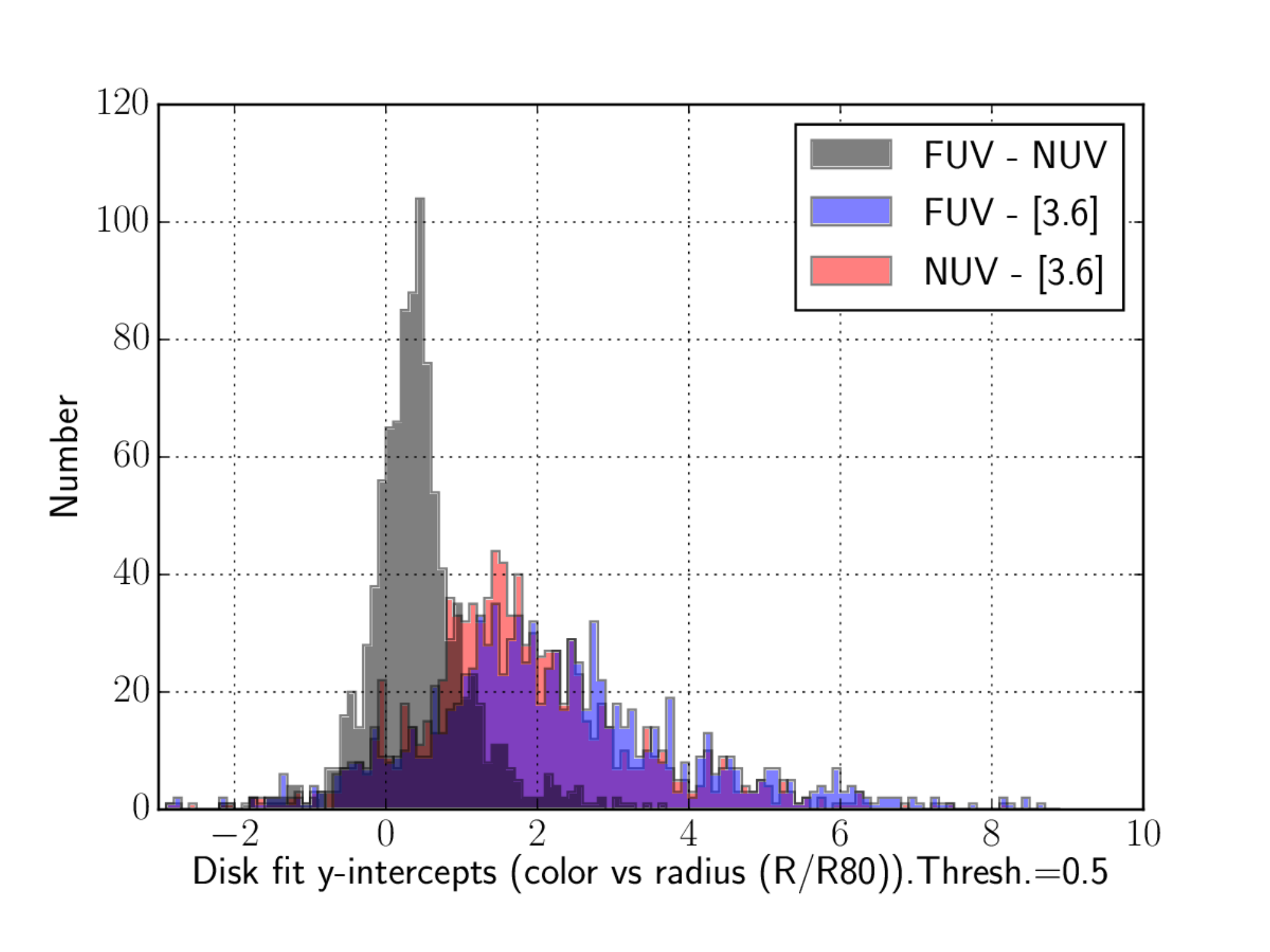}
\caption{
Distributions of the best-fitting coefficients to the surface brightness and color profiles of disks. 
The fitting is performed beyond a radius
$R/R80$=0.5 using points where the surface brightness is fainter than
$\mu_{[3.6]}$=23.5\,mag\,arcsec$^{-2}$. 
\textit{Top left}: the distribution of slopes obtained in the $\mu_{[3.6]}$ vs kpc plane. 
\textit{Bottom left}: the distribution of $y$-intercepts obtained in that plane. 
These correspond to the central surface brightness contribution of the disks.
The dashed vertical line corresponds to the \citet{Freeman1970} value, or the B-band central surface brightness for spirals $<\Sigma_{0}>$=21.48\,B-mag\,arcsec$^{-2}$,
converted to a 3.6\,$\mu$m value of 18.2\,mag\,arcsec$^{-2}$, 
assuming an average central color of (B$_{\mathrm{Vega}}$\,$-$\,[3.6]$_{\mathrm{AB}}$) = 3.32\,mag.
\textit{Top right}: the distribution of slopes obtained
in the (FUV\,$-$\,NUV), (FUV\,$-$\,[3.6]), and (NUV\,$-$\,[3.6]) vs $R/R80$ planes,
i.e$.$ the color gradients.
\textit{Bottom right}: the distribution of $y$-intercepts
obtained in those planes, or, central colors of the disks. Bin width is 0.1 in all cases [in units of
either mag or mag/($R/R80$)]. \label{fig:histoslope50}
}
\end{center}
\end{figure*}

We also apply oblique cuts in the $\mu_{[3.6]}$ versus $R/R80$ plane instead of a combination of vertical and horizontal cuts.
Table~\ref{Table:chisqtable2} shows the resulting average reduced-$\chi^{2}$ and the number of galaxies for a
combination of cutoff slopes $a$ and cutoff $y$-intercepts $b$.
We tried all the combinations of slopes ranging from $-$7 to $-$1 (in units of mag\,arcsec$^{-2}$/($R/R80$))
and $y$-intercepts between 20 and 30\,mag\,arcsec$^{-2}$. The best
compromise between average reduced-$\chi^{2}$ and number of
galaxies is for slope and y-intercept values of $a$=$-$1 and $b$=24\,mag\,arcsec$^{-2}$
where the average reduced-$\chi^{2}=2.07$ and the number of galaxies is 1233 ($\sim$64\% of the $\galex$/$\sfg$ sample; see Figure~\ref{fig:mu36SBprofilesR80}).
Graphical representations of the slopes and $y$-intercepts at these best cutoffs are shown in Figure~\ref{fig:histoslope50}.

The relatively good isolation of the disk component by some of these
sets of criteria opens the door to statistical studies of the
photometric properties of disks in thousands or millions of galaxies
using existing data (SDSS) or data from future facilities and missions
such as LSST or \textit{EUCLID}.

\subsection{Spatially resolved star-forming main sequence from UV and near-IR SB profiles} \label{sec:sfms}
\begin{figure}
\begin{center}
\includegraphics[width=0.49\textwidth]{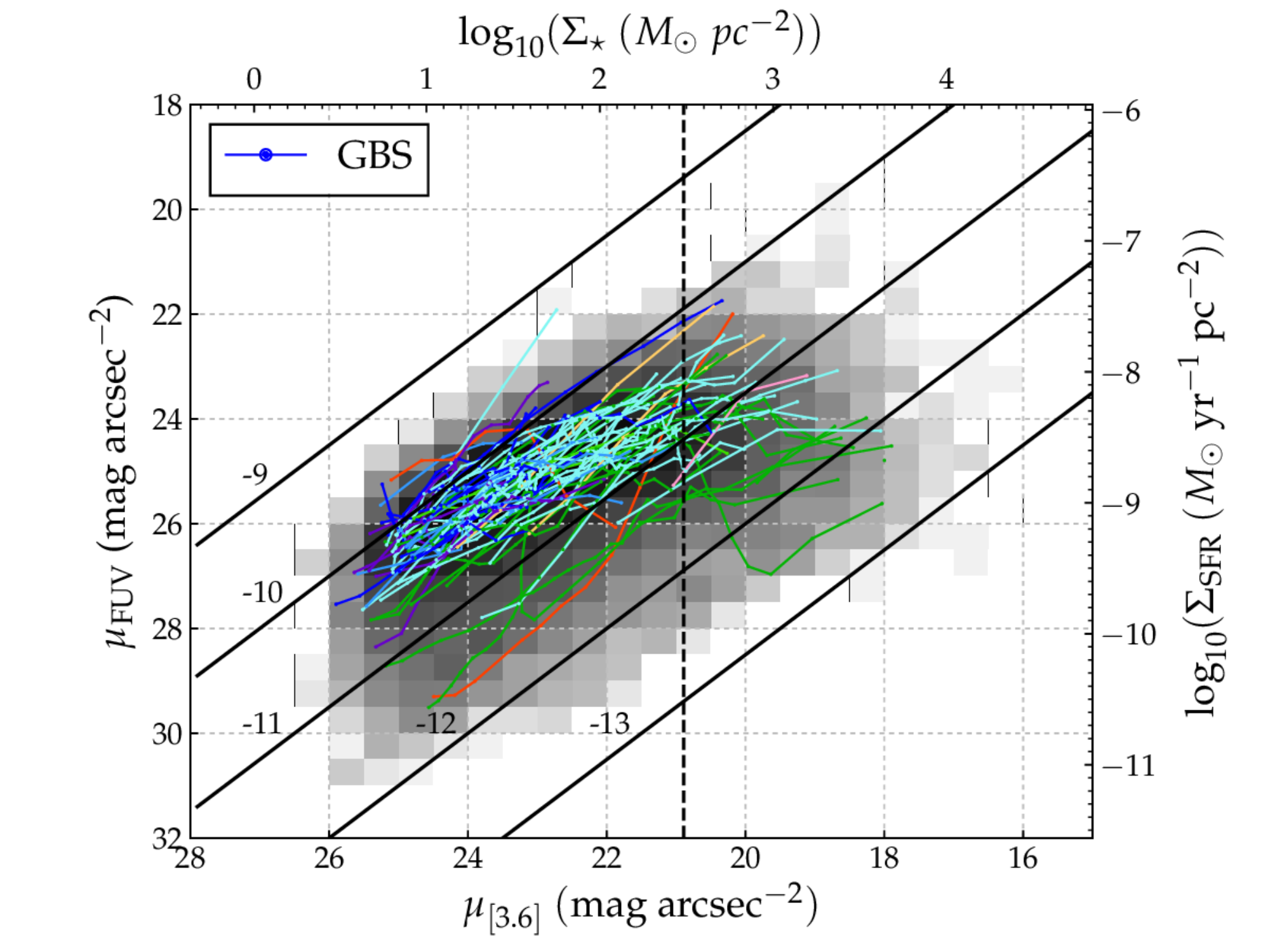}\\
\includegraphics[width=0.49\textwidth]{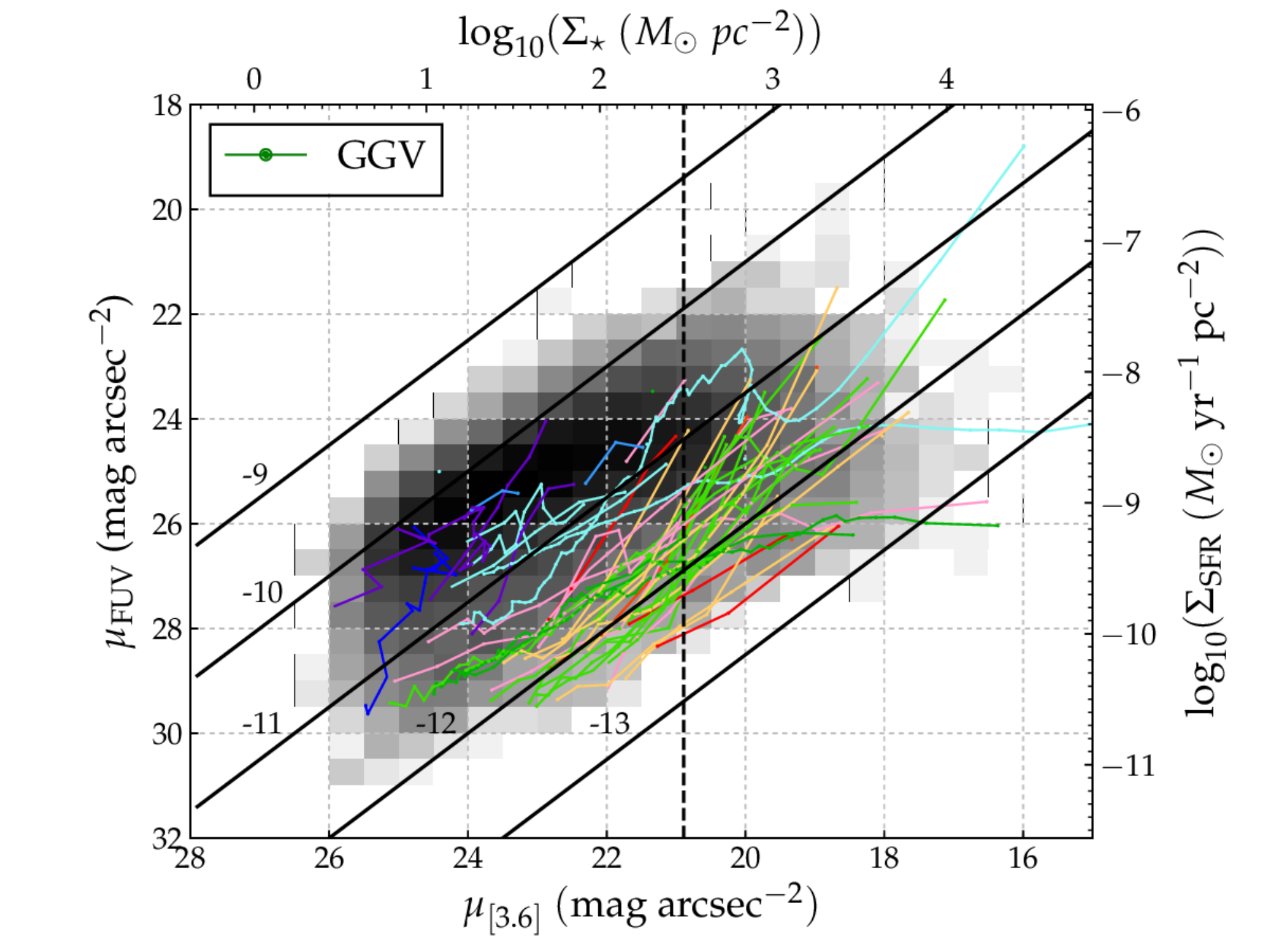}\\
\includegraphics[width=0.49\textwidth]{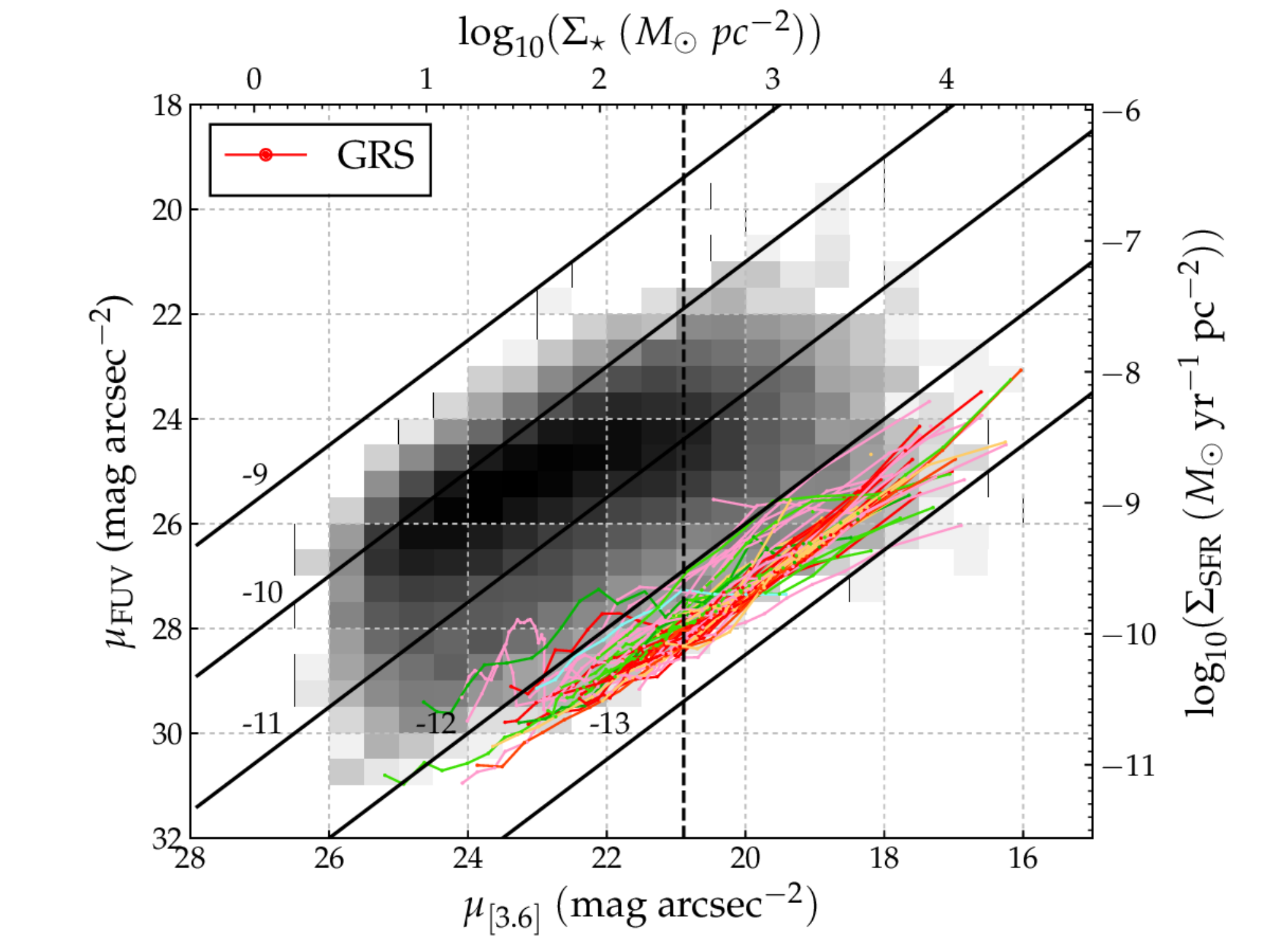}
\caption{
FUV surface brightness versus 3.6\,$\mu$m surface brightness for randomly selected
GBS (\textit{top}), all the GGV (\textit{middle}) and all the GRS galaxies (\textit{bottom})
subsamples. 
The diagonal solid lines represent constant sSFR, and are annotated with the decimal exponent of the logarithm. 
These plots are equivalent to the (observed) Star Formation Main Sequence (SFMS) but in surface brightness.
Both the segregation in sSFR between the GBS, GGV and GRS and the bending at high (surface
density) masses toward lower sSFR values are also clear in this plot.
The vertical black dashed line corresponds to $\Sigma_{\star}$\,=\,$3\times10^{8}$ $M_{\odot}$ $kpc^{-2}$ (or $\mu_{[3.6]}=20.89$ mag arcsec$^{-2}$) \citep{Kauffmann2006}.
The 2D density histogram shown in the background of each panel represents the data points density of GBS galaxies.
\label{fig:SFMS} 
}
\end{center}
\end{figure}

In Figure~\ref{fig:SFMS} we plot the FUV surface brightness
$\mu_{FUV}$ versus the 3.6\,$\mu$m surface brightness $\mu_{[3.6]}$ for
galaxies belonging to the GBS, GGV, and GRS subsamples based on their
integrated colors. Both axes are expressed in mag\,arcsec$^{-2}$. 

This figure can be also seen as a comparison between the observed SFR
(i.e$.$ not-corrected for internal dust extinction) and the stellar mass surface
densities (see Appendix~\ref{appA}), except for those
cases where the FUV emission is not due to young massive stars. In
that regard, this point is equivalent to the star formation main sequence
(SFMS) but in surface brightness \citep[see][]{CanoDiaz2016}.

Each data point is the averaged value within fixed-inclination
elliptical ring apertures of 6$\arcsec$ width.  The innermost ring
has a semi-major axis length of 6$\arcsec$ with a width of 6$\arcsec$,
defined by an inner ellipse with a major axis of 3$\arcsec$ from
the center and an outer ellipse with a semi-major axis of 9$\arcsec$
from the center.  The initial ring does not cover the center of the
galaxy as this could be affected by differences in the PSF
amongst the three bands and by the contribution of an AGN. Subsequent
rings increase in size in 6$\arcsec$ steps, i.e$.$, they have semi-major
axis radii of 12$\arcsec$, 18$\arcsec$, 24$\arcsec$, and so on.

For early-type GRS galaxies, the FUV and 3.6\,$\mu$m surface
brightnesses show a pretty tight correlation, which indicates that the
3.6\,$\mu$m emission traces not only the stellar mass, but also the bulk of the
stars dominating the FUV emission in these objects, mainly
main-sequence turn-off or extreme horizontal branch (EHB) stars, depending
on the strength of the UV-upturn. Despite the large scatter of the GRS
found in \citet{Bouquin2015}, the use of spatially resolved data with
the 3.6\,$\mu$m surface brightness as normalizing parameter leads now to
a very tight GRS in this SB-SB plane (or a very small range in FUV\,$-$\,[3.6] color). The comparison of these profiles
with those of the GGV galaxies shows that in the latter case the
central stellar mass surface density is 1.5-2\,mag fainter than in the
former and that most GGV galaxies (all except the few very late-type
GGVs) have (outer) disks that follow a trend similar to that followed
by the outer regions of GRS galaxies. Finally, late-type galaxies in
the GBS span a large range of values in both $\mu_{FUV}$ and
$\mu_{[3.6]}$.  Irregulars, Sm, and Sd galaxies have the highest SFR
surface densities (for a given stellar mass surface density) amongst
the GBS subsample.

Despite the large scatter of GBS galaxies, they can be clearly distinguished
from the early-type galaxies of the GRS and even GGV galaxies by
looking at the (observed) sSFR values in their disks. Thus, while GBS
disks have sSFR values that are higher than 10$^{-11.5}$\,yr$^{-1}$,
the outer regions of GGV and GRS galaxies are in the majority of
the cases (all in the case of the GRS) below this value. This value
could be used to easily discriminate between star-forming and
quiescent regions within galaxies.

GBS galaxies define a well separated sequence, and with the spatial
information now available, we can now see what parts of the galaxies
are now just leaving the GBS, that is, have their SF suppressed or exhausted.
While a few GGV galaxies show a decrease in the sSFR of
their inner regions, most of these galaxies are within the locus of
the GBS in the inner parts but approach the sequence marked by the GRS
profiles in their outer regions. In other words, the fact that these
galaxies where identified as leaving the GBS in \citet{Bouquin2015}
is mainly due to their outer parts, likely caused by the disks of GGV galaxies
undergoing either an outside-in SF quenching or an inside-out
rebirth.

It is worth emphasizing here that only the combined use of FUV, NUV,
and 3.6\,$\mu$m allows properly separating the "classical blue cloud"
(now blue sequence) and the "classical red sequence" and determining
which galaxies are now leaving (or entering) the GBS and what regions
within galaxies are responsible for it. 

We mark in Figure~\ref{fig:SFMS} the $\mu_{[3.6]}$ value that corresponds to the surface stellar
mass density of $\Sigma_{\star} = 3 \times 10^{8}$ $M_{\odot}$
kpc$^{-2}$ = 300 $M_{\odot}$ pc$^{-2}$ ($\mu_{[3.6]} = 20.89$ mag
arcsec$^{-2}$) proposed by \citet{Kauffmann2006} to separate between
bulge-dominated and disk-dominated objects. 

In the case of our GBS galaxies, this stellar
mass surface density indicates the region inside which the SFR surface
density flattens relative to the stellar mass surface density,
i.e. when the (FUV\,$-$\,[3.6]) color becomes
significantly redder (see Figure~\ref{fig:nine}). A similar change is observed when using
light-weighted age of the stellar population in galaxies instead \citep{GonzalezDelgado2014}.

\begin{figure}
\begin{center}
\includegraphics[width=0.49\textwidth]{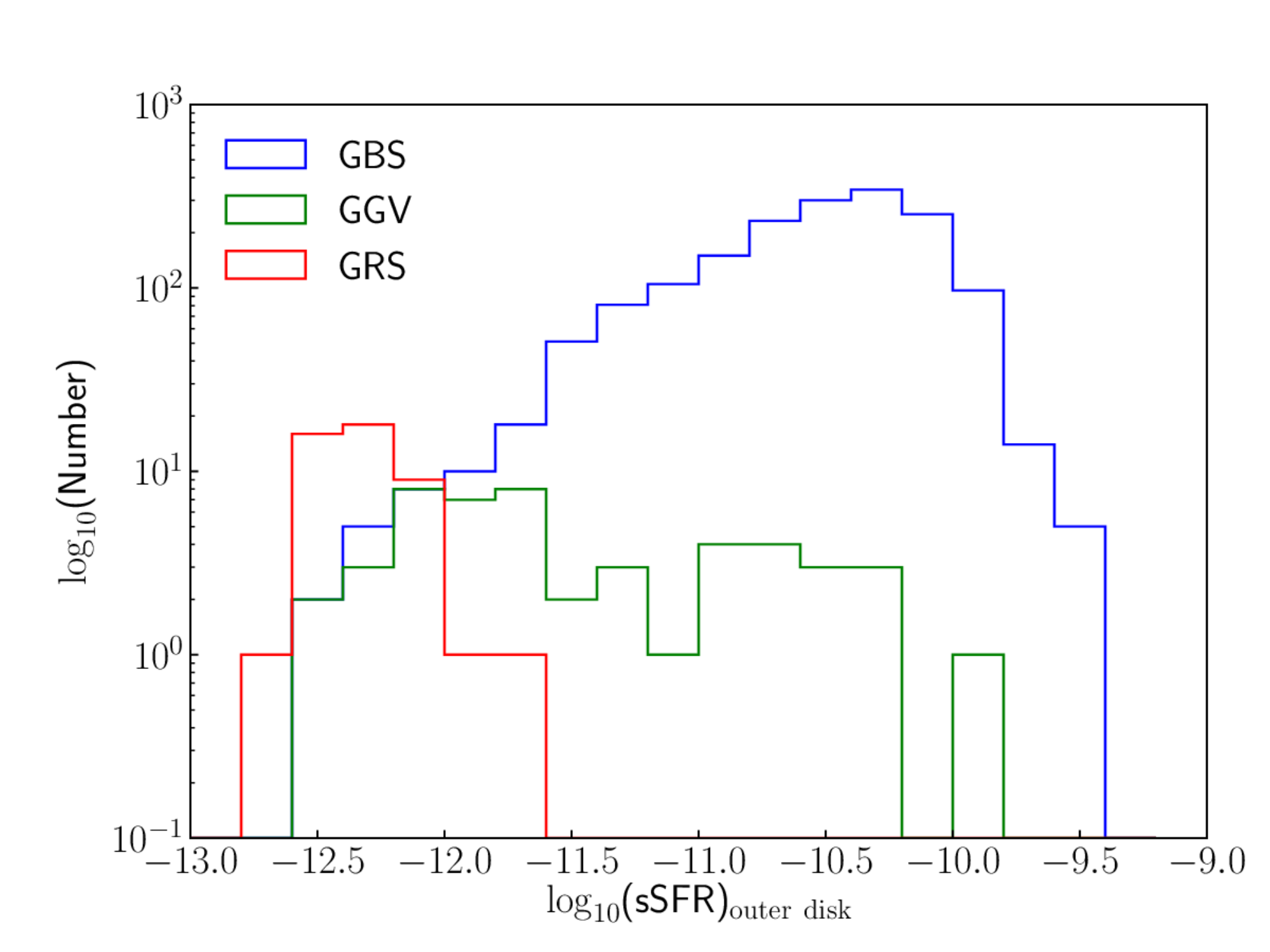}\\\caption{
Decimal log-log histogram of the mean sSFR obtained by fitting the outer disk part (beyond $\mu_{[3.6]}=20.89$ mag\,arcsec$^{-2}$) of GBS, GGV, and GRS galaxies. The bin size is 0.2 dex.
 \label{fig:histoplot6}
}
\end{center}
\end{figure}

The sSFR of the outer parts (beyond $\mu_{[3.6]}$ = 20.89 mag\,arcsec$^{-2}$) is shown in
Figure~\ref{fig:histoplot6} for GBS, GGV, and GRS galaxies. 
This is done simply by calculating the linear scale sSFR of one galaxy, 
at each point that are in the outer parts, and averaging these sSFR values (not light/mass weighted)
in order to get a single sSFR value per galaxy (and expressing them in the logarithmic scale at the end).
We find the following specific star formation rate density range:
$-$12.5 $<$ $\log_{10}(\mathrm{sSFR})$ $<$ $-$9.5 for GBS galaxies,
$-$12.4 $<$ $\log_{10}(\mathrm{sSFR})$ $<$ $-$9.8 for GGV galaxies, and
$-$12.6 $<$ $\log_{10}(\mathrm{sSFR})$ $<$ $-$11.7 for GRS galaxies.
Since we do not correct for internal dust attenuation, these values should be viewed as lower limits of the true sSFR.
Previous studies of the impact of dust on the (FUV\,$-$\,[3.6]) colors \citep{MunozMateos2007,MunozMateos2009a,MunozMateos2009b}
have shown that dust attenuation A$_{\mathrm{FUV}}$ decreases as we move outward in the disks, although the dust content differs from one morphological type bin to another,
for example, Sb-Sbc galaxies have higher A$_{\mathrm{FUV}}$ at all radii than the other types, whereas Sdm-Irr have relatively very low dust content.
It should be noted, however, that besides dust, the reddening in the outer parts of quiescent galaxies is due to their older stars.
There is a clear difference between the outer parts of GRS galaxies having low sSFR and a narrow range of values,
and those of GBS galaxies with a wide range of sSFR but in general not as low as the outer parts of the GRS.
For our sample, we have a distribution in outer disks sSFR with the mean at 
$-$10.6 dex and $\sigma$=0.5 dex (rms) for GBS, 
$-$11.5 dex and $\sigma$=0.7 dex for GGV, and 
$-$12.3 dex and $\sigma$=0.2 dex for GRS galaxies.
The sSFR of the outer parts of GGV galaxies in our sample covers a wider range of values but is not as high as some GBS galaxies, and not as low as some GRS galaxies.
Note that in the case of the GRS galaxies,
the UV emission might not be due to recent SF but to the light from low-mass evolved stars.


\subsection{Color and sSFR profiles} \label{sec:colorandssfr}
\begin{figure*}
\begin{center}
\includegraphics[width=0.49\textwidth]{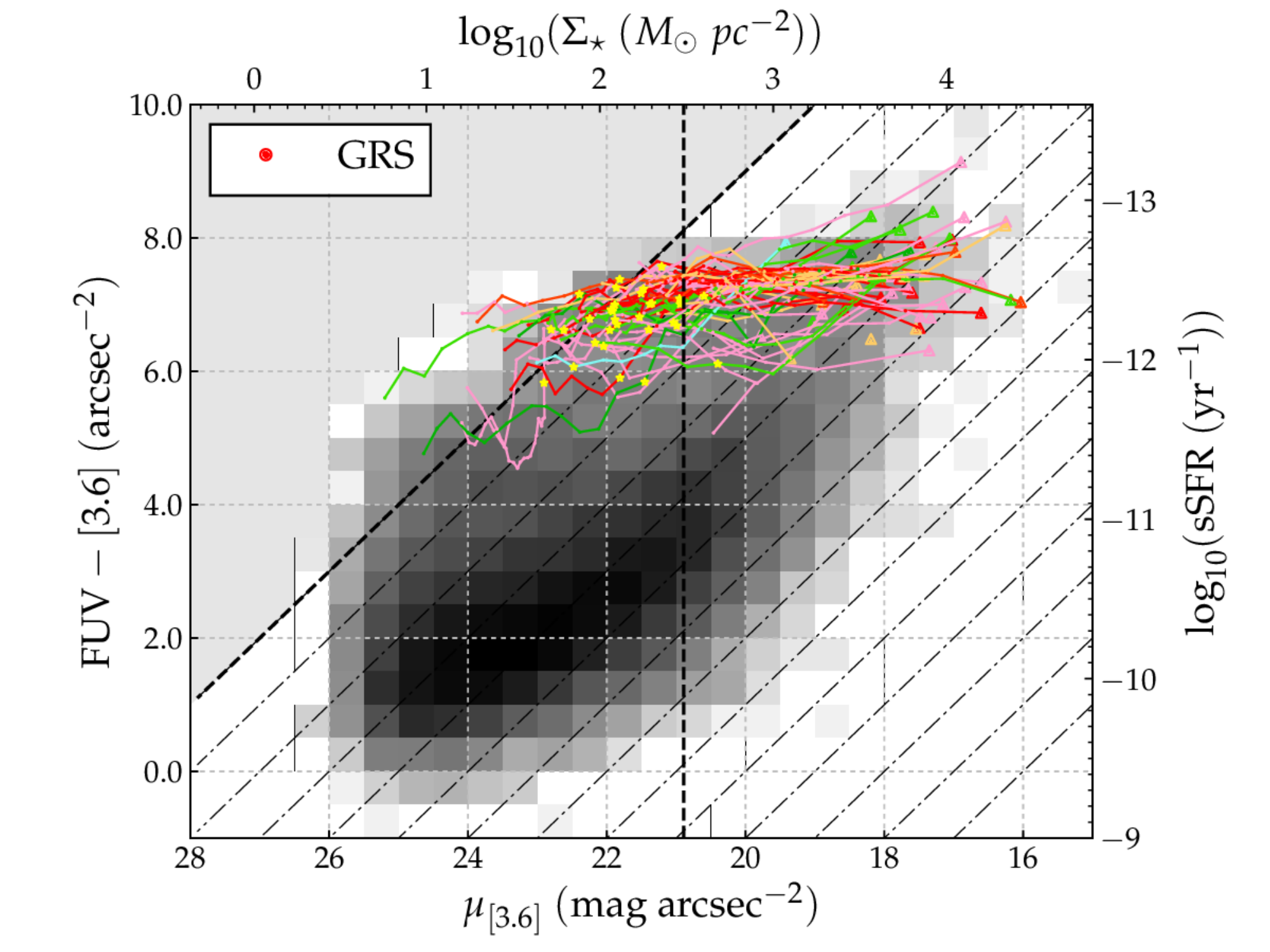}\\
\includegraphics[width=0.49\textwidth]{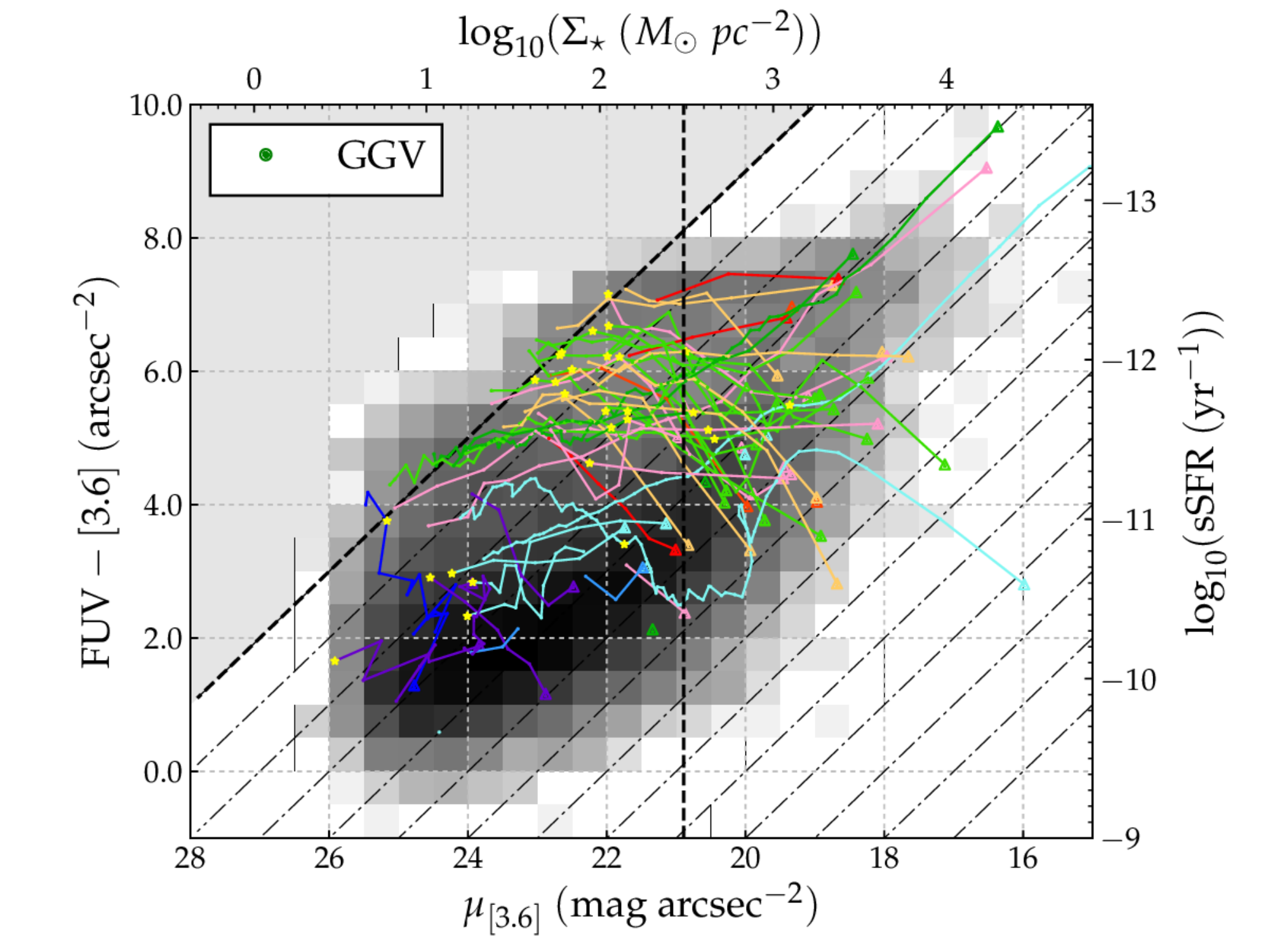}\\
\includegraphics[width=0.49\textwidth]{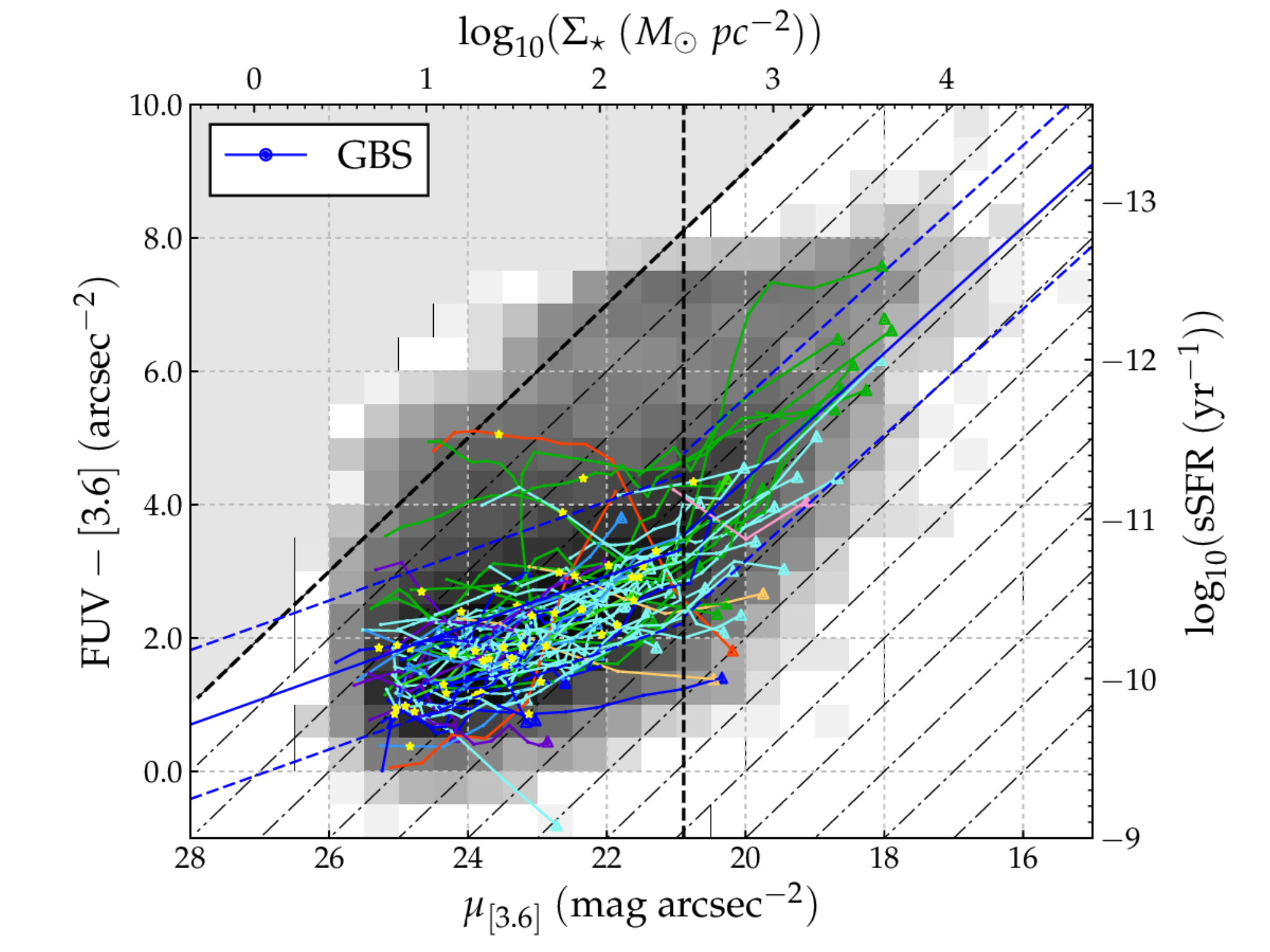}
\caption{
(FUV\,$-$\,[3.6]) color versus $\mu_{[3.6]}$
surface brightness for the radial profiles of GRS \textit{top row},
GGV \textit{middle row}, and GBS \textit{bottom row} galaxies. Each
galaxy's center (within 6$\arcsec$ of the central most aperture) is
represented by a triangle, and subsequent values are taken at every
6$\arcsec$ and are represented by smaller dots if these values
exist. 
Values (dots) belonging to the same galaxy are connected by a
line of the same color as the dots. 
The yellow star shows the radial distance at which 80\% of the 3.6\,$\mu$m
light is enclosed. 
Diagonal dot-dashed lines are lines of constant
$\mu_{\mathrm{FUV}}$\,arcsec$^{-2}$ (i.e$.$ observed SFR surface density), with the left-most
dashed line corresponding to $\mu_{\mathrm{FUV}}$\,=\,29 ABmag arcsec$^{-2}$
(corresponding to $\Sigma_{\mathrm{SFR}} = 4.36 \times 10^{-5}$ $M_{\odot}/yr/kpc^{2}$ for a Kroupa IMF),
the approximate sensitivity limit of our $\galex$ observations. 
The vertical black dashed line
corresponds to $\Sigma_{\star}$\,=\,$3\times10^{8}$ $M_{\odot}$
$pc^{-2}$ (or $\mu_{[3.6]}=20.89$ mag arcsec$^{-2}$)
\citep{Kauffmann2006}. The solid blue lines that go through the data
points in the GBS plot are the fits to all the data points
on each side of $\mu_{[3.6]}=20.89$. The parallel dashed blue lines 
show the $\pm1\sigma$ (rms) of the distribution.
In the case of the GBS plot, we show randomly-selected galaxies to better illustrate how GBS galaxies behave. In all cases, the entirety of the data is shown as a logarithmic 2D density histogram with 0.5\,$\times$\,0.5 binning, where darker (brighter) shades mean higher (lower) data point density.
\label{fig:nine}
}
\end{center}
\end{figure*}

Figure~\ref{fig:nine} shows the GRS (\textit{top row}), GGV
(\textit{center row}), GBS (\textit{bottom row}) galaxies'
(FUV\,$-$\,[3.6]) color profiles versus 3.6
$\mu$m surface brighntess $\mu_{[3.6]}$, with the same color-coding
per morphological type as in previous plots. Again, the 3.6\,$\mu$m
surface brightness corresponds to the stellar mass per area (see
eq.~\ref{eq:stelmass} in Appendix~\ref{appA}), and the
(FUV\,$-$\,[3.6]) color is equivalent to the
observed (not corrected for internal extinction) sSFR (units
$yr^{-1}$) (see eq.~\ref{eq:ssfr} in Appendix~\ref{appB}).

The yellow \textit{star} symbol corresponds to the radial measurement where the 
cumulative magnitude at 3.6\,$\mu$m reaches 80\% of the enclosed light
at these wavelength. 

The (FUV\,$-$\,[3.6])
color profiles are very different for the GRS, GGV, and GBS
subsamples. In the case of GRS galaxies, which are mostly early-type but not
exclusively, the color is tightly constrained within a range from 6 to
8 mag but gets a bit bluer to the outer regions, especially for
GRS galaxies of S0, Sa, Sb, and Sc morphological types.

On the other hand, in the case of GBS galaxies, 
their (FUV\,$-$\,[3.6]) color ranges from $-$1 to 10 mag, 
corresponding to a sSFR value ranging from $10^{-10}$ to $10^{-13}$\,yr$^{-1}$.
Regarding the differences in the color profiles for each
galaxy type, Sa, Sb, and Sc galaxies go from red to blue inside-out, while
Sd, Sm, and Irregulars are much bluer than Sa, Sb, and Sc at a given
stellar mass surface density but their color gradients are somewhat flatter.
Again, it should be noted that we are not correcting for dust and that the effect of dust
is to redden the (FUV\,$-$\,[3.6]) color \citep{MunozMateos2007} and therefore yields a lower limit to the sSFR.

The fact that most profiles of GBS galaxies become bluer from
inside-out indicates that the lower the surface stellar mass density (the greater the galactocentric distance) the greater the sSFR, 
i.e$.$, the higher the SFR for a given surface stellar mass density, 
the more stars are born in the outskirts.
Correcting for internal dust extinction, assuming that dust extinction and reddening effects are stronger in the inner regions than the outer parts, would yield bluer centers compared to the outer disk. This has the effect of increasing the slope of the gradient, where negative color gradients would become flatter, and positive color gradients even more positive. Such effect would translate to a less-pronounced degree of inside-out growth.
It should be noted that while the internal dust-correction would affect the color profiles of the galaxies, it is not enough to explain why most galaxies are becoming bluer inside-out \citep[see Figure 2 in][]{MunozMateos2007}.
Studies by \citet{MunozMateos2007,MunozMateos2011,Pezzulli2015} on nearby galaxy samples have shown
that mass growth and radial growth of nearby spiral disks, growing inside-out, 
have timescales on the order of $\sim$10\,Gyr and \,30 Gyr, respectively.
Isolating a few galaxies actively forming stars in their outer regions,
reveals that their outer regions fall indeed near $\log_{10}$(sSFR) $\sim$ $-$10 yr$^{-1}$,
or $\sim$10 Gyr, in agreement with the above work (see GBS plot in Figure~\ref{fig:nine}).
In the case of profiles reddening in the outskirts, the lower
$\Sigma_{\star}$ becomes, the smaller the sSFR.

Remarkably, a clear color flattening is observed in the outer parts of the
profiles of most GBS galaxies when $\Sigma_{\star}$$<$ 300 $M_{\odot}$ pc$^{-2}$.
Applying a weighted linear fit to the left-hand-side and right-hand-side of 
$\mu_{[3.6]} = 20.89$ mag arcsec$^{-2}$, 
we get $(\mathrm{FUV} - [3.6]) = (-0.395\pm0.023)\cdot \mu_{[3.6]} + (11.537\pm0.453)$, 
and $(-0.438\pm0.009)\cdot \mu_{[3.6]} + (12.210\pm0.193)$ respectively.
These are shown in Figure~\ref{fig:nine} as solid blue lines for the mean value, 
accompanied by parallel dashed blue lines corresponding to the 1$\sigma$ uncertainty.

The galaxies falling into the GGV category globally are clearly distinct from
the GBS ones also in terms of their spatially resolved properties. 
They show flat or even inverted color (and sSFR)
profiles as a function of stellar mass surface density \citep[hardly due to
radial variations in the amount of dust reddening; see][]{MunozMateos2007}, 
which indicates either a decline in the observed SFR (oblique lines in
Figure~\ref{fig:nine}) in their outskirts or, alternatively, a recent
enhancement of the SFR in the inner regions of an otherwise passively
evolving system. In the latter case, the low fraction of
intermediate-type spirals in the GRS (compared to the GGV) suggests that this rebirth should
be accompanied by a morphological transformation from ETGs towards later galaxy types. 
There are, indeed, post-starburst (E+A) or (K+A) galaxies that are in the classical green valley \citep{French2015} that did have centrally concentrated star formation \citep{Norton2001}.

In the more likely case of a decline of
the SFR in the outer disks of GGV intermediate-type-spirals we should
then invoke the presence of a quenching (or, at least, damping) mechanism for the star formation acting primarily in these regions. 

\begin{figure}
\begin{center}
\includegraphics[width=0.49\textwidth]{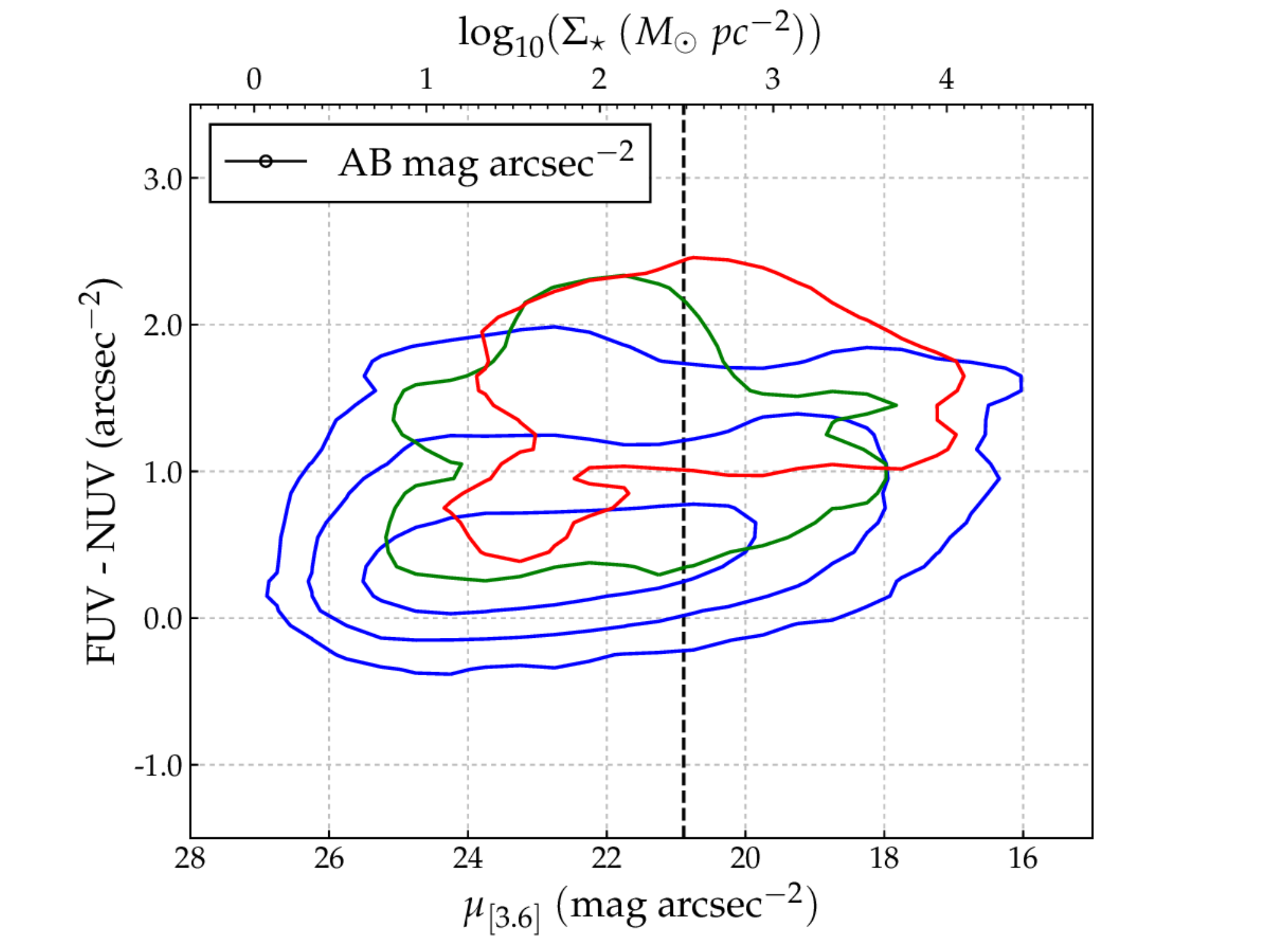}\\
\includegraphics[width=0.49\textwidth]{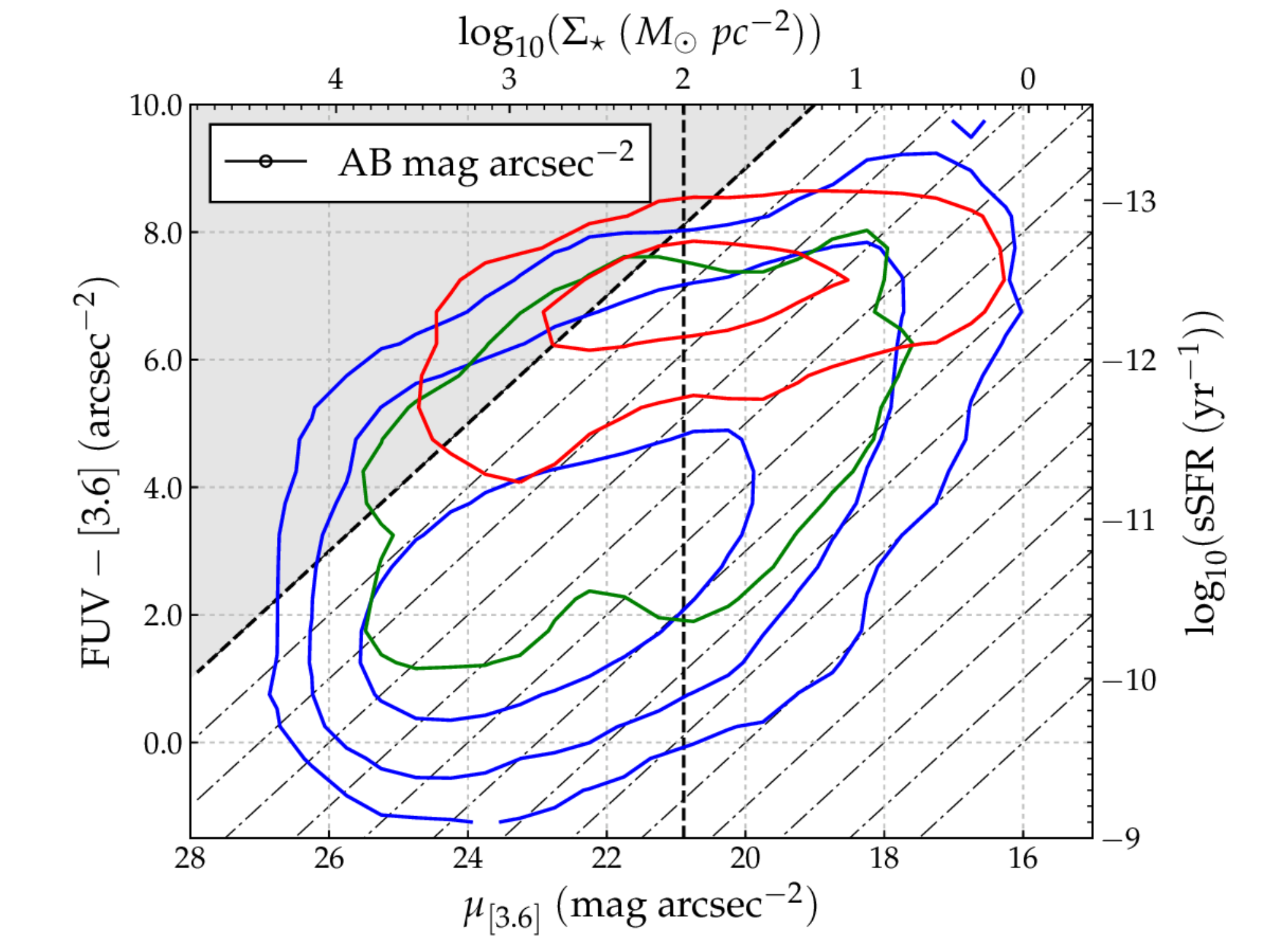}\\
\includegraphics[width=0.49\textwidth]{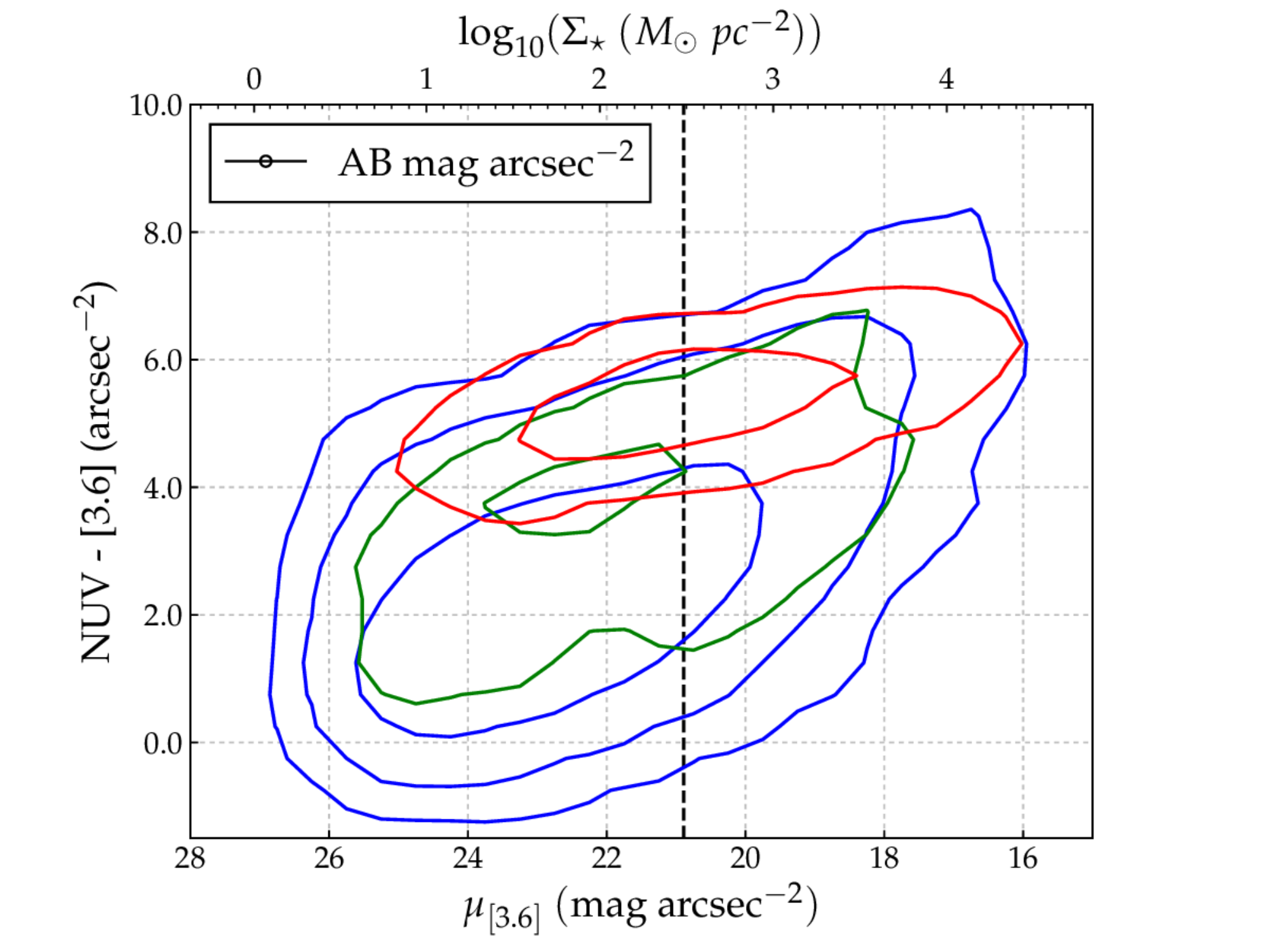}\\
		\caption{(FUV\,$-$\,NUV), (FUV\,$-$\,[3.6]) and (NUV\,$-$\,[3.6])
                  colors vs $\mu_{[3.6]}$ surface brightness profiles contours for GBS, GGV, and GRS galaxies.
  		 Contours levels were slightly smoothed with a gaussian kernel and describe the number density of SB profiles data points.
		 The outer most level corresponds to number densities of 0 dex (i.e$.$ at least one data point) in each 2D bins (the binning is 0.1 mag for (FUV\,$-$\,NUV) color,
		 and 0.5 mag for everything else). Then each contour level corresponds to an increase in number density by 1 dex.
                  In the case of the (FUV\,$-$\,[3.6]) vs
                  $\mu_{[3.6]}$ diagram, diagonal lines represent
                  constant FUV surface brightness $\mu_{FUV}$, with
                  the right-most dashed line corresponding to
                  $\mu_{FUV}$\,=\,29 ABmag\,arcsec$^{-2}$ (at the grey
                  boundary), and decreasing by unity for each diagonal
                  to the left. These are equivalent to lines of
                  constant observed SFR surface density. The vertical
                  black dashed line corresponds to
                  $\log_{10}$($\Sigma_{\star}$ ($M_{\odot}$
                  $pc^{-2}$))=2.477 (or $\mu_{[3.6]}=20.89$ mag
                  arcsec$^{-2}$) \citep{Kauffmann2006}. The entire
                  sample is shown in this case. \label{fig:ccpEmu36} }
\end{center}
\end{figure}

Figure~\ref{fig:ccpEmu36} shows the color profiles of 
(FUV\,$-$\,NUV), 
(FUV\,$-$\,[3.6]), and 
(NUV\,$-$\,[3.6]) vs $\mu_{[3.6]}$ surface brightness. Linear fits to
these color profiles were performed for each individual galaxy and are included in
Table~\ref{Table:slopestable2}. 
The fits were performed for SB fainter than $\mu_[3.6]$ = 20.89\,mag arcsec$^{-2}$ in these cases.

While a positive gradient seems to be more pronounced in (FUV\,$-$\,NUV) color compared to the other two, 
it is not clear what is driving it. Since dust reddening is rarely increasing toward the outer parts, 
those objects with positive
(FUV\,$-$\,NUV) color gradients are
likely suffering changes in the recent SF history of their outer
regions. 
The dominant morphological type of positive (FUV\,$-$\,NUV) color gradient galaxies are S0-a galaxies. 

The comparison between the (NUV\,$-$\,[3.6])
and (FUV\,$-$\,[3.6]) color profiles (both
shown in Figure~\ref{fig:ccpEmu36}) is also important to determine
whether the UV emission is coming from newly formed O and B stars, or
from evolved UV-upturn sources \citep[likely associated to extreme horizontal branch
stars; see also Section~\ref{sec:CCD}, which mainly contribute to the FUV band;][]{Oconnell1999}.

\begin{table*}
	\begin{center}
	\caption{COLOR AND SURFACE BRIGHTNESS GRADIENTS FROM RADIAL PROFILES NORMALIZED TO $R80$ (except for $\mu_{[3.6]}$)}
	\label{Table:slopestable}

		\begin{threeparttable}
                		\begin{tabular}{ l|rr|rr|rr|rr }
                		&
                                \multicolumn{2}{ |c| }{FUV\,$-$\,NUV}& 
                                \multicolumn{2}{ |c| }{FUV\,$-$\,[3.6]} & 
                                \multicolumn{2}{ |c| }{NUV\,$-$\,[3.6]}&
                                \multicolumn{2}{ |c }{$\mu_{[3.6]}$}\\
                                \hline
			       \multicolumn{1}{l}{unit}&
                                \multicolumn{2}{ |c| }{mag/($R/R80$)}& 
                                \multicolumn{2}{ |c| }{mag/($R/R80$)} & 
                                \multicolumn{2}{ |c| }{mag/($R/R80$)}&
                                \multicolumn{2}{ |c }{mag/kpc}\\
                                \hline
                                \multicolumn{1}{l}{name\tablenotemark{1}}&
                                \multicolumn{1}{|c}{a\tablenotemark{2}}&
                                \multicolumn{1}{c|}{b\tablenotemark{3}}&
                                \multicolumn{1}{c}{a}&
                                \multicolumn{1}{c|}{b}&
                                \multicolumn{1}{c}{a}&
                                \multicolumn{1}{c|}{b}&
                                \multicolumn{1}{c}{a}&
                                \multicolumn{1}{c}{b}\\
				\hline
ESO293-034 & -0.03$\pm$0.06 & 0.48$\pm$0.07 & -1.31$\pm$0.12 & 4.30$\pm$0.13 & -1.37$\pm$0.09 & 3.91$\pm$0.09 & 0.55$\pm$0.02 & 21.41$\pm$0.14 \\
NGC0007 & 0.15$\pm$0.42 & 0.21$\pm$0.34 & -0.11$\pm$0.48 & 1.64$\pm$0.39 & -0.25$\pm$0.14 & 1.42$\pm$0.11 & ---$\pm$--- & ---$\pm$--- \\
IC1532 & -0.76$\pm$0.92 & 0.96$\pm$0.66 & -0.12$\pm$0.95 & 2.44$\pm$0.67 & 0.63$\pm$0.01 & 1.48$\pm$0.01 & ---$\pm$--- & ---$\pm$--- \\
NGC0024 & 0.00$\pm$0.02 & 0.34$\pm$0.02 & -0.54$\pm$0.05 & 3.04$\pm$0.04 & -0.58$\pm$0.05 & 2.74$\pm$0.04 & 1.04$\pm$0.03 & 20.81$\pm$0.12 \\
ESO293-045 & 0.05$\pm$0.07 & 0.08$\pm$0.06 & -0.83$\pm$0.41 & 1.20$\pm$0.33 & -0.92$\pm$0.36 & 1.15$\pm$0.29 & 0.64$\pm$0.03 & 23.15$\pm$0.12 \\
UGC00122 & 1.01$\pm$0.21 & -0.63$\pm$0.16 & 1.16$\pm$0.30 & -0.23$\pm$0.23 & 0.15$\pm$0.25 & 0.40$\pm$0.19 & 0.81$\pm$0.04 & 24.24$\pm$0.10 \\
NGC0059 & 0.09$\pm$0.11 & 1.60$\pm$0.09 & 0.18$\pm$0.11 & 4.82$\pm$0.09 & 0.10$\pm$0.05 & 3.21$\pm$0.04 & 2.23$\pm$0.07 & 21.38$\pm$0.09 \\
ESO539-007 & -1.11$\pm$1.14 & 0.97$\pm$0.90 & -5.36$\pm$0.44 & 4.58$\pm$0.33 & -3.73$\pm$0.71 & 3.22$\pm$0.50 & 0.24$\pm$0.04 & 24.37$\pm$0.14 \\
ESO150-005 & -0.27$\pm$0.18 & 0.36$\pm$0.14 & -0.77$\pm$0.50 & 1.89$\pm$0.35 & -0.56$\pm$0.40 & 1.57$\pm$0.28 & 0.24$\pm$0.04 & 24.19$\pm$0.14 \\
NGC0100 & 0.87$\pm$0.18 & 0.02$\pm$0.12 & -1.15$\pm$0.85 & 4.10$\pm$0.57 & -2.05$\pm$0.68 & 4.10$\pm$0.45 & ---$\pm$--- & ---$\pm$--- \\
NGC0115 & 0.06$\pm$0.04 & 0.16$\pm$0.04 & -0.48$\pm$0.13 & 1.83$\pm$0.11 & -0.51$\pm$0.13 & 1.63$\pm$0.11 & 0.52$\pm$0.03 & 21.27$\pm$0.17 \\
UGC00260 & 0.04$\pm$0.07 & 0.29$\pm$0.09 & -1.29$\pm$0.13 & 3.76$\pm$0.16 & -1.43$\pm$0.14 & 3.57$\pm$0.16 & 0.27$\pm$0.02 & 22.96$\pm$0.26 \\
NGC0131 & 0.11$\pm$0.08 & 0.34$\pm$0.08 & 0.00$\pm$0.12 & 2.86$\pm$0.10 & -0.15$\pm$0.08 & 2.55$\pm$0.06 & ---$\pm$--- & ---$\pm$--- \\
UGC00320 & 0.12$\pm$0.17 & 0.20$\pm$0.15 & 0.15$\pm$0.46 & 1.61$\pm$0.36 & -0.01$\pm$0.31 & 1.44$\pm$0.23 & 0.50$\pm$0.03 & 22.88$\pm$0.17 \\
                                ... & & & & & & & & \\
Total\tablenotemark{4} & 1541 & & 1541 & & 1541 & & 992 & \\
                                \end{tabular}
\begin{tablenotes}
    \item[1] Same nomenclature as the $\sfg$.
    \item[2] Slope of linear fit and 1-$\sigma$ uncertainty obtained
      with \textit{scipy.optimize.curve} package. We applied the
      cutoffs at $R/R80$=0.5 and $\mu_{[3.6]}$=23.5 mag\,arcsec$^{-2}$ only to the linear fit to the $\mu_{[3.6]}$ vs kpc data. 
      In the case of colors, only the radial cutoff at $R/R80$=0.5 is applied. We also applied a simple inclination correction to the data by adding $-2.5\log_{10}(b/a)$ where $a$ and $b$ are the semi-major and semi-minor axes respectively.
    \item[3] Y-intercept of linear fit with uncertainty.
    \item[4] Total number of successful fits for each column. There are three galaxies with $T < -3.5$ (E galaxies, namely ESO548-023, NGC4278, and NGC5173) that are included in the $\mu_{[3.6]}$ vs kpc column, bringing the total to 992 galaxies, but are removed from the subsample for further analysis.
\end{tablenotes}
                \end{threeparttable}
	\end{center}
\end{table*}

\begin{table*}
	\begin{center}
	\caption{GRADIENTS OF COLOR VERSUS 3.6 MICRON SURFACE BRIGHTNESS PROFILES}
	\label{Table:slopestable2}

		\begin{threeparttable}
                		\begin{tabular}{ l|rr|rr|rr }
                		&
                                \multicolumn{2}{ |c| }{(FUV\,$-$\,NUV)/$\mu_{[3.6]}$}& 
                                \multicolumn{2}{ |c| }{(FUV\,$-$\,[3.6])/$\mu_{[3.6]}$} & 
                                \multicolumn{2}{ |c| }{(NUV\,$-$\,[3.6])/$\mu_{[3.6]}$}\\
                                \hline
			       \multicolumn{1}{l}{unit}&
                                \multicolumn{2}{ |c| }{mag/(mag/arcsec$^{2}$)}& 
                                \multicolumn{2}{ |c| }{mag/(mag/arcsec$^{2}$)} & 
                                \multicolumn{2}{ |c| }{mag/(mag/arcsec$^{2}$)}\\
                                \hline
                                \multicolumn{1}{l}{name\tablenotemark{1}}&
                                \multicolumn{1}{|c}{a\tablenotemark{2}}&
                                \multicolumn{1}{c|}{b\tablenotemark{3}}&
                                \multicolumn{1}{c}{a}&
                                \multicolumn{1}{c|}{b}&
                                \multicolumn{1}{c}{a}&
                                \multicolumn{1}{c|}{b}\\
				\hline
UGC00017 & -0.03$\pm$0.08 & 0.94$\pm$1.92 & -0.69$\pm$0.18 & 18.83$\pm$4.39 & -0.63$\pm$0.10 & 17.21$\pm$2.30 \\
ESO409-015 & 0.21$\pm$0.03 & -4.78$\pm$0.64 & 0.93$\pm$0.07 & -21.89$\pm$1.64 & 0.72$\pm$0.05 & -17.13$\pm$1.08 \\
ESO293-034 & -0.01$\pm$0.02 & 0.79$\pm$0.48 & -0.40$\pm$0.04 & 11.81$\pm$0.88 & -0.42$\pm$0.03 & 11.61$\pm$0.62 \\
NGC0210 & -0.09$\pm$0.04 & 2.27$\pm$0.96 & -1.01$\pm$0.18 & 26.28$\pm$4.03 & -0.79$\pm$0.13 & 20.81$\pm$2.84 \\
ESO079-005 & -0.04$\pm$0.03 & 1.33$\pm$0.75 & -0.40$\pm$0.10 & 10.96$\pm$2.41 & -0.40$\pm$0.08 & 10.69$\pm$1.76 \\
NGC0216 & 0.26$\pm$0.01 & -5.20$\pm$0.22 & 0.39$\pm$0.03 & -5.96$\pm$0.74 & 0.13$\pm$0.03 & -0.74$\pm$0.60 \\
PGC002492 & -0.09$\pm$0.03 & 2.30$\pm$0.64 & -0.50$\pm$0.05 & 13.43$\pm$1.23 & -0.46$\pm$0.05 & 12.20$\pm$1.18 \\
IC1574 & 0.19$\pm$0.05 & -4.10$\pm$1.20 & 0.64$\pm$0.16 & -13.28$\pm$3.86 & 0.41$\pm$0.13 & -8.35$\pm$2.98 \\
NGC0244 & 0.24$\pm$0.09 & -4.96$\pm$1.99 & 0.38$\pm$0.03 & -6.72$\pm$0.75 & 0.13$\pm$0.06 & -1.45$\pm$1.32 \\
PGC002689 & -0.08$\pm$0.05 & 1.99$\pm$1.13 & -0.08$\pm$0.19 & 2.80$\pm$4.53 & 0.03$\pm$0.16 & 0.11$\pm$3.78 \\
UGC00477 & -0.04$\pm$0.05 & 1.17$\pm$1.21 & -0.68$\pm$0.06 & 17.59$\pm$1.43 & -0.63$\pm$0.03 & 16.10$\pm$0.71 \\
ESO411-013 & -0.25$\pm$0.15 & 6.24$\pm$3.58 & -0.50$\pm$0.07 & 13.91$\pm$1.60 & -0.35$\pm$0.19 & 10.11$\pm$4.56 \\
NGC0247 & -0.06$\pm$0.02 & 1.76$\pm$0.51 & -0.23$\pm$0.11 & 7.87$\pm$2.33 & -0.10$\pm$0.09 & 4.73$\pm$1.95 \\
                                ... & & & & & &\\
Total\tablenotemark{4} & 1650 & & 1650 & & 1650 & \\
                                \end{tabular}
\begin{tablenotes}
    \item[1] Same nomenclature as the $\sfg$. Sorted by right ascension.
    \item[2] Slope of linear fit and 1-$\sigma$ uncertainty obtained
      with \textit{scipy.optimize.curve} package. We applied the
      cutoff $\mu_{[3.6]}$=20.89 mag\,arcsec$^{-2}$. No inclination correction is applied in these cases.
    \item[3] Y-intercept of linear fit with uncertainty.
    \item[4] Total number of successful fits for each column.
\end{tablenotes}
                \end{threeparttable}
	\end{center}
\end{table*}

\begin{table*}
\begin{center}
\caption{$\galex$/$\sfg$ sample radial ranges}
\label{Table:ranges}
\begin{tabular}{lrrrrr}
\hline
  \multicolumn{1}{l}{Morph\tablenotemark{a}} &
  \multicolumn{1}{r}{N\tablenotemark{b}} &
  \multicolumn{1}{c}{$R80$ range\tablenotemark{c}} &
  \multicolumn{1}{c}{max range\tablenotemark{d}} &
  \multicolumn{1}{c}{$<$$R80$$>$\tablenotemark{e}} &
   \multicolumn{1}{c}{$<$max$>$)\tablenotemark{f}}\\
\hline
& & kpc & kpc & kpc & kpc\\
\hline
E & 11 & 1.86 --- 7.66 & 13.48 --- 61.95 & 4.73 & 39.84 \\
E-S0 & 10 & 0.94 --- 5.03 & 5.00 --- 47.92 & 3.21 & 24.48 \\
S0 & 21 & 1.01 --- 12.22 & 6.41 --- 57.85 & 3.45 & 25.47 \\
S0-a & 47 & 1.44 --- 15.14 & 10.91 --- 85.28 & 4.34 & 30.74 \\
Sa & 133 & 1.46 --- 13.35 & 9.25 --- 86.54 & 5.27 & 36.63 \\
Sb & 289 & 0.97 --- 17.21 & 5.49 --- 107.31 & 5.76 & 35.88 \\
Sc & 553 & 0.88 --- 14.94 & 3.67 --- 58.02 & 6.16 & 31.26 \\
Sd & 120 & 1.24 --- 13.76 & 4.14 --- 59.61 & 5.41 & 22.74 \\
Sm & 114 & 0.59 --- 10.93 & 2.95 --- 62.56 & 4.96 & 19.86 \\
Irr & 101 & 0.64 --- 11.62 & 1.67 --- 80.25 & 3.83 & 15.36 \\
\hline
Total & 1399 & & & & \\
\end{tabular}
\tablenotetext{1}{RC2 morphological types}
\tablenotetext{2}{Number of galaxies. The total number is small due to the }
\tablenotetext{3}{Smallest and largest $R80$ distance in kiloparsec}
\tablenotetext{4}{Smallest and largest maximum size of galaxies in kiloparsec}
\tablenotetext{5}{Average $R80$}
\tablenotetext{6}{Average maximum}
\end{center}
\end{table*}

\subsection{Color-color diagrams} \label{sec:CCD}
\begin{figure*}
\begin{center}
\includegraphics[width=1.0\textwidth]{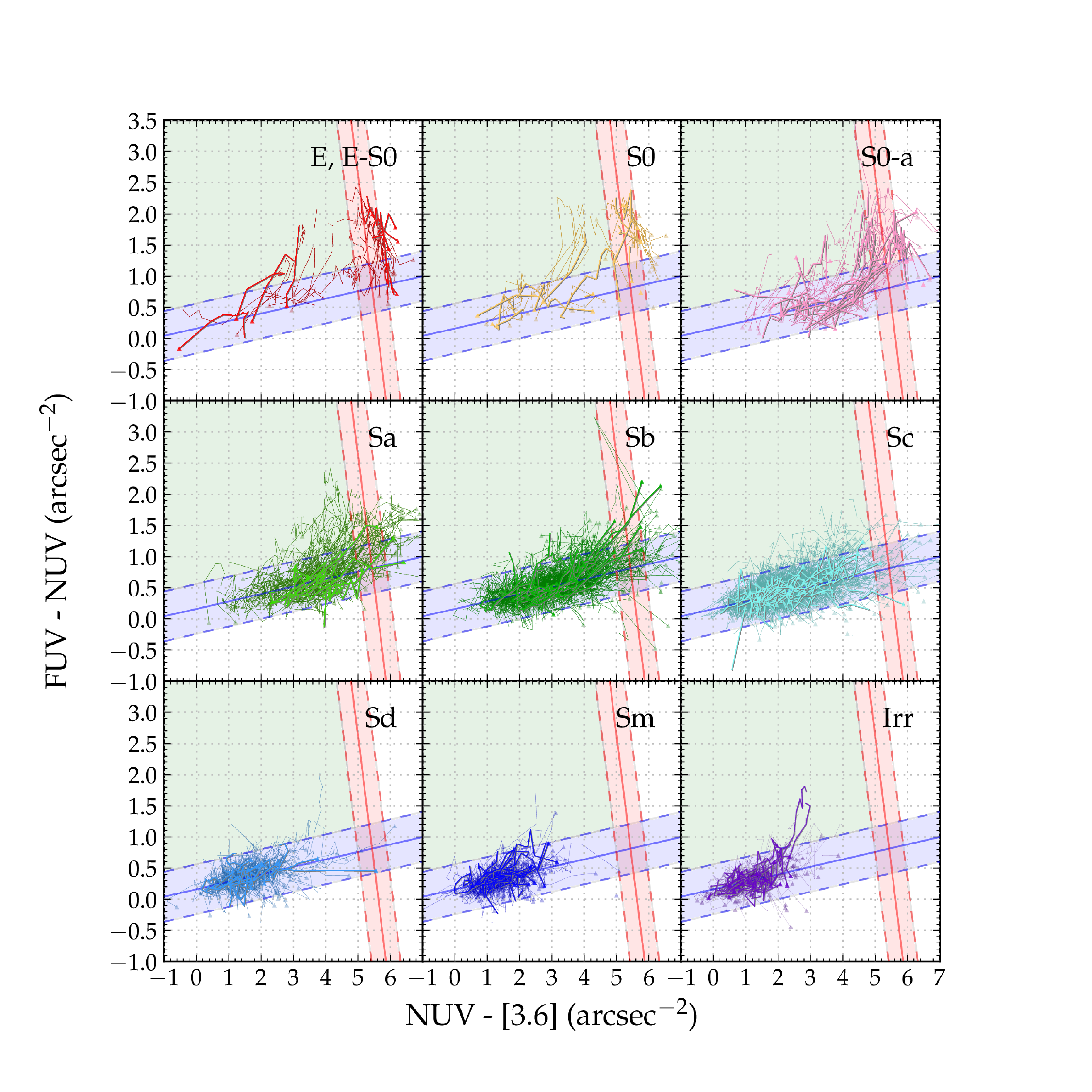}
\caption{
(FUV\,$-$\,NUV) vs (NUV\,$-$\,[3.6]) spatially resolved color-color diagrams per morphological type. The regions delineated by a solid line and two parallel dashed lines are the GBS in blue and the GRS in red, and the region in green in the upper-left quadrant is the GGV, as defined in \citet{Bouquin2015}. Measurements at the center are represented by triangles, and other measurements, as we move radially outward every 6\arcsec, are represented by dots connected by a line for each galaxy. Randomly selected galaxies are emphasized in each panel for better visualization.
\label{fig:ccdFNN36m}
}
\end{center}
\end{figure*}

\begin{figure*}
\begin{center}
\includegraphics[width=1.0\textwidth]{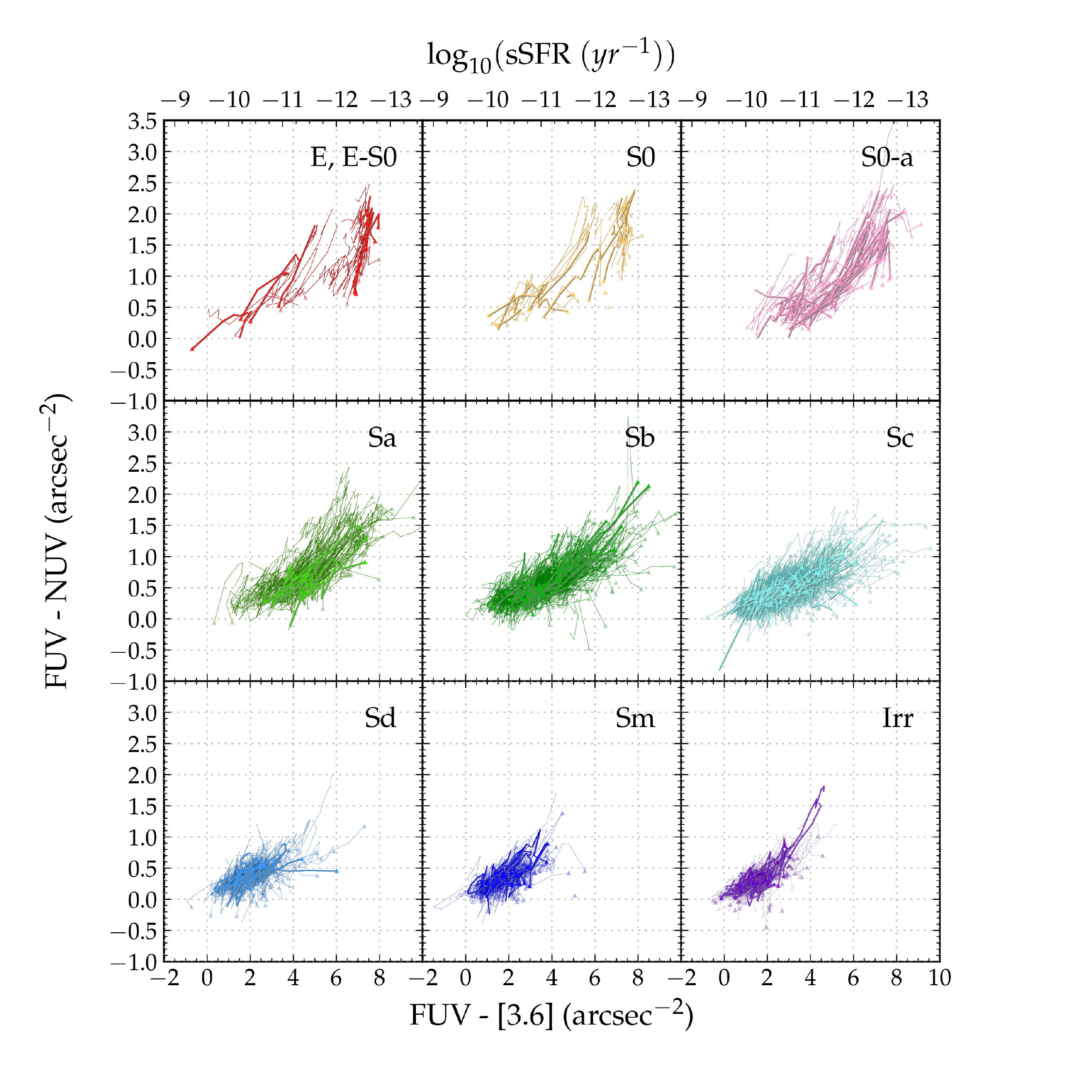}
\caption{
(FUV\,$-$\,NUV) vs (FUV\,$-$\,[3.6]) spatially resolved color-color diagrams per morphological type.
\label{fig:ccdFNF36m}
}
\end{center}
\end{figure*}

From the colors measured above, we formed three color-color diagrams,
namely (FUV\,$-$\,NUV) vs (NUV\,$-$\,[3.6]) (Figure~\ref{fig:ccdFNN36m}), 
(FUV\,$-$\,NUV) vs (FUV\,$-$\,[3.6]) (Figure~\ref{fig:ccdFNF36m}), and 
(FUV\,$-$\,[3.6]) vs (NUV\,$-$\,[3.6]) (not shown). 
The color-color diagrams presented here show the galaxies separated into 9 panels of separate morphological type.

Comparing the (FUV\,$-$\,NUV) vs (NUV\,$-$\,[3.6]) and the
(FUV\,$-$\,NUV) vs (FUV\,$-$\,[3.6]) color-color diagrams, we
can see that the two sequences are more distinguishable in the former. 
This is mainly caused by the fact that the GRS is orthogonal
to the GBS in the case of the (FUV\,$-$\,NUV) vs (NUV\,$-$\,[3.6]) diagram.
This is due to the fact that the strength of the UV upturn also increases with the stellar mass surface density. 
This is also the case when considering the total galaxy mass \citep{Boselli2005}.
We cannot determine here whether this is due to the stellar
populations at high stellar mass surface densities hosting either an
important helium rich or metal poor HB population \citep{Yi2005,Yi2011},
or whether it is related to changes in the IMF \citep[as suggested by][]{Zaritsky2014,Zaritsky2015a}.

The (FUV\,$-$\,NUV) vs (NUV\,$-$\,[3.6]) color-color diagram is where we defined the GBS, GRS, and GGV subsamples
from the galaxies' integrated (asymptotic) magnitudes,
by visually separating the distribution into two regions and 
fitting an error-weighted least-square line to each region \citep{Bouquin2015}.
With our current spatially resolved data, we can see the
spatially resolved (radially, at least) color evolution of galaxies in these three categories.
While ETGs such as E, E-S0, S0, S0-a, and Sa galaxies span across both the GBS and GRS regions,
LTGs such as Sb, Sc, Sd, Sm, and Irregular galaxies have this color much more constrained,
and have their entire profile mostly located within the GBS region
(mean $\pm$ 2$\sigma$).

In the panels for the E, E-S0, S0, and S0-a types (top row) of the 
(FUV\,$-$\,NUV) vs (NUV\,$-$\,[3.6]) (Figure~\ref{fig:ccdFNN36m}) and (FUV\,$-$\,NUV) vs (FUV\,$-$\,[3.6]) (Figure~\ref{fig:ccdFNF36m}) color-color diagrams, 
the galaxies are distributed into two regions,
the bottom-left (blue-blue) and the top-right (red-red) parts in both color-color diagrams: 
the ones with the bluest central region have redder disks in
(FUV\,$-$\,NUV), as well as in both (NUV\,$-$\,[3.6]) and (FUV\,$-$\,[3.6]) colors; 
the others with the reddest central region also have redder disks in
(FUV\,$-$\,NUV), but not much in (NUV\,$-$\,[3.6]) or (FUV\,$-$\,[3.6]).
In both cases, their central regions (triangles) are bluer in (FUV\,$-$\,NUV) color than their outer parts.
If the blueing were caused by residual star formation (RSF), which contributes in both FUV and NUV,
the observed data points would be bluer in all three colors.
This is indeed the case for the ETGs seen in the bottom-left (well
within the GBS) in both color-color diagrams,
where RSF is more prominent in their central regions. Note that the
innermost 6\,arcsec (in semi-major axis, i.e$.$, 12\,arcsec in major axis) are
excluded so the potential contribution of AGN should not be affecting these results in a direct way.

For the ETGs in the top-right of these plots, there is
a difference between the (NUV\,$-$\,[3.6]) and (FUV\,$-$\,[3.6]) colors.  While in
the (FUV\,$-$\,NUV) vs (NUV\,$-$\,[3.6]) color-color diagram the distribution of
these reddest systems has a negative slope (which provides a better
isolation of the GRS), it has a positive slope in the (FUV\,$-$\,NUV) vs
(FUV\,$-$\,[3.6]) color-color diagram.  The central regions of these
galaxies are bluer in (FUV\,$-$\,[3.6]) than in (NUV\,$-$\,[3.6]), which is the
sign of a weaker contribution from the emitter of the UV radiation in
these systems in the NUV than in the FUV, compared to GBS galaxies.
This can probably be attributed to evolved (UV-upturn) stars.

Our color-color diagrams are, thus, able to segregate and allow us to
extract the properties of a whole range of galaxies, from star-forming
LTGs, to ETGs with and without RSF. For ETGs, they allow us to
directly see the effect of UV-upturn stars, which can only be done in
the UV-to-IR colors. In this regard we find that RSF in ETGs seems to be
concentrated in the center and the UV-upturn is also stronger as we
move to the inner regions of red (in NUV\,$-$\,[3.6]) ETGs. 
However, it should be noted that recent study by \citet{Yildiz2017} 
have shown that a not-insignificant fraction, 20\%, of field (non-Virgo)
nearby galaxies have disks or rings of HI gas around them, and that
their UV profiles are closely tied to their HI gas reservoir.

This color-color
diagram does not allow us to clearly determine whether the UV-upturn
is also present in the bulges of early-type spiral galaxies (such as
in the case of M31; Brown 2004) as they are located in a position
similar to that expected for turn-off stars in these bulges. We can
nevertheless conclude that in galaxies with morphological types later
than Sc the light from HB stars is clearly overshone by these turn-off
stars of progressively higher masses (statistically speaking) as we move to later types.

\subsection{$\galex$ Green Valley galaxies} \label{sec:GGV}
A subsample of 70 $\galex$ Green Valley (GGV) galaxies was identified in the 
(FUV\,$-$\,NUV) vs
(NUV\,$-$\,[3.6]) integrated color-color
diagram by \citet{Bouquin2015}.

As already pointed out in that paper, these objects can be interpreted
as galaxies that have either left the GBS and are ``transitioning'' to
eventually reach the GRS or were previously in
the GRS and are now experiencing a modest rebirth or rejuvenation (in terms
of the light-weighted ages of their stellar populations) and are evolving back to
the GBS. 

In the former scenario star formation would have been suppressed (or, at least, damped), either by
starvation from having used up all the gas or by ram-pressure
stripping, or by quenching due to the perturbations induced from AGN,
merger events, or some other gas-heating process. OB stars would not
form any longer and the FUV and NUV emissions decrease, with the FUV
emission evolving faster than the NUV because of the shorter lifespan of the most
massive stars, resulting in a progressive reddening of their
(FUV\,$-$\,NUV) color. 

In the case of the latter (rejuvenation) scenario, these galaxies
would have started to form stars on top of relatively passively
evolving galaxies either by the accretion of new gas or by cooling gas
that was already present in the galaxy in a hotter phase. 

\begin{figure}[ht!]
\begin{center}
\includegraphics[width=0.49\textwidth]{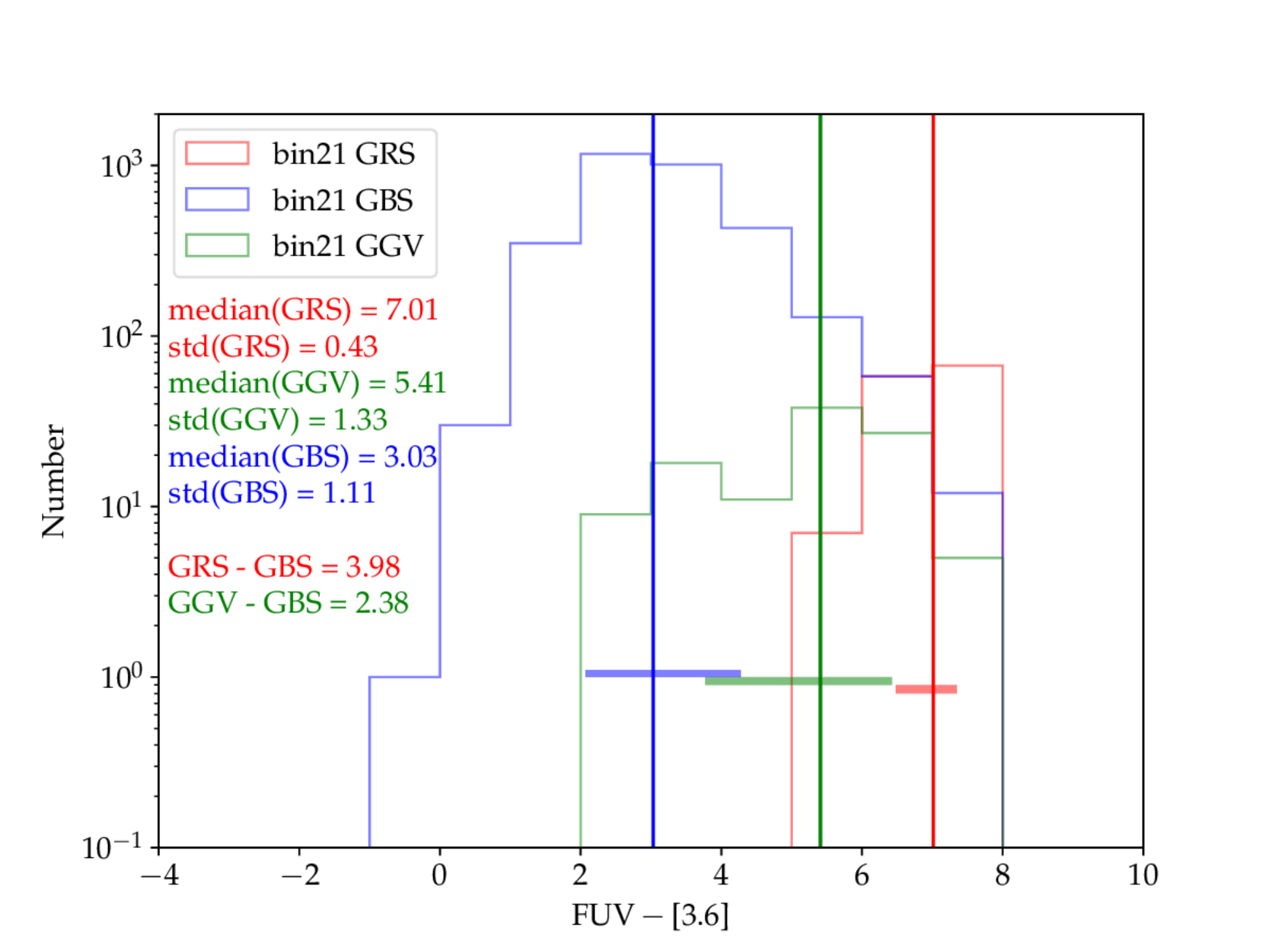}
\includegraphics[width=0.49\textwidth]{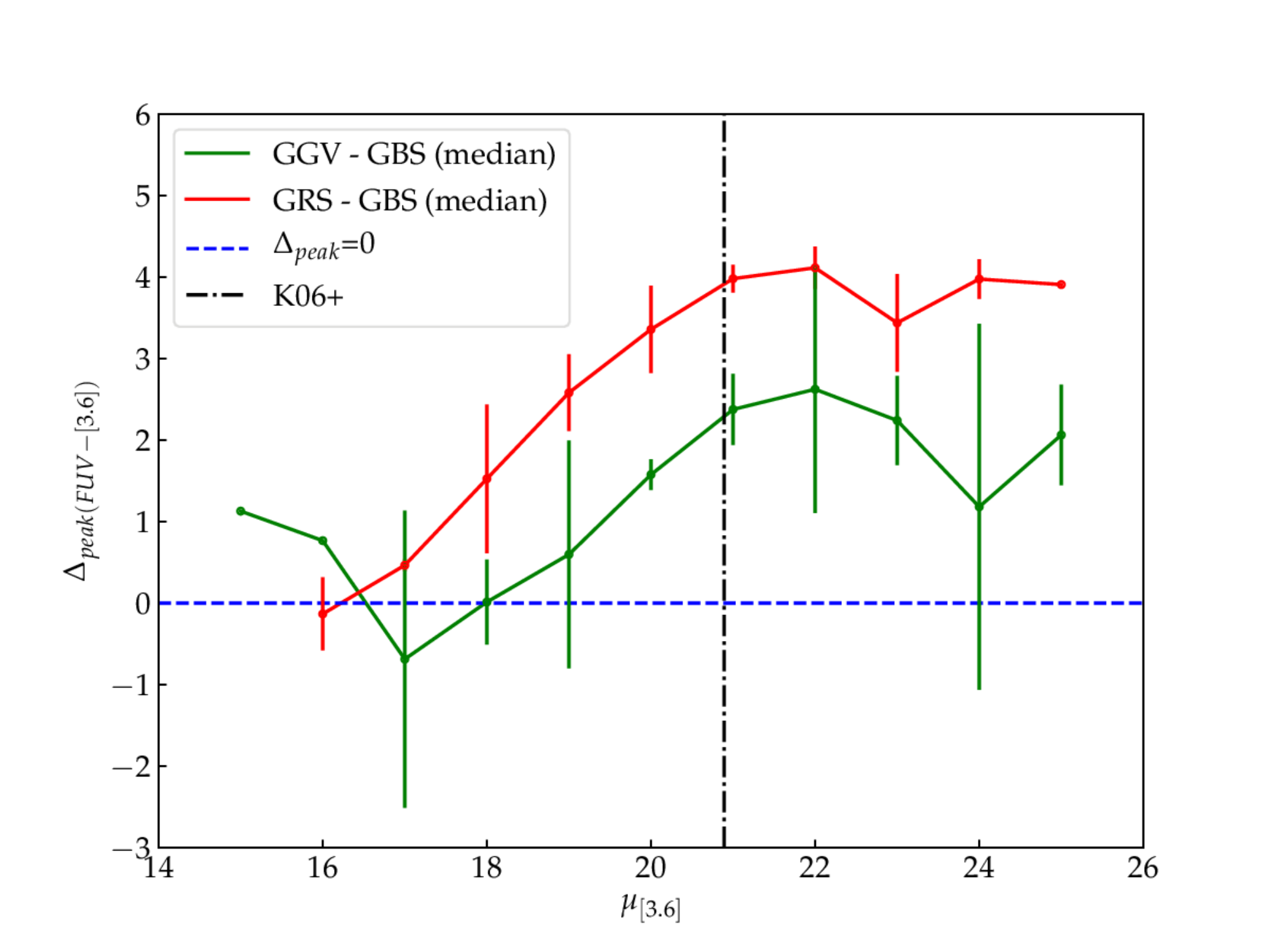}
\caption{Top: the number counts of data points of GBS (blue), GGV (green), and GRS (red) galaxies within a 3.6\,$\mu$m surface brightness bin 21\,$\leq$\,$\mu_{[3.6]}$\,$<$22, in the (FUV\,$-$\,[3.6]) color versus $\mu_{[3.6]}$ surface brightness plot. Solid lines correspond to the median and the horizontal shaded area represents the extent of one standard deviation above and below the mean (not shown), all in their respective colors.
Bottom: in red, the difference of the peak (FUV\,$-$\,[3.6]) color (i.e$.$ the difference of the median) of the GRS distribution and the GBS distribution for each $\mu_{[3.6]}$ bins. The same in green for the difference between the peaks of the distributions of GGV and GBS. The horizontal dashed blue line at $\Delta_{peak}$=0, and the vertical dot-dashed black line at $\mu_{[3.6]}$=20.892 mag\,arcsec$^{-2}$ of \citet{Kauffmann2006}, are shown for references. Errorbars represent the 15.865 and 84.135 percentiles of the GRS and GGV distributions obtained by using the IQR method.
\label{fig:histocounts}
}
\end{center}
\end{figure}

The results presented above provide another fundamental piece of
evidence for the origin of these transitioning objects. In particular,
we have shown that the outer parts of most GGV galaxies are redder
than their inner parts and that this reddening is progressive (see,
e.g$.$, Figure~\ref{fig:histocounts}). In the case of the quenching
scenario this implies that the mechanism responsible for the quenching
is acting in an outside-in fashion. Should the rejuvenation scenario
be happening then these galaxies would be starting to form stars from
inside-out. As the associated blue colors are not limited to the
very central regions this would likely imply the growth of a disk,
again, in an inside-out fashion.  

With regard to the mechanism(s) that could potentially lead to the
supression of the star formation in the outskirts we showed in
\citet{Bouquin2015} that the GGV has the highest fraction of Virgo
cluster galaxies, with 20 (out of 70) GGV objects in the Virgo
cluster, i.e$.$ $\sim$29\%, in comparison to a fraction of Virgo cluster
galaxies in the GBS of only $\sim$7\% (124/1753) and in the GRS of $\sim$18\% (14/79).
For example, one ram pressure model in Virgo \citep{Boselli2006} creates an inverted color gradient
compared to late-type field galaxies, with redder outer disks and bluer inner parts.

We also analyze whether the GGV objects are mainly
located in groups where environmental effects might start to occur
\citep[in particular, strangulation;][]{Kawata2008}. 
Amongst the 70 GGV galaxies of our sample,
28 (40\%) are field galaxies, and 42 (60\%) are in groups or clusters.
We see that the fraction of field galaxies decreases to 30\% 
while the fraction of group galaxies increases to 70\% (56/79) in the case of GRS galaxies.
In contrast, the fraction of field/group galaxies is 51\%/49\% in the case of GBS galaxies,
and that of the overall sample is 50\%/50\%.
That is, we see an increase in fraction of galaxies belonging to groups as we go
from the GBS to the GRS.
This result hints that the disk-reddening that we see in GGV galaxies is 
likely due to a mechanism that is favored in dense environments.
We note that this result does not exclude rejuvenation scenarios,
as many ETGs with extended star formation are recently now being identified \citep{Salim2012,Fang2012,Yildiz2017}.

%
\section{Modeling 3.6\,$\mu$m exponential disks} \label{sec:modeling}
The linear disk fits were compared to the profiles of BP00 disk models, generated with various circular velocities and spin parameters. 
These are simple disk models, without any bulge, bar, or mass outflow features, calibrated on the Milky Way (MW), with the assumption that our Galaxy is a typical spiral galaxy \citep{BP99} (BP99), and using simple scaling relations to extend the initial model to other spirals (BP00). These models grow inside-out with an infall of primordial gas (i.e$.$ low-metallicity) with radially varying and exponentially decreasing infall rate with time. They include realistic yields and lifetimes from stellar evolution models and metallicity-enhancement by SNIa, and adopt a Kroupa IMF. The local SFR varies with the gas surface density and the angular velocity. The chemical and photometric evolution of the disk is then followed within this self-consistent framework. 
The rotational velocity, $v_{\mathrm{c}}$, is related to the total baryonic mass \citep{Mo1998} and is implemented in a relative way with respect to the  Milky Way model: 
\begin{equation}
\frac{v_{c}}{220} = \left(\frac{M}{M_{MW}}\right)^{1/3}
\end{equation}
and the dimensionless spin parameter $\lambda$ is defined as \citep{Peebles1969}
\begin{equation}
\lambda = J \vert E \vert^{1/2} G^{-1} M^{-5/2}
\end{equation}
where $M$ is the total baryonic mass, $M_{MW}$ is the total baryonic mass of the Milky Way, and 220 (km s$^{-1}$) is the circular velocity of the Milky Way, $J$ is the angular momentum, $E$ is the energy, of the halo, and $G$ is the gravitational constant. In the BP00 models, the spin parameter only influences the scalelength of the disk with respect to the MW: 
\begin{equation}
\frac{R}{R_{MW}} = \frac{V}{V_{MW}}  \frac{\lambda}{\lambda_{MW}}
\end{equation}
where $R$ and $\lambda$ are the scalelength and the spin parameter of the considered model, 
and $R_{MW}$ and $\lambda_{MW}$ are those of the MW.
We show that we are able to obtain circular velocities and spin for the galaxies of our sample from this method (Section~\ref{sec:velandspin}).
Finally, we show color gradients against circular velocity, spin parameter, and stellar mass of our sample (Section~\ref{sec:colorgradient}).
In particular, gradients are positive at $\sim$50 km/s , the average is flat at $\sim$75 km/s , while above $\sim$100 km/s, most galaxies have negative gradients in all three colors.

\subsection{Obtaining circular velocity and spin}\label{sec:velandspin}
\begin{figure}[ht!]
\begin{center}
    	\includegraphics[width=0.5\textwidth]{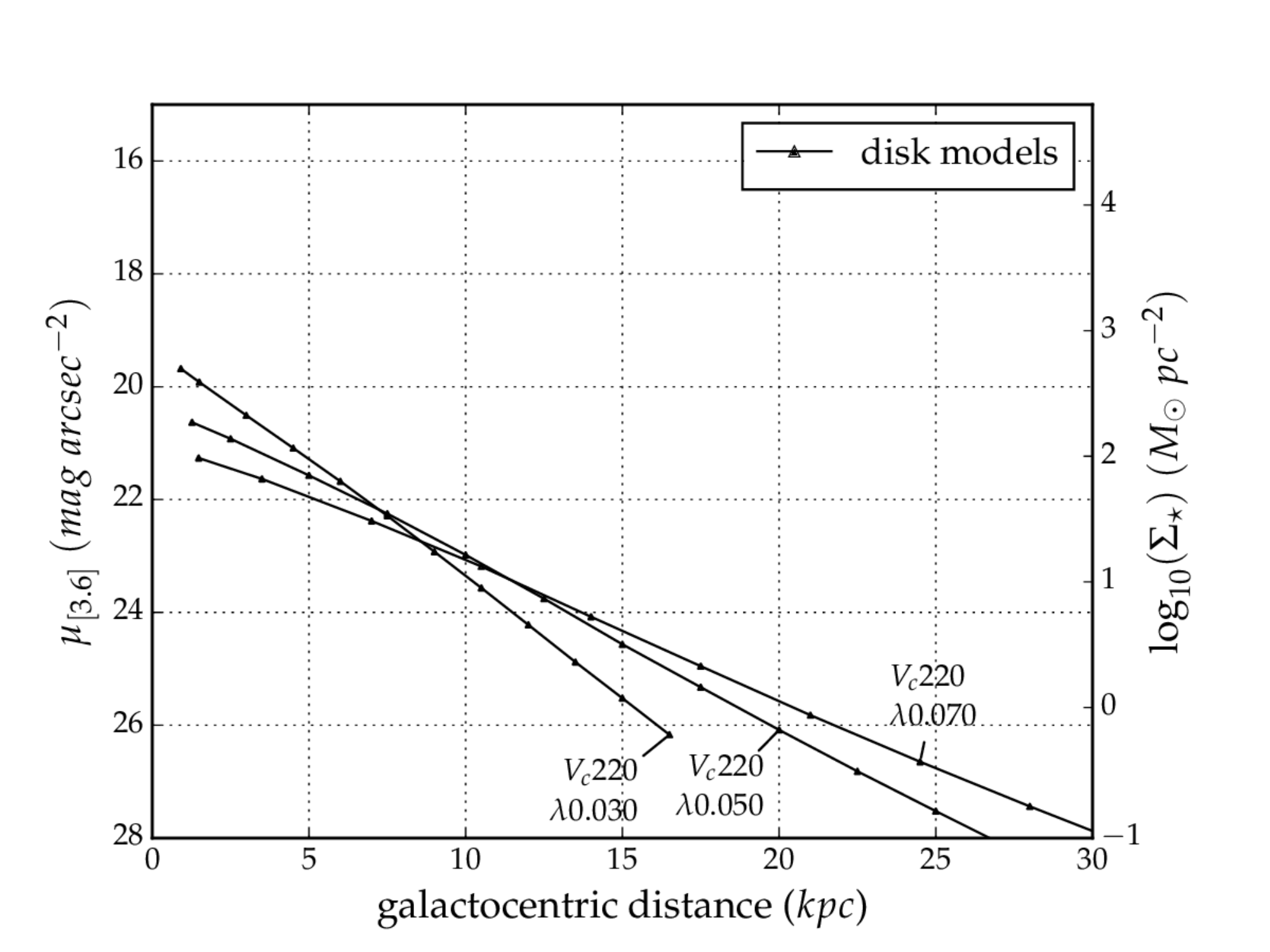}
    	\includegraphics[width=0.5\textwidth]{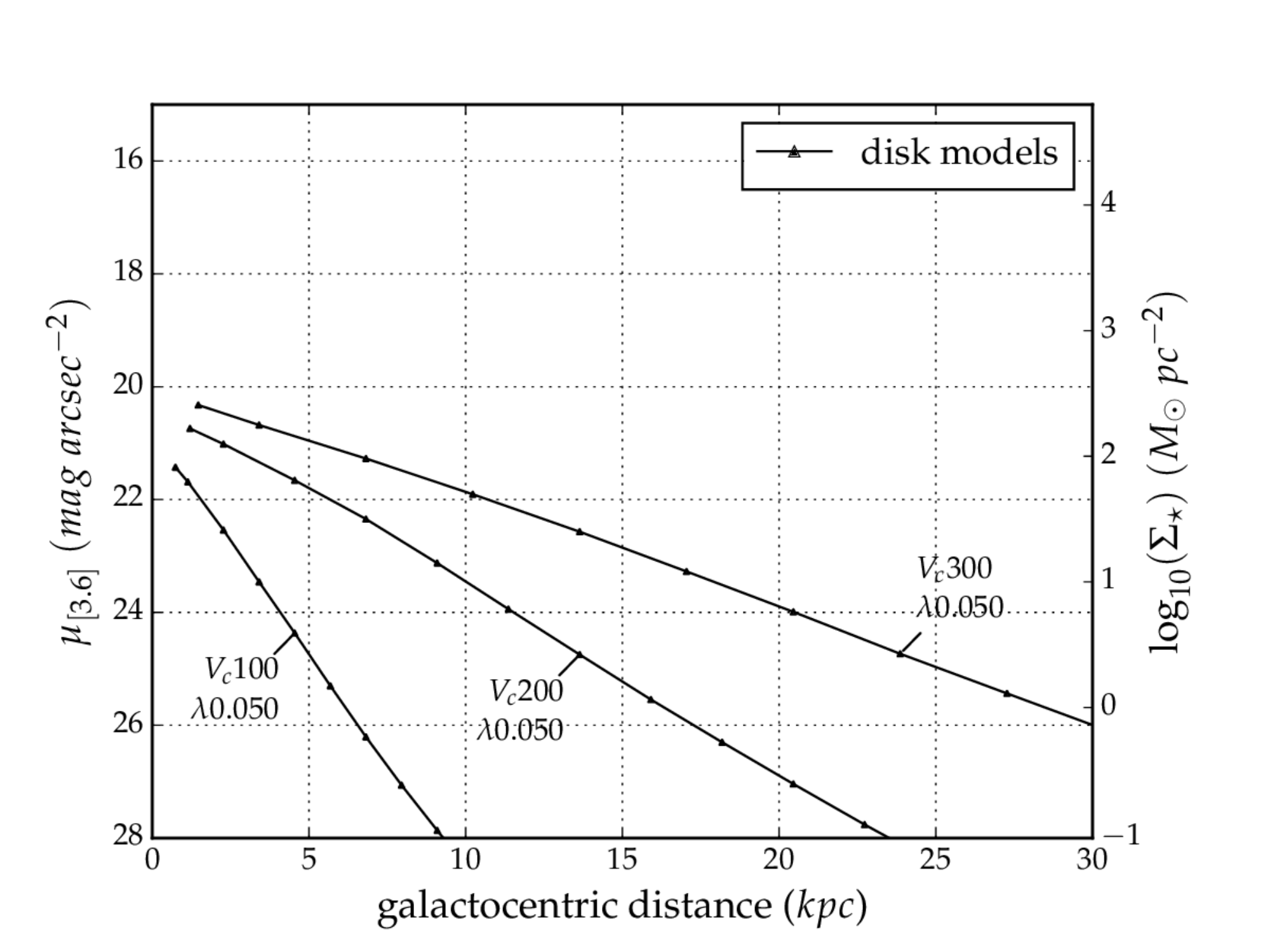}
\caption{
Examples of 3.6\,$\mu$m surface brightness profiles of BP00 disk-models with fixed circular velocity $v_{\mathrm{c}}$ and variable spin parameter $\lambda$ (\textit{top}), and with fixed spin parameter and variable circular velocity (\textit{bottom}).
\label{fig:samprofiles}
}
\end{center}
\end{figure}

In this study, we use the disk models of BP00 as in the version presented in \citet{MunozMateos2011},
but increasing the sampling and range spanned by the model parameters,
namely circular velocity $v_{\mathrm{c}}$ and spin $\lambda$.
As mentioned above, these are bulgeless, disk-only models, that naturally grow \textit{inside-out}
from gas infall and are left to run for $T$=13.5 Gyr to the
present. 
They include scaling laws so that mass scales as $v^{3}$ and scale length as $\lambda \times v$ \citep{Mo1998}.
As can be seen in Figure~\ref{fig:samprofiles},
an increase in circular velocity $v_{\mathrm{c}}$ leads to an
increase in both the total stellar mass and the disk scale-length, 
whereas increasing the spin parameter $\lambda$ only increases the scalelength.
Correcting our observed galaxies for inclination (see Section~\ref{sec:diskseparation}) 
leads to a dimming in surface brightness at all radii, and thus eventually, would yield a lower circular velocity
and a larger spin than when not applying the correction.

These models are aimed to reproduce the multiwavelength SB profile by varying only those two parameters. Other assumptions were calibrated in the Milky Way model (BP99) and in nearby disks (BP00). 
Predictions for disks with different spins and velocities are based on $\Lambda$CDM scaling laws.
Disk models were generated for various spin parameters and circular velocity combinations:
the spin ranges from 0.002 to 0.15, inclusively, and varying by a step of 0.001, i.e$.$, 149 different spins,
while the velocity ranges from 20 to 430\,km\,s$^{-1}$, inclusively, and varying by a step of 10\,km\,s$^{-1}$, i.e$.$, 42 different velocities.
The total number of models generated is 6258.
We fit these models with an error-weighted linear fit (in surface brightness scale) in a similar manner to what we do with our data points.
It is, however, necessary to insert an uncertainty to the data point of each model
in order to compute the reduced-$\chi^{2}$ of the fit and to
determine whether an exponential law properly describes also the
radial distribution of the UV through near-infrared light in these
models. A reasonable assumption in this regard is 0.10 - 0.15\,mag (see e.g \citet{MunozMateos2011}).
Indeed, a value of 0.15\,mag yields a reduced-$\chi^{2}$ close to unity for most of the models.

\begin{figure*}[ht!]
\begin{center}
\includegraphics[width=1.0\textwidth]{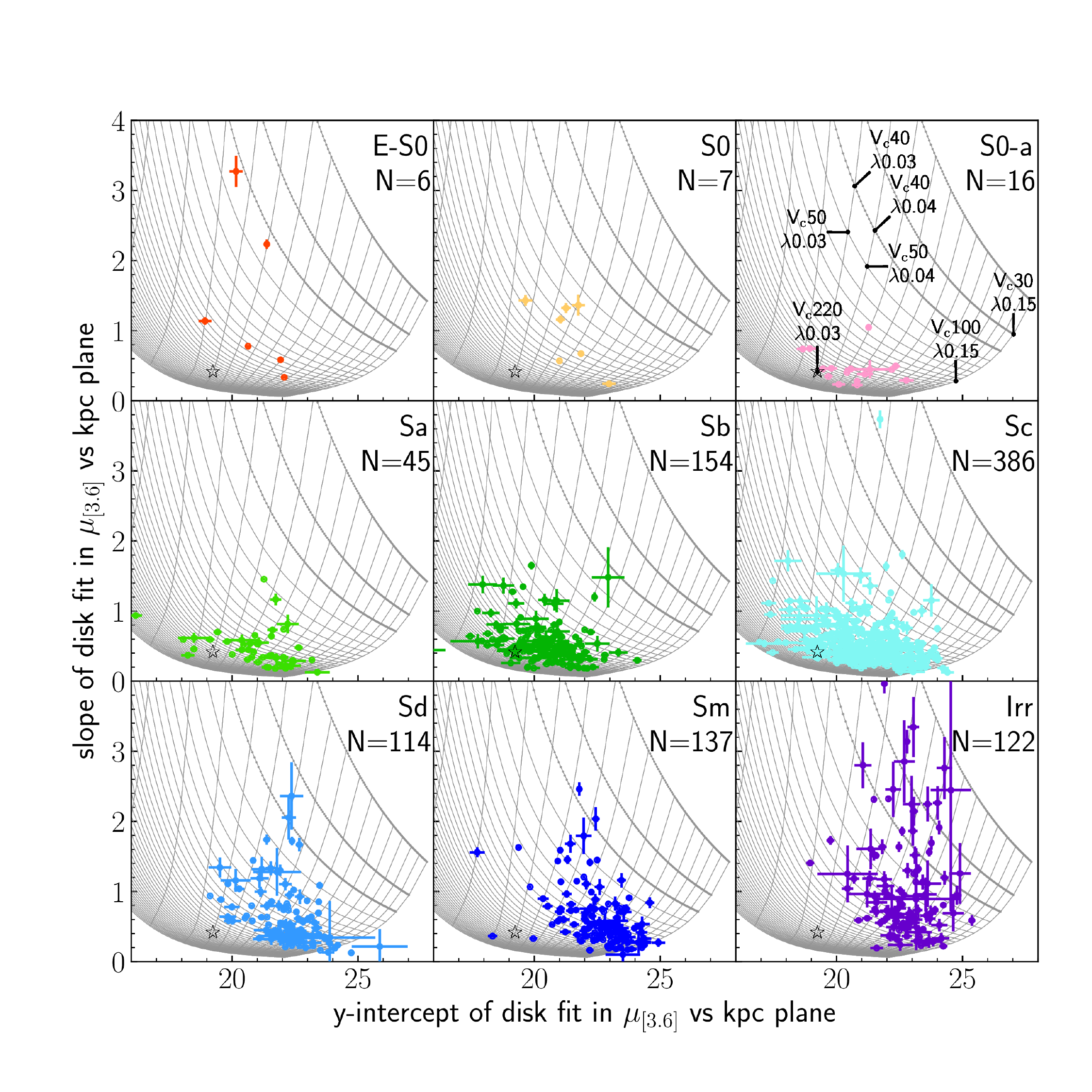}
\caption{
Slope (i.e$.$ scale length) versus y-intercept (i.e$.$ central SB) of the linear fits to the disks of the $\sfg$
galaxies in the [3.6] micron radial profile $\mu_{[3.6]}$ (in surface brightness scale) vs kpc plane, 
and where the different panels correspond to different morphological
types. Cutoffs for the linear fit in surface brightness scale were set
at $R/R80$=0.5 and $\mu_{[3.6]}$=23.5\,mag\,arcsec$^{-2}$. 
Color-coding by morphological type is the same as in the previous figures.
The total number of galaxies in this plot is 987.
The star marker, at $v_{\mathrm{c}}=220$ km s$^{-1}$ and $\lambda=0.03$, represents the circular velocity and spin parameter of the Milky Way and is present in all panels as reference.
\label{fig:samgridm}
}
\end{center}
\end{figure*}

Figure~\ref{fig:samgridm} shows the slopes and $y$-intercepts obtained
from the fits to the IR surface brightness profiles (corrected for
inclination) of our galaxy sample plotted along with the grid of
slopes and y-intercept obtained from fits to the BP00 models described
above. Data errorbars are coming from the slope and y-intercept
fitting errors obtained from the weighted fits to our surface
brightness profiles. The errorbars in
the slopes and $y$-intercepts of the models are omitted for
simplicity. They are separated by morphological type.

\begin{figure*}[ht!]
\begin{center}
\includegraphics[width=0.49\textwidth]{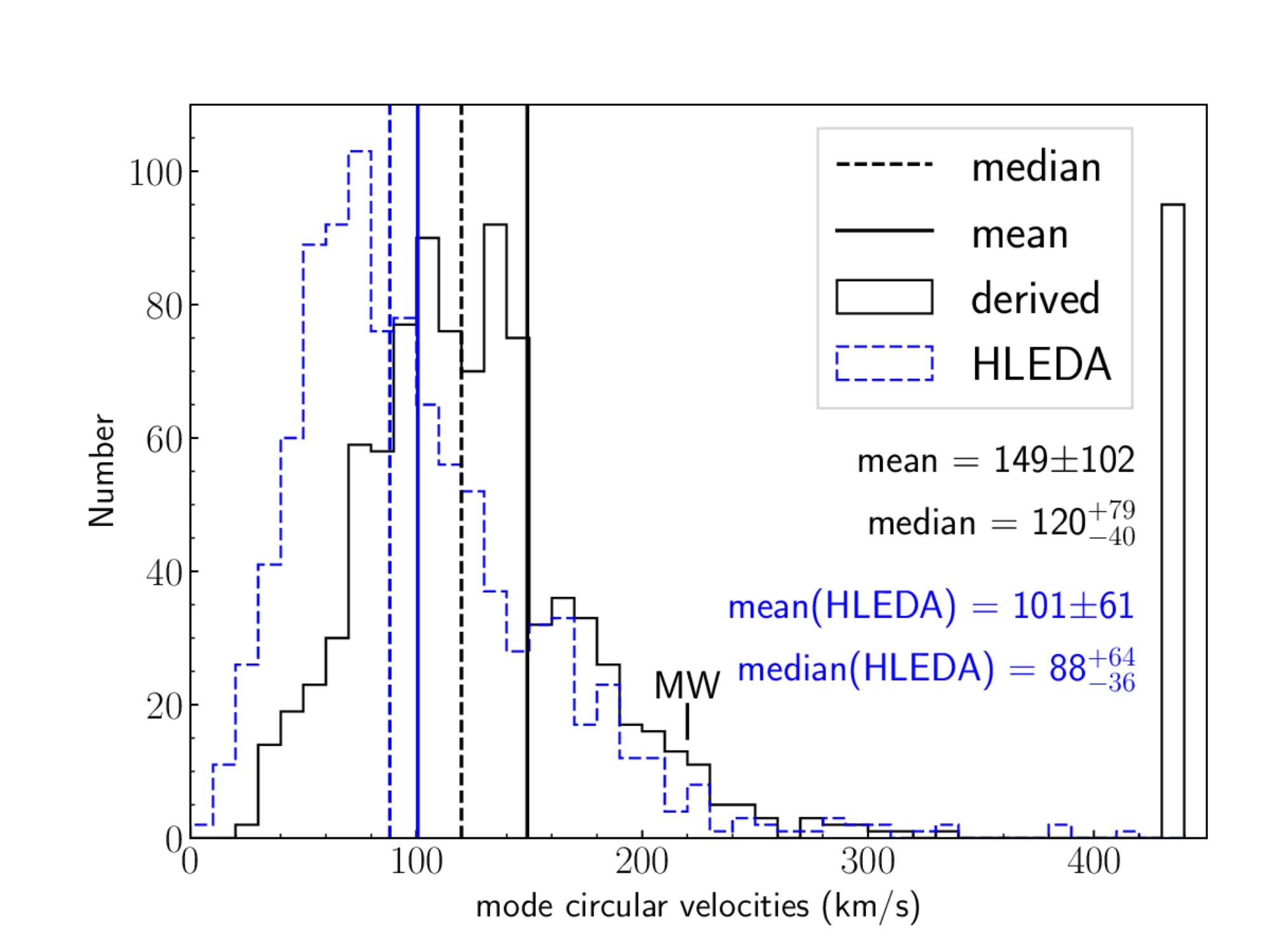}
\includegraphics[width=0.49\textwidth]{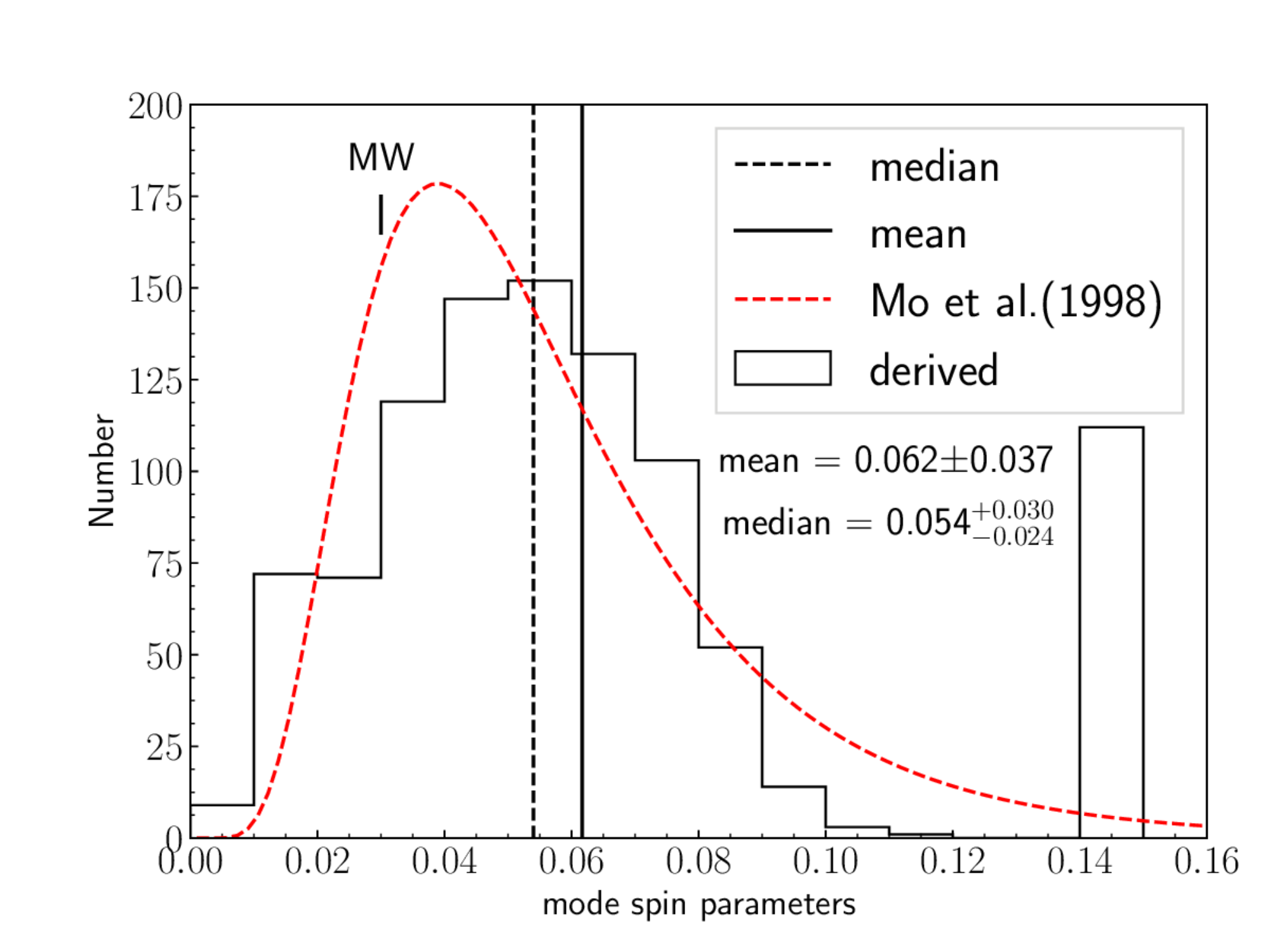}\\
\vspace{-0.1cm}
\includegraphics[trim={0 1cm 0 2cm},clip,width=0.95\textwidth]{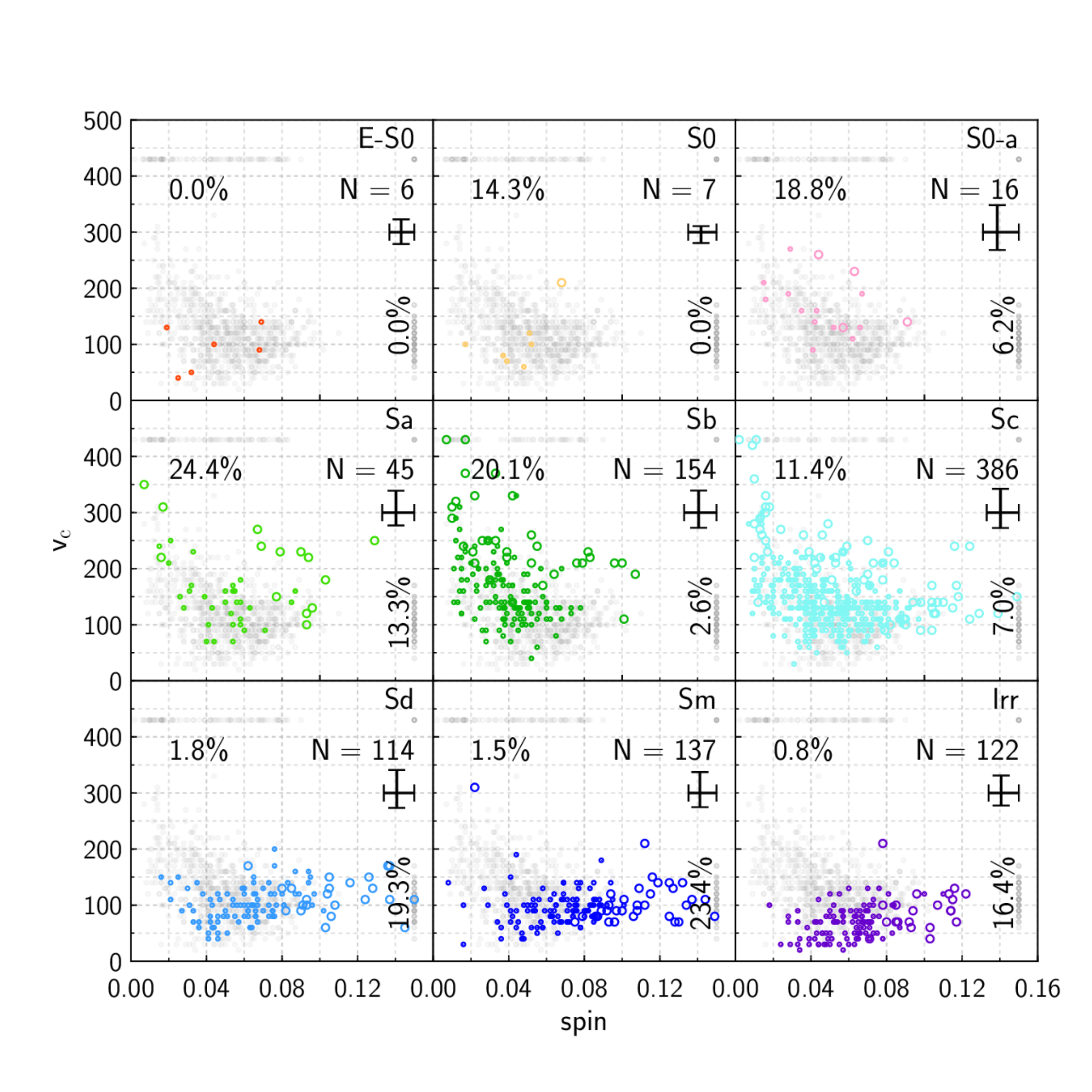}  
\caption{
Distribution of the mode of the best-fitting models circular velocity $v_{\mathrm{c}}$ (\textit{top left}) and spin
parameter $\lambda$ (\textit{top right}) for the $\sfg$ sample when cutoffs of $R/R80$=0.5 and
$\mu_{[3.6]}$=23.5 mag\,arcsec$^{-2}$ are used to isolate the disk
component of these galaxies' profiles. 
For the spin parameter distribution, we also show the
probability distribution (scaled to our distribution so that both distributions have the same area) of
the spin parameter derived by \citet{Mo1998} for comparison (\textit{red}). For the circular velocity distribution, we compare it
with circular velocities obtained from HyperLeda (\textit{blue dashed histogram}).
The deduced circular velocities distribution (\textit{black solid lines}) 
is compared with the circular velocities available in HyperLeda (\textit{blue dashed lines}). The mean (\textit{solid vertical line}) and the median (\textit{dashed vertical line}) positions are shown for each distribution, annotated with the value and the 1-$\sigma$ uncertainty.
The circular velocity plotted against the spin parameter, split by morphological type, is shown in the \textit{bottom panel}. 
Average uncertainties are shown in the top-right corner. The Milky Way (MW) values are shown in both panels.
Open circles indicate galaxies that have extreme values, either in circular velocity or in spin parameter, or both, and are shown using their central value instead. The percentage shown in the upper-left is the fraction of outliers in $v_{\mathrm{c}}$, whereas the one shown to in the lower-right and rotated is the fraction of outliers in spin parameters. Outliers in both $v_{\mathrm{c}}$ and spin are included in both fractions.
\label{fig:velvsspin}
}
\end{center}
\end{figure*}

This approach allows us to assign to a given galaxy disk a specific
3.6\,$\mu$m central surface brightness and scale length along with the corresponding closest model.
That way we are able to deduce circular velocities and spin parameters
for the entire $\sfg$ sample. In Figure~\ref{fig:velvsspin} we show the
circular velocity and spin distributions and the comparison between
both parameters for the entire sample.
We split these parameters by morphological type.

\begin{figure*}[ht!]
\begin{center}
\includegraphics[width=0.35\textwidth]{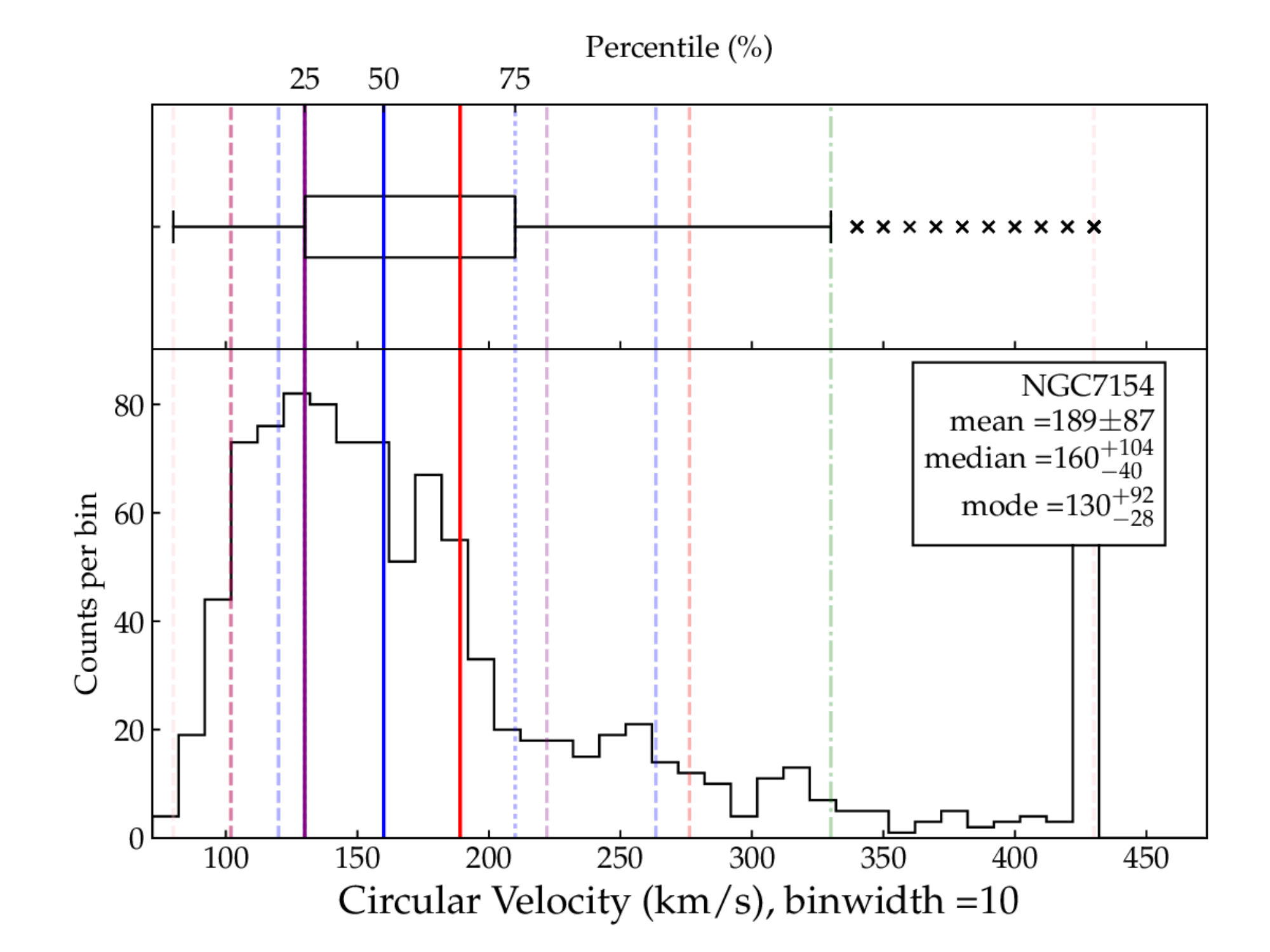}
\includegraphics[width=0.35\textwidth]{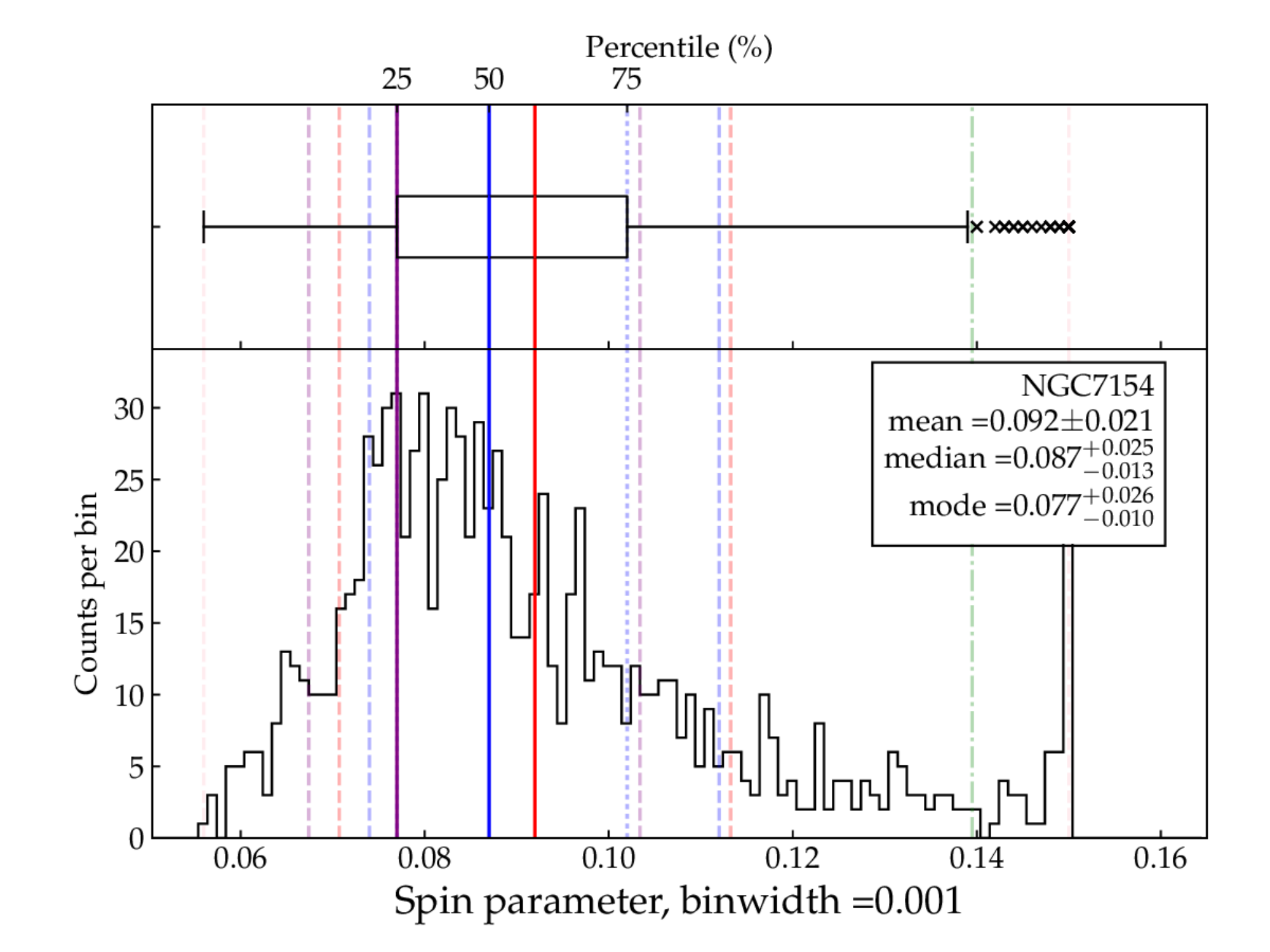}
\includegraphics[width=0.26\textwidth]{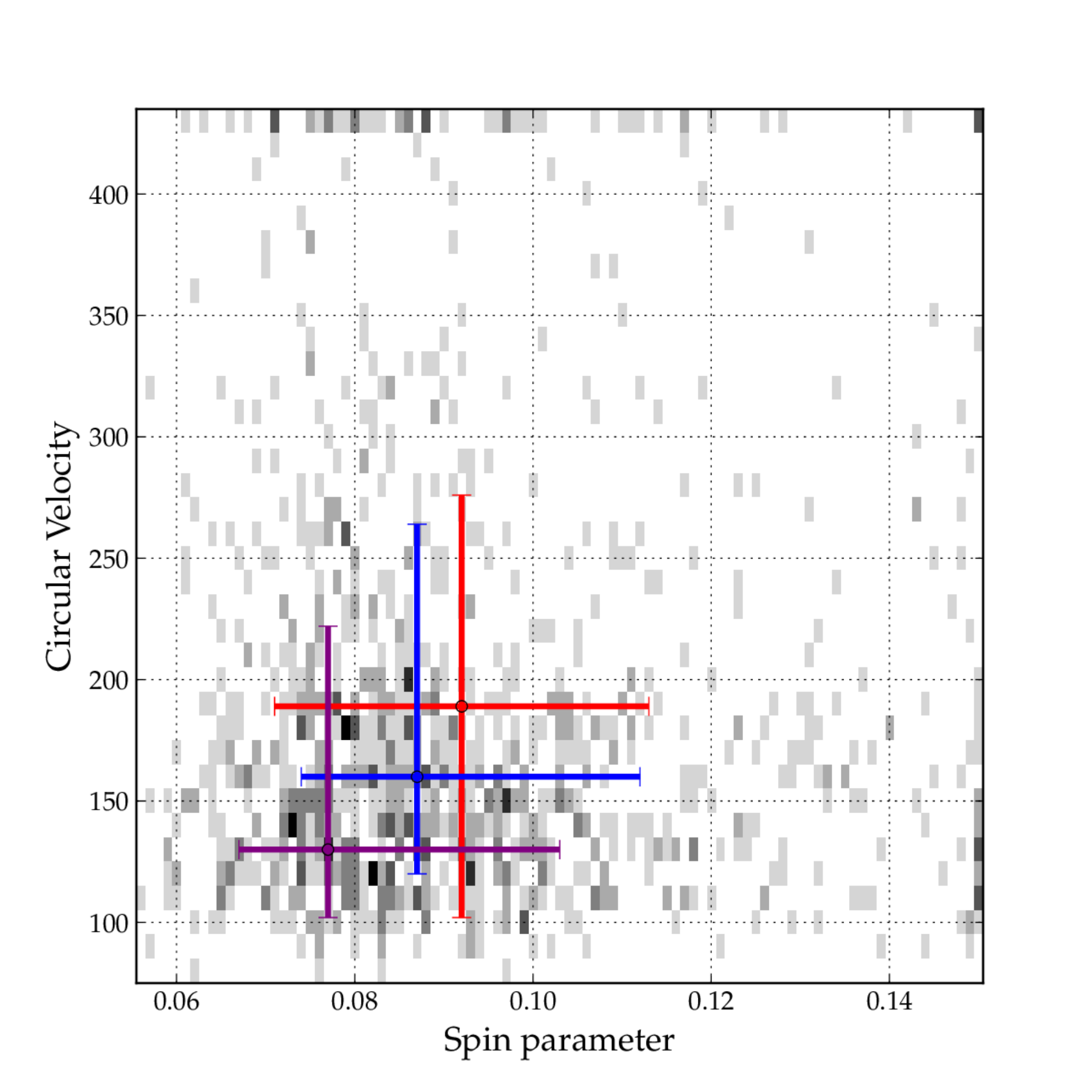}\\

\includegraphics[width=0.35\textwidth]{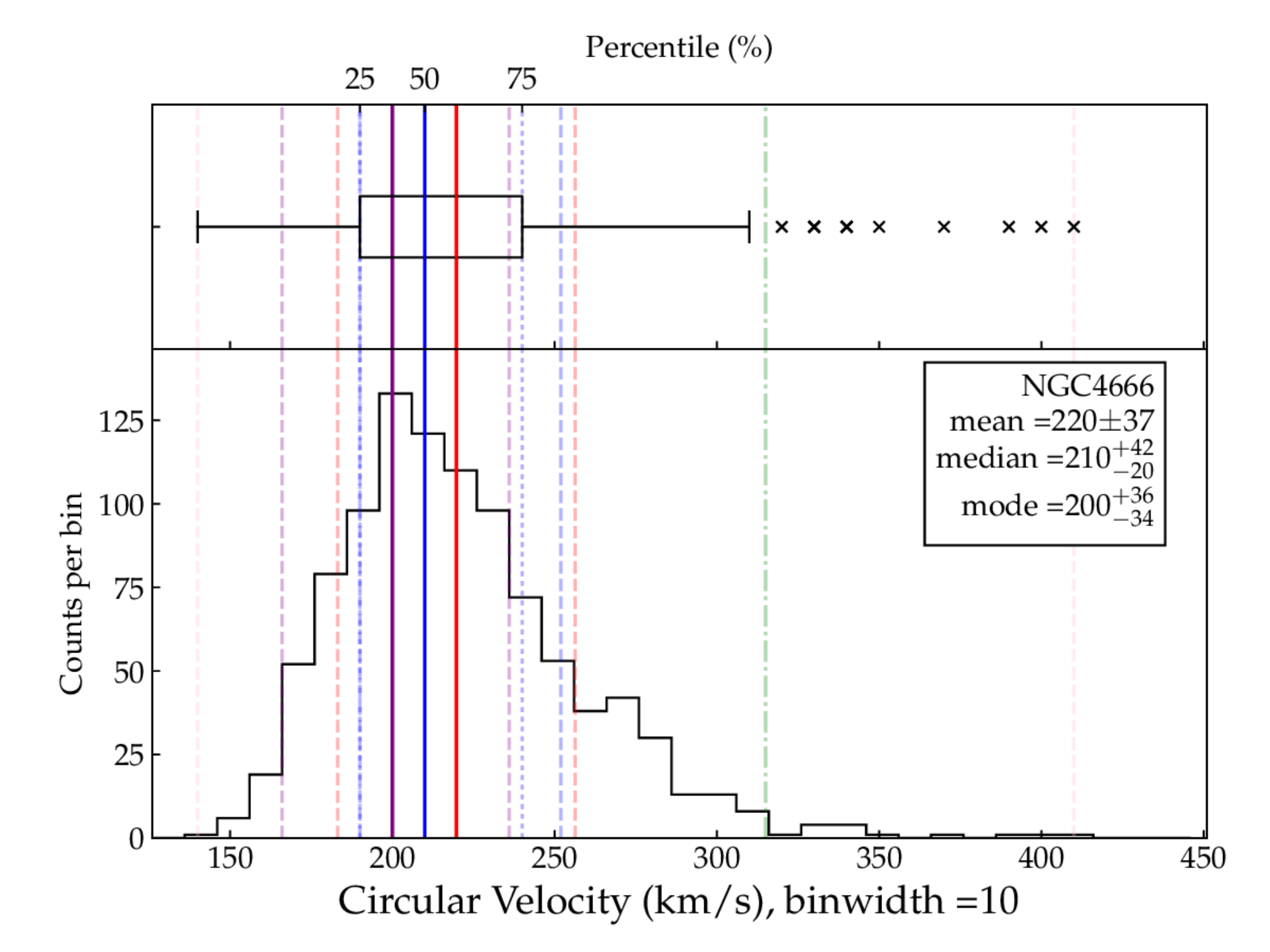}
\includegraphics[width=0.35\textwidth]{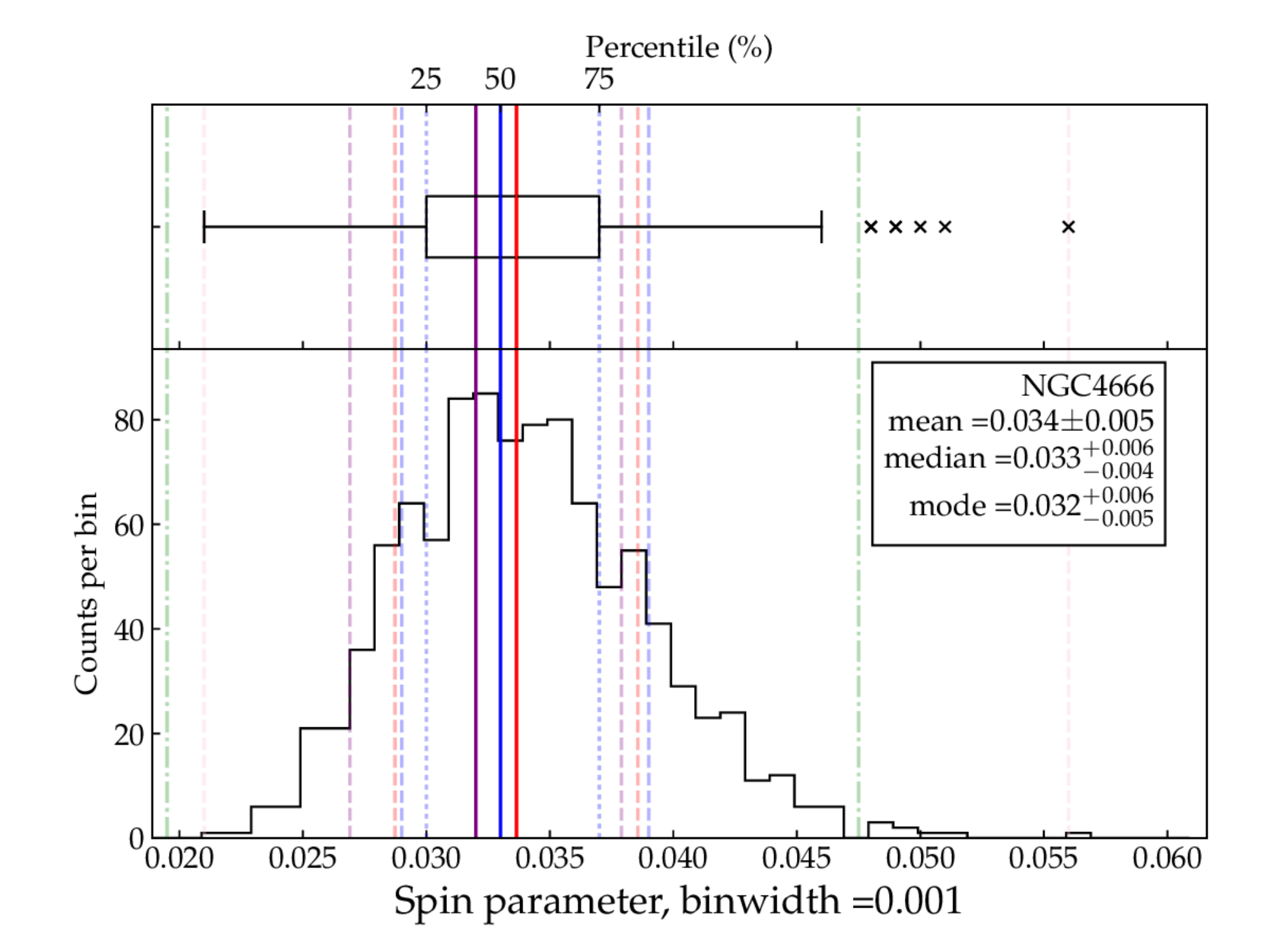}
\includegraphics[width=0.26\textwidth]{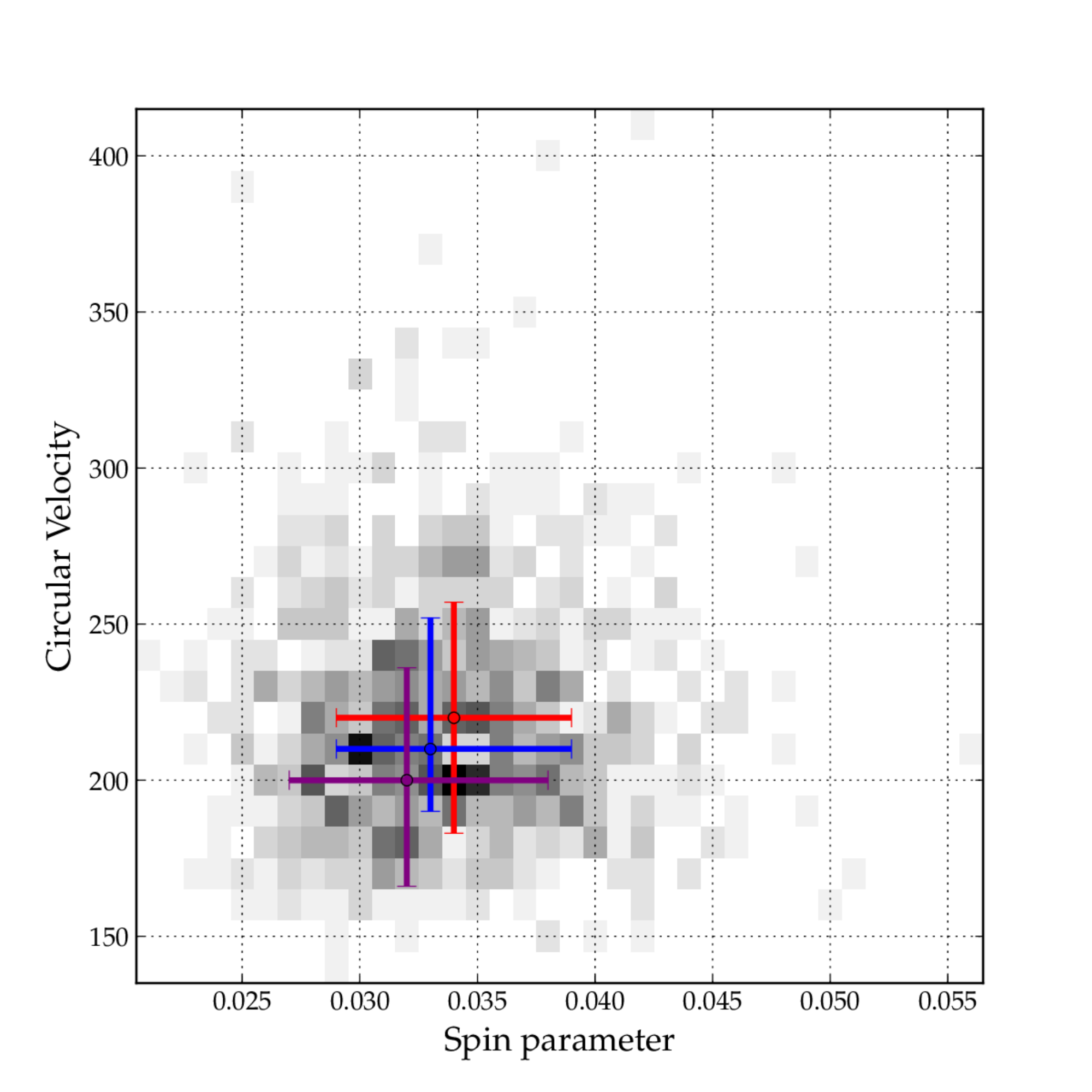}\\

\includegraphics[width=0.35\textwidth]{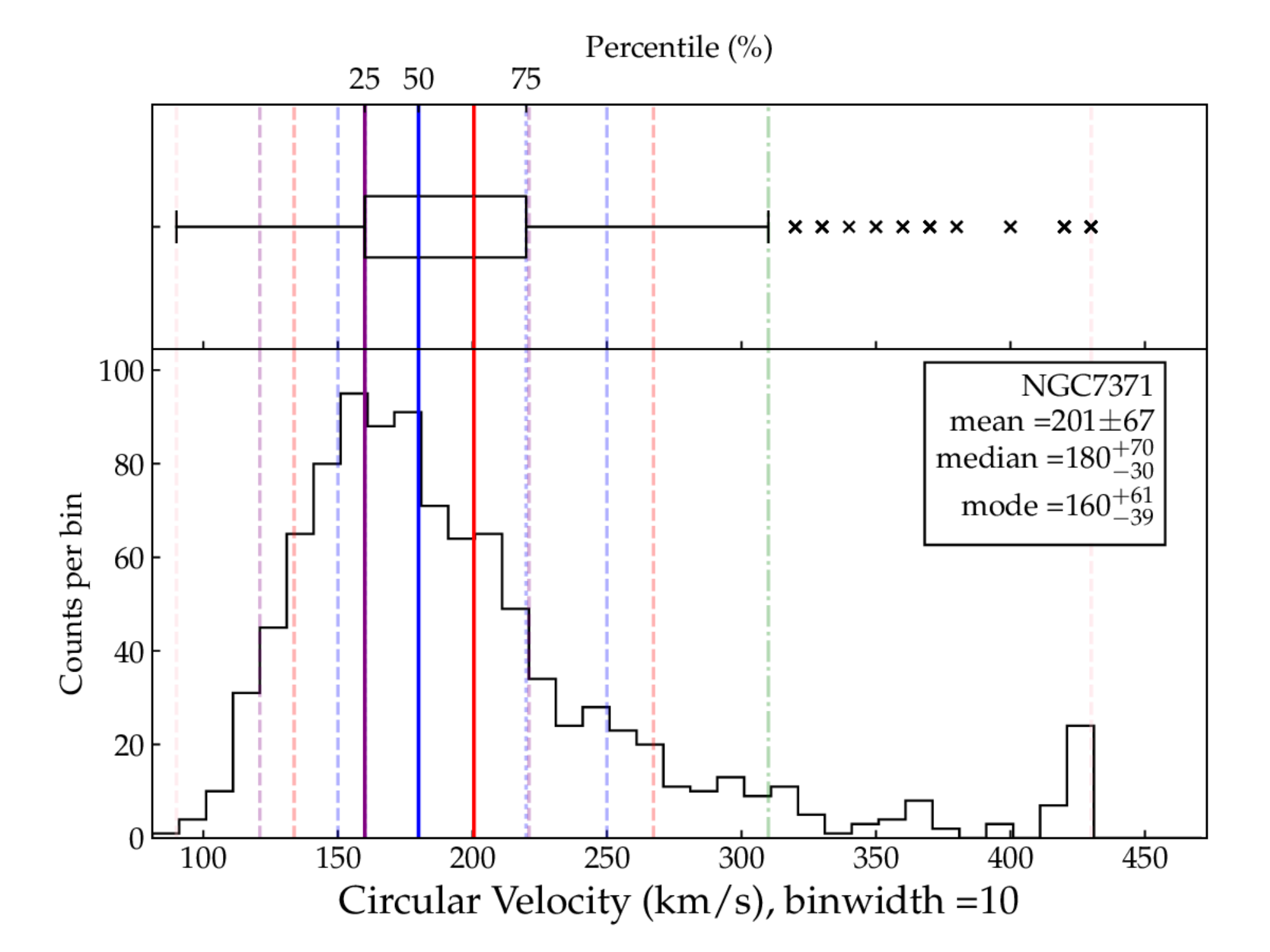}
\includegraphics[width=0.35\textwidth]{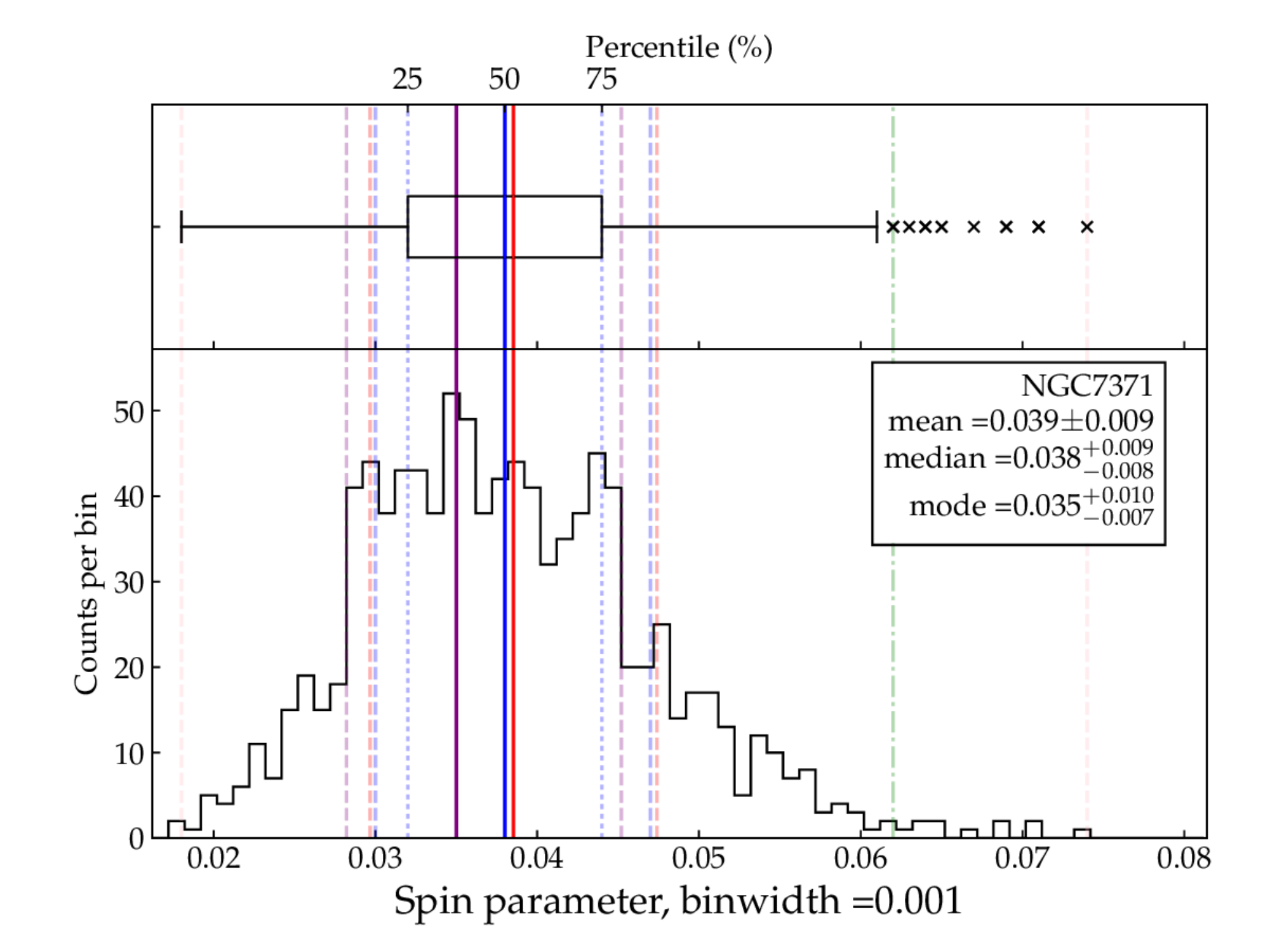}
\includegraphics[width=0.26\textwidth]{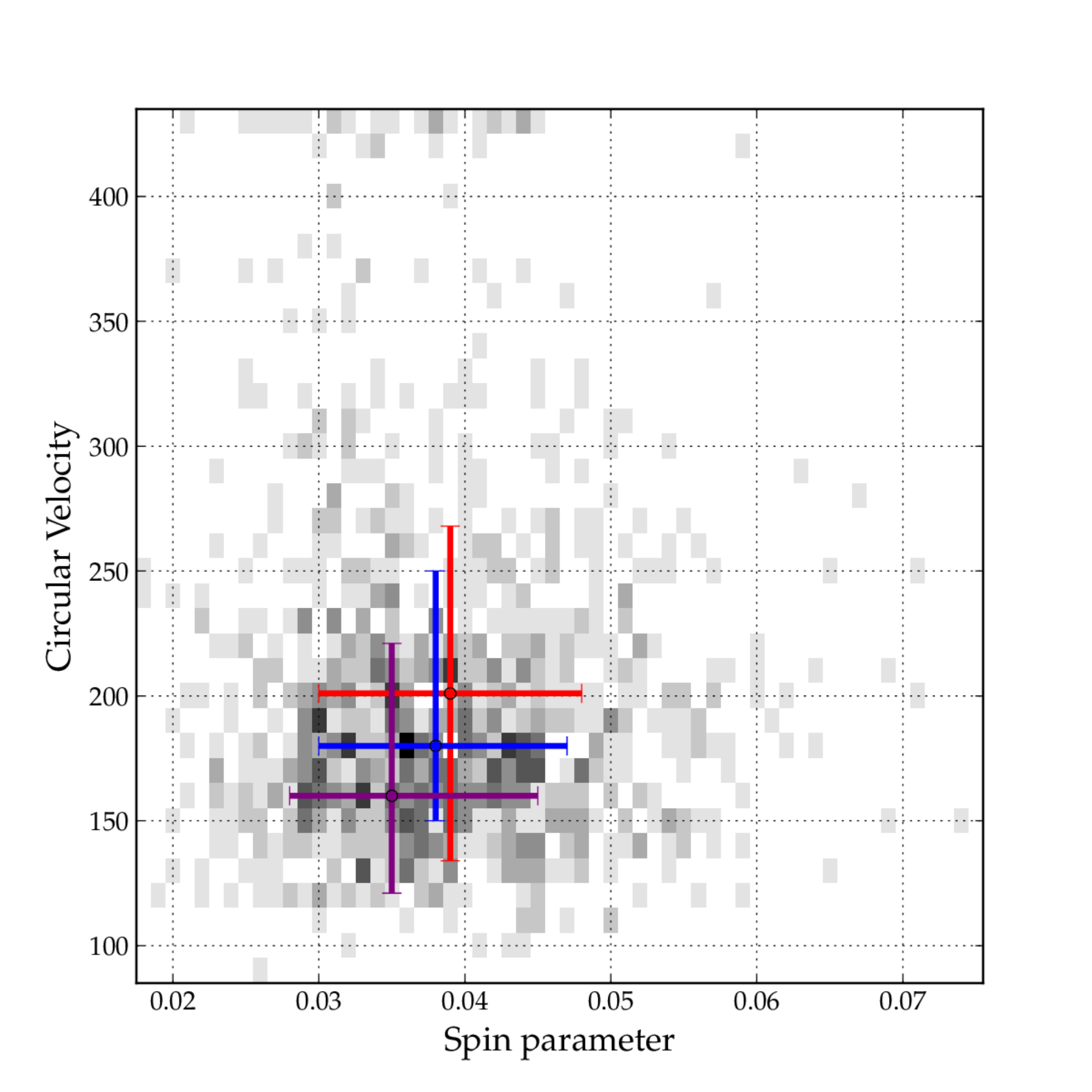}\\

\caption{
Examples of typical circular velocities $v_{\mathrm{c}}$ (\textit{left column}) and spin parameters $\lambda$ (\textit{middle column}) distributions 
(namely NGC7154, NGC4666, and NGC7371
that are classified as Sm, Sc, and S0-a galaxies, 
having absolute AB magnitude $\mathrm{M}_{[3.6]}= -19.62$, $-21.90$, and $-20.77$ mag)
for a sampling of 1000 slope and y-intercept values.
The plots in the \textit{right column} show the distribution of the 1000 simulated MC particles in circular velocity versus spin. The spin and circular velocity mean, median, and mode values and corresponding derived errors (assuming the two quantities are derived separately) are also shown.
The red solid line is the mean, the red dashed lines are $\pm1\sigma$
from the mean, the blue solid line is the median, the blue dashed
lines are $\pm1\sigma$ from the median (15.865\% and 84.135\% lower
and upper percentiles), the blue dotted lines are Q1 (25\%) and Q3
(75\%), the green dot-dashed lines are Q1-1.5$\cdot$IQR and
Q3+1.5$\cdot$IQR, where IQR is the interquartile range Q3 - Q1, and
the pink dashed lines are the distribution's range. Crosses are used
to show the bins that were excluded when computing the
percentiles.
\label{fig:egveldistrib}
}
\end{center}
\end{figure*}

For each pair of best-fitting slope (i.e$.$ scale-length) and y-intercept (i.e$.$ central surface brightness) measurements we generated a thousand random points
using elliptical 2D-gaussian probability distribution functions with
the 1-$\sigma$ being the uncertainties in these measurements and
obtained the closest model for each Monte-Carlo particle.
Thus, for each data point (i.e$.$, for each galaxy) we obtained a
distribution in best-fitting circular velocity and spin parameter. 
Typical distributions of circular velocities from the sampling of 1000 points are shown in Figure~\ref{fig:egveldistrib}.
This figure also shows the distribution of the individual 1000 points in the circular velocity versus spin diagram for three example galaxies. 
There is a mild degeneracy between the two parameters (although we show galaxies with very skewed distributions) in some of these objects that is in the same direction as the correlation seen in Figure~\ref{fig:velvsspin} for late type galaxies. 
Note, however, that such correlation is not driving the whole distribution of points in Figure~\ref{fig:velvsspin} and that the latter spans a wider range of spins and circular velocities than the 1$\sigma$ errors found for the individual galaxies. 
Thus, although the degeneracy between the two parameters certainly contributes to the morphology of the different panels of Figure~\ref{fig:velvsspin}, it also reflects the bona fide distribution of physical properties of the disks of galaxies in the Local Universe.
We see that this method of sampling produces circular velocity distributions with long tail towards high $v_{\mathrm{c}}$. 
These asymmetric distributions, 
for which the median or the mode give a 
better estimate of the peak of the distribution (rather than the mean) for the corresponding parameter, 
are a consequence of the non-regularity of the coverage of the model grid in Figure~\ref{fig:samgridm}. 
In this work, we make use of the mode values and the percentiles obtained from these
distributions to get the data points and average errorbars in Figure~\ref{fig:velvsspin}.
We also list the results obtained in Table~\ref{Table:velspin}.
\begin{table}
\begin{center}
\caption{$\galex$/$\sfg$ sample circular velocity and spin obtained from a grid of BP00 disk models}
\label{Table:velspin}
\begin{tabular}{lrrr}
\hline
  \multicolumn{1}{c}{galaxy name\tablenotemark{a}} &
  \multicolumn{1}{c}{$v_{\mathrm{c}}$\tablenotemark{b}} &
  \multicolumn{1}{c}{$\lambda$\tablenotemark{c}} &
  \multicolumn{1}{c}{T\tablenotemark{d}} \\
\hline
ESO293-034 & 130$^{+40}_{-28}$ & 0.041$^{+0.008}_{-0.007}$ & 6.2\\
NGC0024 & 110$^{+13}_{-27}$ & 0.027$^{+0.003}_{-0.006}$ & 5.1\\
ESO293-045 & 90$^{+24}_{-16}$ & 0.066$^{+0.009}_{-0.009}$ & 7.8\\
UGC00122 & 70$^{+15}_{-25}$ & 0.067$^{+0.008}_{-0.007}$ & 9.6\\
NGC0059 & 50$^{+...}_{-...}$ & 0.032$^{+0.004}_{-0.004}$ & -2.9\\
ESO539-007 & 110$^{+25}_{-45}$ & 0.150$^{+...}_{-0.029}$ & 8.7\\
ESO150-005 & 110$^{+45}_{-35}$ & 0.150$^{+...}_{-0.033}$ & 7.8\\
NGC0115 & 130$^{+41}_{-29}$ & 0.044$^{+0.010}_{-0.007}$ & 3.9\\
UGC00260 & 430$^{+...}_{-288}$ & 0.070$^{+0.023}_{-0.012}$ & 5.8\\
  ... & & & \\
\hline
\end{tabular}
\tablenotetext{1}{same as the $\sfg$ nomenclature.}
\tablenotetext{2}{circular velocity (mode) $v_{\mathrm{c}}$ plus-minus 1$\sigma$ uncertainty in km\,s$^{-1}$}.
\tablenotetext{3}{spin parameter (mode) $\lambda$ plus 1$\sigma$ uncertainty.}
\tablenotetext{4}{numerical morphological type.}
\tablecomments{$v_{\mathrm{c}}$ and spin obtained from BP00.}
\end{center}
\end{table}
\begin{figure*}
\begin{center}
\includegraphics[width=1.0\textwidth]{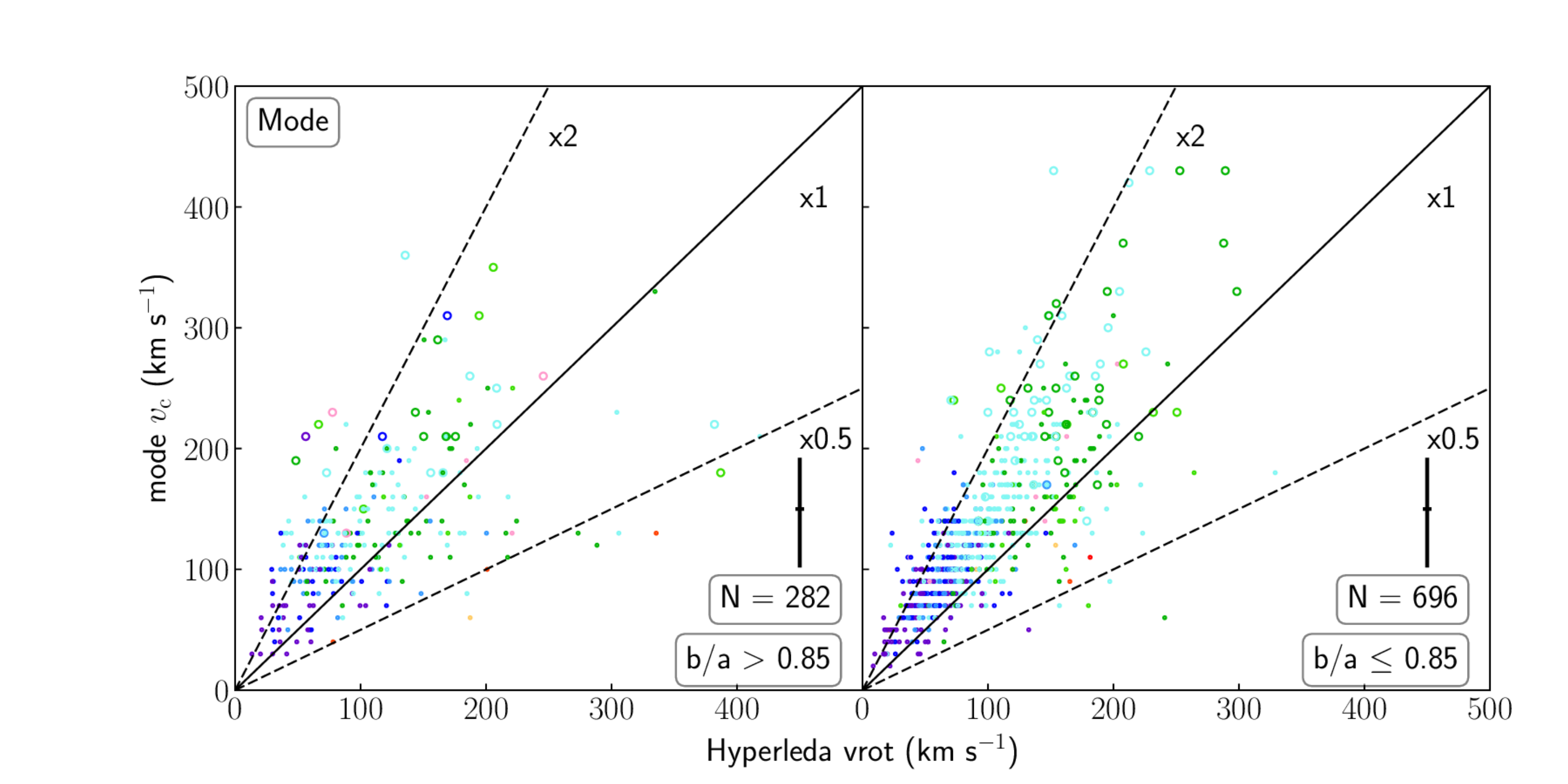}
\includegraphics[width=1.0\textwidth]{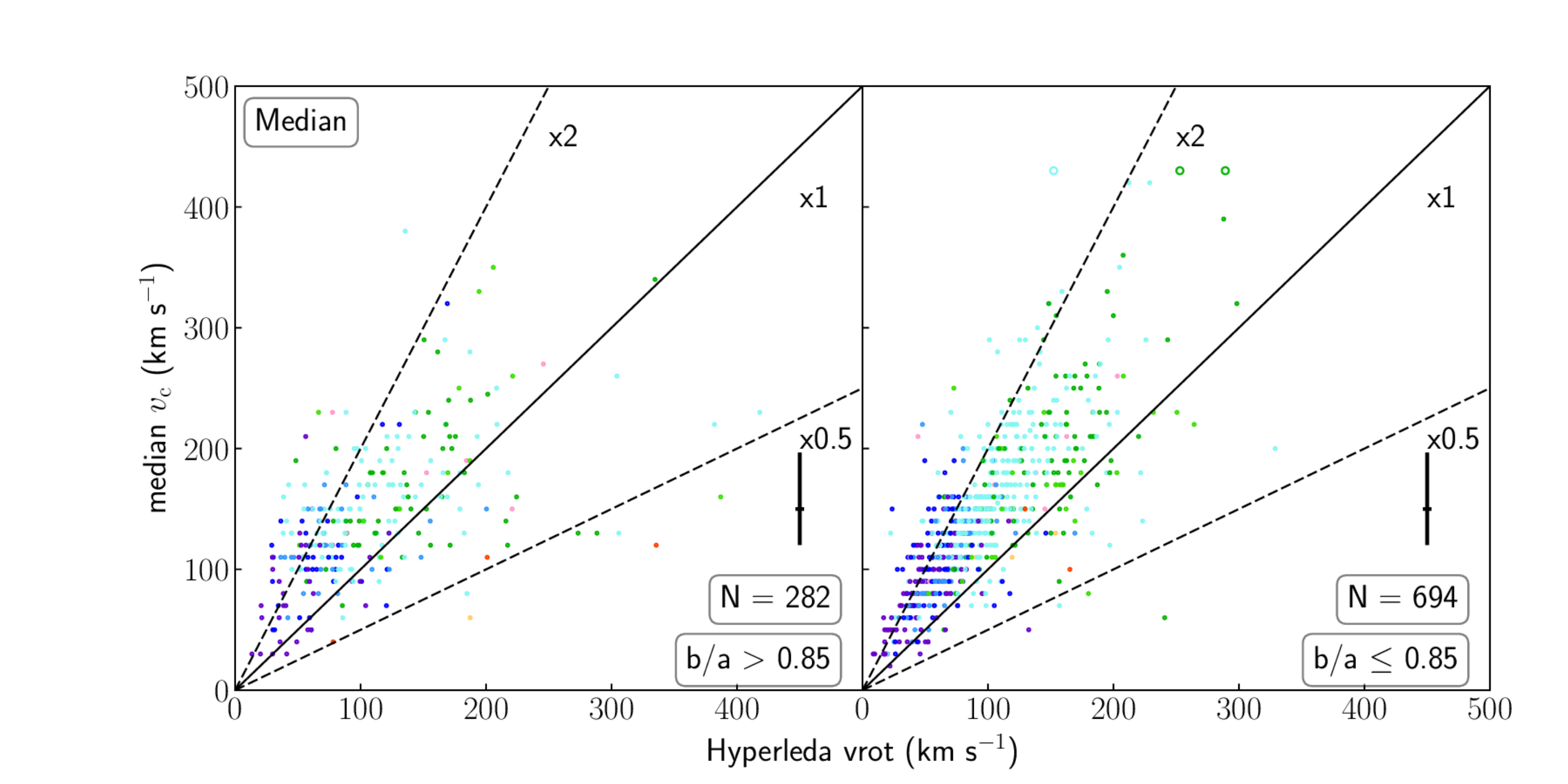}\\
\caption{
Rotational velocity $v_{\mathrm{c}}$ obtained from BP00 grid models for our
sample galaxies once the $\mu_{[3.6]}$ profiles are corrected for
inclination compared to the maximum rotational velocity obtained from
HyperLeda (corrected for inclination) $v_{\mathrm{rot}}$. The 1:1 and
  2:1 \& 1:2 ratios are shown as dashed lines as visual guides. Here, we use
  the mode (\textit{top}) and median (\textit{bottom}) of $v_{\mathrm{c}}$ obtained for each galaxy (see
  Figure~\ref{fig:egveldistrib}). We distinguish between galaxies of
high and low inclination using the minor-to-major axis ratio $b/a$
($\leq 0.85$ for highly inclined galaxies in the \textit{left panel}, and $0.85 < b/a \leq 1$ for
low-inclination galaxies in the \textit{right panel}). Both axes are in units of km\,s$^{-1}$.
Open circles indicate galaxies that have extreme values, as in Figure~\ref{fig:velvsspin}, and for which we use their central value.
We do not take their uncertainties into account for the computation of the average uncertainties.
\label{fig:velvel}
}
\end{center}
\end{figure*}

We then compare our values with observed values of the circular
velocity for galaxies for which we have data (see
Figure~\ref{fig:velvel}). We obtained the inclination-corrected
maximum rotational (i.e$.$, circular) velocity and its associated
uncertainty from HyperLeda, \textit{vrot} and \textit{e\_vrot}.  These
observed values are computed from the apparent maximum rotation
velocity obtained from the width of the 21\,cm line at various levels,
or from H$\alpha$ rotation curves.  They are homogenized using a large
sample ($>$50000) of measurements and are corrected for inclination
\citep{Paturel2003b}. 
We do not aim here to provide a fully coherent set of circular velocity measurements 
but to see whether or not the values that we obtain from our method are similar to the observed ones.
In the case of our `best'
$\chi^{2}$ fit with cutoffs of $R/R80$\,$\simeq$\,0.5 and
$\mu_{[3.6]}$\,$\simeq$\,23.5 mag\,arcsec$^{-2}$, 976 galaxies when using the median, and 978 when using the mode
out of 987 have actual measurements in HyperLeda. 
In the case of the mode, we quantify the 1-$\sigma$ (68.269\%) distribution range to the left (right) of the mode 
by counting only the bins on the left-hand-side (right-hand-side) distribution starting from the bin of the mode, but excluding it from the counts.
Also, in case of multimodal distributions, we chose the bin with the smallest associated value.
When we compare the two, we see that
most of our values are larger than the observed ones, but rarely above twice the observed rotational velocity. 
This effect comes partly from the accuracy of extracting the peak value over the 
skewed distributions of the circular velocities and spin parameters that we obtained from our method as can be seen in Figure~\ref{fig:velvel}. 
When the distribution is skewed to the left, the mean is systematically larger than the median, and the median is larger than the mode, and vice-versa when the skew is to the right, then the mean is smaller than the median, and the median is smaller than the mode.
This comes from our grid of models (Figure~\ref{fig:samgridm}) and the sampling that we use to extract the best model. 
For given observational uncertainties and as the slope flattens out, 
higher circular velocity models that match the observations largely increase.
The same holds true for larger values for the y-intercept (i.e$.$, fainter): the higher the spin, the more models that match the observations.
Hence the sampling distributions show a tail toward larger circular velocities and spin parameters. 
Using either the median or the mode gives similar results.
This is shown in Figure~\ref{fig:velvel} where using the median yields a similar scatter that when using the mode.

Our values are consistent with the observed values within a factor of 0.5 to 2, especially if the very large uncertainties present are taken into account. 
The distributions given in Figures~\ref{fig:velvsspin } provide powerful tools to test the predictions for numerical simulations of disks in a
cosmological context.

\subsection{Color Gradient versus circular velocity, spin, and stellar mass} \label{sec:colorgradient}
\begin{figure*}
\begin{center}
\includegraphics[width=1\textwidth]{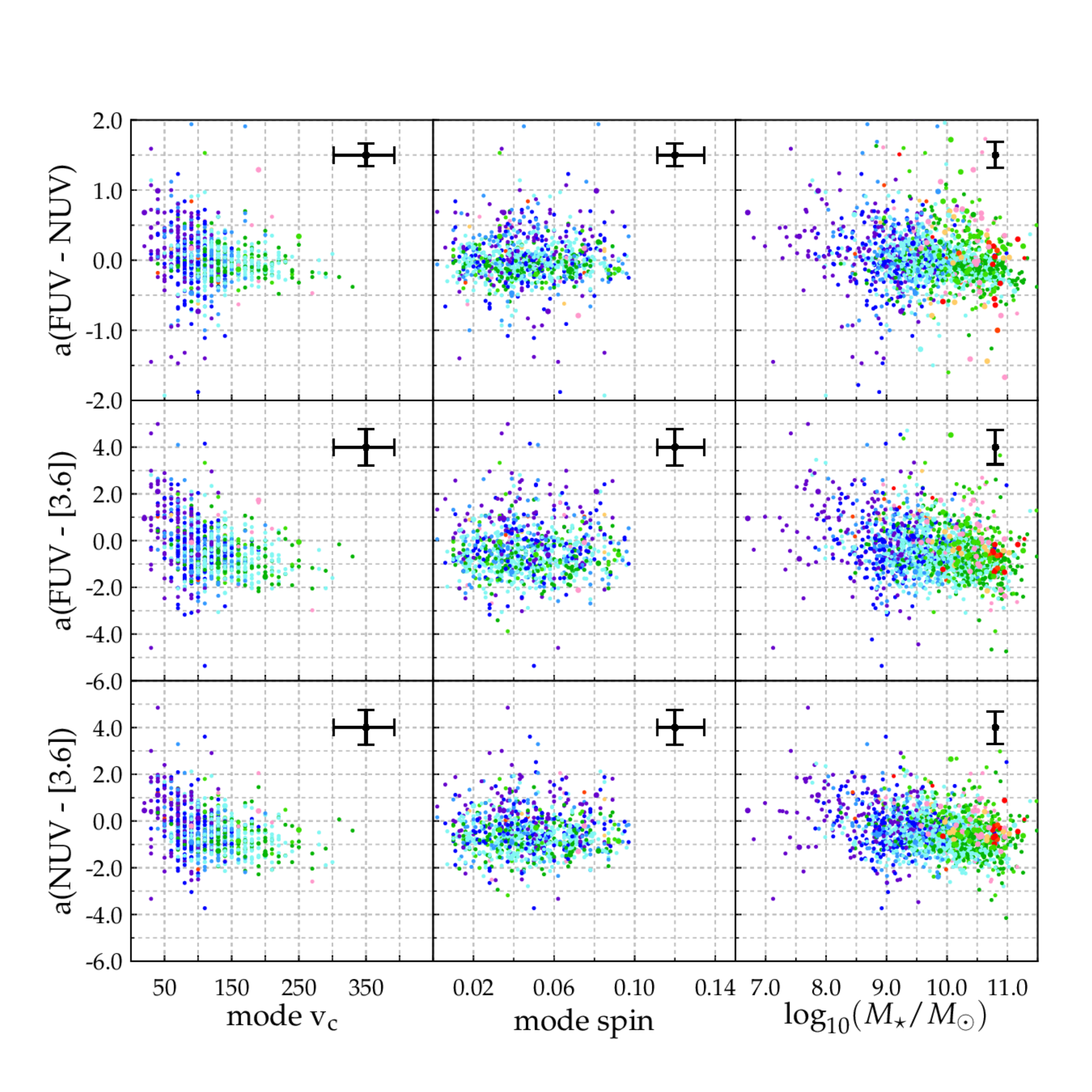}
\caption{
Color gradients (slopes of least-square linear fit) in the (FUV\,$-$\,NUV), (FUV\,$-$\,[3.6]), and (NUV\,$-$\,[3.6]) colors versus circular velocities (mode), spins (mode), and stellar masses, respectively. A positive color gradient indicates a reddening as we move to the outskirts. A negative color gradient indicates a blueing. Average 1-$\sigma$ uncertainties in both axes are shown in the upper-right corner of each panel.
\label{fig:colorgradient}
}
\end{center}
\end{figure*}

Finally, we compare the color gradients (the slopes) obtained in the (FUV\,$-$\,NUV), (FUV\,$-$\,[3.6]), and (NUV\,$-$\,[3.6]) color profiles (with a cutoff at $R/R80$=0.5 but no cutoff in SB; see Table~\ref{Table:slopestable}) against the mode circular velocities, mode spins that we derived with the method described in Section~\ref{sec:velandspin}, and stellar masses calculated from the 3.6\,$\mu$m SB. This is shown in Figure~\ref{fig:colorgradient}. The panels showing the circular velocity and spin comprises 987 galaxies, whereas the panels showing the stellar mass comprises 1541 galaxies.
For the mode circular velocities, a large scatter is seen especially for low-mass systems, in all three colors. In the case of the mode spin parameters, the scatter is very much the same throughout the entire range of spins, for all morphological types, and for all three colors. Then, the color gradient versus the stellar mass plots show a large scatter for low-mass galaxies with stellar mass of around 10$^{8}$--10$^{9}$\,M$_{\odot}$. On average, there is a trend toward more negative gradient as we move to larger masses, and therefore indicating bluer outer disks. However, most low-mass galaxies and a non-negligible fraction of massive galaxies show positive color gradients.

\section{DISCUSSION} \label{sec:discussion}
We discuss here about circular velocity and spin of galaxies in the local Universe in Section~\ref{sec:vcandspin}.
It is crucial to understand where this UV emission is coming from within the galaxies (Section~\ref{sec:radialdistrib}).
In this regard, an important point that needs to be addressed is the fact that the UV emission not only comes from newly born massive stars, but also from evolved low-mass stars. We also discuss galaxy evolution and the effects of the environment on populating the green valley (Section~\ref{sec:evolutionofGGV}).

\subsection{Circular velocity and spin of galaxies in the local Universe} \label{sec:vcandspin}

The results shown in Section~\ref{sec:modeling} show that it is possible
to derive (albeit with relatively large uncertainties in specific
cases) the statistical distribution of circular velocity (total mass)
and spin (specific angular momentum) of galaxies from the analysis of
deep near-infrared photometry of their disks. Besides, the fact that
we can impose some simple criteria to isolate the disk component of
the profiles makes this kind of analysis a very powerful tool for
its application to upcoming surface photometry data from LSST, \textit{EUCLID} or \textit{WFIRST}.

Our analysis reveals that up to the current surface brightness
detection limits (we note that all $\sfg$ galaxies are detected by
\textit{Spitzer} but many low-surface brightness objects might still be missing
from the catalogs) nearby galaxies show a wide distribution in spin
with a maximum at $\lambda$$\sim$0.06 and a relatively high fraction
(24\%) of galaxies with $\lambda$$>$0.08 (see top-right panel of
Figure~\ref{fig:velvsspin}).

The comparison of these values with those derived for the SINGS sample
\citep{Kennicutt2003} by \citet{MunozMateos2011} using a similar
method and a similar set of models indicate a larger number of high-$\lambda$ systems in our sample,
in terms of the mean, median, and mode of the distribution. This is
expected in the case of the SINGS sample as this is biased towards
high surface brightness systems with low angular momentum content
relative to their mass. In addition, the SINGS sample (75 galaxies) was constructed
to sample physical parameters (morphological type, luminosity, and FIR/optical luminosity ratio)
and therefore, is not representative in numbers of different kind of galaxies.
With respect to the predictions of
semi-analytic models \citep[e.g.,][]{Mo1998}, 
we find a median value
that is displaced towards larger spins (0.06) relative to
recent simulations \citep[$\sim$0.036, quite independently of the galaxy mass
and method of determining $\lambda$;][]{RodriguezPuebla2016}. 
To quantify this distribution, for the $\sfg$ sample we derive
mean (with 1-$\sigma$ distribution width) and median values of 0.062$\pm$0.037,
0.054$^{+0.030}_{-0.024}$, respectively, while,
for Milky Way mass halos, the models of \citet{RodriguezPuebla2016} yield a mean of
0.036 with a dispersion of 0.24 dex \citep[for the spin parameter $\lambda_{P}$ of][]{Peebles1969}.

With regard to the circular velocities, we find a relatively narrow
mode distribution peaking at 120$\sim$$v_{\mathrm{c}}$$\sim$149\,km\,s$^{-1}$ (ignoring the outliers.)
(top-left panel of Figure~\ref{fig:velvsspin}). This indicates a
lack of low-mass (dwarf) systems, which is probably occurring both at
the high surface brightness (because of our diameter selection for $\sfg$) and
low surface brightness ends (because of the limiting central surface brightness present in the catalogs of nearby galaxies). Determining a volume- and diameter-corrected circular velocity (i.e$.$, halo mass) function is beyond the scope of this paper. It could, however, be an interesting test for the models complementary to the halo mass functions derived from dynamical masses obtained from the modeling of 21\,cm velocity maps and line profiles \citep{deBlok2008,Papastergis2013} \citep[see also][which connected the kinematics to the baryon fractions using the $\sfg$]{Zaritsky2014b}. 

Finally, the distribution of circular velocity vs spin (bottom panel of Figure~\ref{fig:velvsspin}) shows a larger dependence of the spin on the circular velocity than that found by \citet{RodriguezPuebla2016}. 
In those cases where the data point, or a significant portion of the input probability distribution of the y-intercept and slope of the 1000 sampled points, is outside of the grid, the mode is biased towards the maximum value models in spin or circular velocity, or both, we use the central value instead of the mode value. These are shown as open circles. The percentages are the fractions of outliers in circular velocity (top-left) and spin parameter (bottom-right). Outliers in both parameters are included in both fractions. Uncertainties for the outliers are not included in the average uncertainty.
The lack of objects of high spin and low circular velocity could be due to the surface brightness limit involved in defining our sample. However, this would make our distribution even wider towards high-$\lambda$ values \citep[see also][]{Mo1998}. Besides, the lack of low-mass, low-spin galaxies (high surface brightness dwarfs) is attributable to our diameter selection ($>$1\,arcmin), as these would be very compact dwarf systems. 

Thus, we conclude that the strong dependence of the spin on the circular velocity and, in particular, 
the lack of low-mass galaxies at both extremes of the distribution in spin, 
might be due to selection effects in the $\sfg$ survey (at least we cannot conclude otherwise). 
The relatively flat distribution of spin values in the range between $\lambda$=0.03-0.11, 
which would be even more extended towards high-$\lambda$ values, 
used to pose a challenge to current models of galaxy formation.
However, \citet{Amorisco2016} have recently shown that the properties and abundances in clusters of large, ultra-diffuse galaxies (UDGs) can be reproduced from within a standard cosmological framework and classical disk-formation models.
It will be interesting to see, once catalogs of low-surface brightness disk galaxies will be available \citep[including UDGs such as those found by][in Coma]{Koda2015} how they are distributed in terms of spin and circular velocity.

\subsection{Radial distribution of UV emission: UV-upturn and star formation} \label{sec:radialdistrib}

The UV emission found in the 1931 galaxies within the $\sfg$ sample
is clearly aligned in two sequences of UV-to-IR colors. These two
sequences, which are called the $\galex$ Red (GRS) and Blue (GBS)
sequences and which are best isolated in the (FUV\,$-$\,NUV) and
(NUV\,$-$\,3.6\,$\mu$m) color-color diagram and when galaxies are previously
split by morphological type (see Figure~\ref{fig:ccdFNN36m}),
correspond each to a different mechanism responsible for the UV
emission. In the case of the GRS, the big change in (FUV\,$-$\,NUV) color
($\sim$1.5\,mag) with a change in (NUV\,$-$\,3.6\,$\mu$m) of $<$1\,mag can
only be attributed to the UV-upturn phenomenon \citep{Oconnell1999}, which
is believed to be caused by very hot EHB stars 
\citep[see][and references therein]{Zaritsky2015a}.

On the other hand, the GBS has a slope of
0.12 (see Equation~\ref{eq:GBS}), which implies a
change of only 1.2\,mag in the (FUV\,$-$\,NUV) color for a change of 10\,mag
in (NUV\,$-$\,3.6\,$\mu$m). As shown in \citet{Bouquin2015} this slope
agrees well with the color correlation predicted by spectro-photometric
models for the evolution of galaxy disks (see, e.g., BP00), 
so the UV emission of these objects can be interpreted as
due to emission from relatively massive stars in the turn-off of the
main sequence.  

The GRS is mainly populated by E, E-S0, S0 and S0-a morphological
types and it is clearly isolated in its (FUV\,$-$\,NUV) blue end only in
ETGs, i.e$.$, E, E-S0 and S0 galaxies. This isolation is possible thanks
to the dichotomy in the (NUV\,$-$\,3.6\,$\mu$m) colors of the central
regions of ETGs (triangles in
Figure~\ref{fig:ccdFNN36m}). They are either very blue, indicating
(residual) star formation in these innermost regions, or very red,
which points towards a very old (light-weighted) stellar
population. The fact that these very old populations in the centers of
ETGs are also the ones showing the strongest UV-upturn is something
that has been explained in the past as being related to either the
older age or higher metal (helium, at least) abundance of the
horizontal branch stars responsible for the UV emission in these
regions \citep{Boselli2005}, but may also be tied to the observations
supporting differences in IMF \citep[e.g.][]{Conroy2012,Cappellari2012,Cappellari2013a, Cappellari2013b}.

In early-to-intermediate spirals (S0-a through Sc) the central
regions of many galaxies appear in the locus where the GBS and GRS
overlap. Besides, the entire GBS is well populated by measurements
obtained both in the inner and outer regions of galaxies. Thus, these
colors are of no use to determine whether the UV emission from
the bulges of these galaxies are dominated by emission from young
massive or evolved low-mass EHB stars or by a combination of
both. Both the outer parts of early-to-intermediate spirals (except
for regions populating the GGV; see below) and the late-type spirals (Sd
and beyond) at all radii follow a narrow GBS. According to the color
profiles shown in Figure~\ref{fig:SBprofileskpc}, the majority of the galaxies
that are found to populate the GBS show negative color gradients,
which is in agreement with the global scenario of inside-out formation of
their disks. The study of the most extreme cases of inside-out disk
formation will be the subject of a future communication. At the
surface brightness levels reached by our data we do not find clear
signs of color upbending, at least in the bands considered in this work
\citep[see][for studies of
reversed optical color profiles and ionized-gas chemical
abundance gradients in outer disks]{Bakos2008,Marino2016}.

Finally, it is also worth mentioning that the central regions of
galaxies (despite having the highest signal-to-noise ratios) show the
widest dispersion in all three (FUV\,$-$\,NUV), (FUV\,$-$\,3.6\,$\mu$m) and
(NUV\,$-$\,3.6\,$\mu$m) colors among galaxy types and as a whole, covering
$\sim$10\,mag in the case of the latter two colors. This, of course,
indicates that nuclear regions are the least homogeneous within the
population of local galaxies in terms of their stellar population and
dust content.

\subsection{Galaxy evolution through the Green Valley} \label{sec:evolutionofGGV}

Here we focus on discussing the properties of those galaxies that were
identified in \citet{Bouquin2015} as being globally included in
the so-called $\galex$ Green Valley and also of those regions within
galaxies that are now found to be located in the GGV even though they are part of the GBS or GRS
when considered as a whole. 

In Figures~\ref{fig:SFMS} and \ref{fig:histoplot6} we showed that
galaxies that belong (globally) to the GGV are mainly lenticulars and
early-type spirals (S0-a through Sb), showing a relatively narrow
distribution of (observed) sSFR around 10$^{-12}$\,yr$^{-1}$. 
Furthermore, Figure~\ref{fig:nine} shows that the outer regions of GGV galaxies
behave differently from the outer regions of most GBS systems, with
the FUV$-$3.6\,$\mu$m color getting redder as we move progressively
towards their outer disks. This clearly indicates that the reason why
these objects are in the GGV is that their disks are redder, for
the same morphological type and surface brightness, than those of most
GBS galaxies. Exploration of Figure~\ref{fig:nine} in the case of
the GBS shows that the region of red disks is populated by a number of
ETGs with profiles similar in shape to those found in the GRS but that they probably
show a very blue nucleus that places them in the GBS when considered as
a whole. 

A small fraction of GBS galaxies (mainly early-type
spirals, but only a fraction of them) have disks that also redden with
radius. These are objects that are likely to evolve into GGV galaxies
or objects which GGV galaxies will evolve into, depending on whether
GGV galaxies are quenching their star formation or regrowing a disk. 

Our analysis shows that the fraction of galaxies belonging to dense environments is higher
for GGV galaxies than for GBS galaxies, but is less than for the GRS.
This result, combined with the fact that GGV galaxies have redder outer disks,
hints at the direction of the evolution, from GBS to GRS, 
which favors star formation quenching due to environmental effects.
Similar results have been obtained in the analysis of galaxies in clusters \citep[see][]{Wolf2009, Bamford2009, Cibinel2013, Head2014}.

Moreover, a study of galaxies in transition in different environments \citep{Vulcani2015} shows that 
galaxies in groups have a higher quenching efficiency than field galaxies.
Their results show that color transformation is due to the overall decrease in SFR, both in bulges and disks, while maintaining the morphology.
They also show that morphological transformation is due to an increase in bulge-to-disk ratio because of disk removal, and not because of the growth of the bulge, in disagreement to a bulge enhancement and absence of disk-fading scenario \citep{Christlein2004}.
What is presented in Section~\ref{sec:results} is in agreement with the former, where disk-fading occurs,
resulting in an increase in the bulge-to-disk ratio.

\section{SUMMARY AND CONCLUSIONS} \label{sec:summaryandconclusions}
We have gathered $\galex$ FUV and NUV images for the $\sfg$ sample, and have measured their FUV and NUV magnitudes.
Our UV subsample comprises 1931 galaxies, and has an identical distribution in morphological type, distance, and 3.6 micron absolute magnitude
as the $\sfg$ sample of 2352 galaxies (Figure~\ref{fig:histogramsALL}).
Our $\galex$ subsample is compatible with being a random subsample of the entire $\sfg$ sample,
and can also be considered as representative of the local universe.

The photometry is done within rings with fixed PA and $\epsilon$ 
at every 6$\arcsec$ steps in semi-major axis length and with width of 6$\arcsec$.
The products are the $\mu_{\mathrm{FUV}}$ and $\mu_{\mathrm{NUV}}$ surface brightness profiles, 
(FUV\,$-$\,NUV) color profiles,
along with the asymptotic FUV and NUV magnitudes and (FUV\,$-$\,NUV) color.
Data are partially summarized in Table~\ref{Table:sampletable}, and the full catalog is available online.
We have generated RGB postage-stamp images from UV images only, 
and also obtained the $\mu_{\mathrm{FUV}}$, $\mu_{\mathrm{NUV}}$ surface brightness, 
as well as the (FUV\,$-$\,NUV) color profiles.
We used the RC2 numerical morphological classification to roughly classify the galaxies into narrower morphological type bins (sample demographics are summarized in Table~\ref{Table:basictable}).

These UV products, combined with the near-IR products of the $\sfg$ sample,
form an excellent set of tools to probe nearby galaxies,
as we are directly tracing the current SFR with the former and the stellar mass with the latter, thus the sSFR.
We have thus characterized the radial distributions of young and old stars in galaxies in the local Universe.

We also looked at the spatially resolved colors formed by the three bands.
The (FUV\,$-$\,NUV) color is most suitable for detecting variations in recent star formations on time-scales below 1\,Gyr.
The (FUV\,$-$\,[3.6]) color is equivalent to a measurement of the observed sSFR.
The (NUV\,$-$\,[3.6]) color is useful to construct the (FUV\,$-$\,NUV) vs (NUV\,$-$\,[3.6]) color-color diagram, 
in which the $\galex$ Blue Sequence (GBS) and $\galex$ Red Sequence (GRS) subsamples are defined in the preliminary analysis of \citet{Bouquin2015}. We see that the galaxies are grouped into narrow sequences in this color-color diagram, and do separate very well between star-forming (GBS) and quiescent (GRS) galaxies. 
This allowed us to define an intermediate region, the $\galex$ Green Valley (GGV), where we find galaxies that are either leaving the blue sequence due to some damping of their star formation activity, or leaving the red sequence possibly by rejuvenation.
We also performed the fit in the color profiles, and show the distributions of the resulting slopes and $y$-intercepts (scale length and central SB of disks).

Our main results are the following:

- GBS, GGV, and GRS galaxies are well separated in the $\mu_{\mathrm{FUV}}$ vs $\mu_{[3.6]}$ plane.
Most disks are located in a well defined sequence which we call 
the ``spatially resolved main-sequence of star-forming disks", 
with 3.6 micron surface brightness ranging from 20 to 25 mag\,arcsec$^{-2}$, 
and FUV surface brightness ranging from 24 to 27 mag\,arcsec$^{-2}$.
The GBS galaxies are dominating the highest surface sSFR densities,
while the GRS galaxies are dominating the lowest surface sSFR density.
The early-type galaxies of the GRS have a low surface sSFR density, $\Sigma_{\mathrm{sSFR}}$, 
that stays radially constant at (or below) 10$^{-12}$ yr$^{-1}$ pc$^{-2}$.
The late-type galaxies of the GBS, on the other hand, have higher surface sSFR densities the later the type,
with increasing surface sSFR density (blueing) inside-out.
This is not always the case, since inside-out disk reddening is also seen for some of the galaxies. 
This reddening translates to sudden drops in surface SFR density, 
and indicates a possible quenching \citep[or damping; see][]{CatalanTorrecilla2017} of the star formation in the outskirts.

- Star-forming GBS, quiescent GRS, and intermediate GGV galaxies are 
well separated in the (FUV\,$-$\,[3.6]) vs $\mu_{[3.6]}$ plane, 
especially when one looks at the colors of the isophote that 
encompasses 80\% of the 3.6\,$\mu$m light (equivalent to the same percentage of stellar mass).
The isophotes of the GGV galaxies fill the gap between the locus of the GBS and the GRS ones.
Particularly, most GRS galaxies show very similar radial behavior to each other, and 
most of them end up in a similar locus in the (FUV\,$-$\,[3.6]) vs $\mu_{[3.6]}$ plane,
where the 80\% enclosed-light isophote ends up in a narrow range in 3.6\,$\mu$m surface brightness, 
between 21 and 23 mag\,arcsec$^{-2}$, and in (FUV\,$-$\,[3.6]) color range between 6 and 7 mag.

- We performed an analysis of the 3.6\,$\mu$m surface brightness radial profiles
by linearly fitting the data points using an array of cutoffs both in 
radial distance and in 3.6\,$\mu$m surface brightness,
to approximately exclude the bulge part and only fit the disk part.
We find the best cutoffs values to be $R/R80$ = 0.5 and $\mu_{[3.6]}$ = 23.5 mag\,arcsec$^{-2}$ 
(corresponding to a stellar mass surface density of $3\times 10^{7}$\,$M_{\odot}$\,kpc$^{-2}$),
where the mean reduced-$\chi^{2}$ approaches unity and 
the number of galaxies is maximized ($>$50\% of the sample).
Doing so, we efficiently excluded the bulge parts, as well as massive galaxies,
and obtained a subsample of 987 disk galaxies for further analysis.
These slope and y-intercept of the linear fit translates to circular velocities and central surface brightness (of the disk).

- Finally, we compared the slope and y-intercept of the linear fit to the outer disk parts of our subsample to 
the slope and y-intercept of the linear fit to of over 6258 simulated disk models of BP00 varying based on
the circular velocity $v_{\mathrm{c}}$ and the spin parameter $\lambda$, 
thus obtaining a fine grid of slopes (i.e$.$ scale length) and $y$-intercepts (i.e$.$ central surface brightness).
From this, we deduced the circular velocity for each of our galaxies by finding the closest model matching the slope and the y-intercept of the galaxy. 
We find a distribution for the mode circular velocity with mean $v_{\mathrm{c}} = 149\pm102$\,km\,s$^{-1}$ (standard deviation 1$\sigma$)
and median $v_{\mathrm{c}} = 120^{+70}_{-40}$\,km s$^{-1}$  (1$\sigma$ with IQR method and thus, excluding the outliers), 
and a distribution for the mode spin parameter with mean $\lambda=0.062\pm0.037$ (standard deviation 1$\sigma$) 
and median $\lambda=0.054^{+0.030}_{-0.024}$ (1$\sigma$ with IQR method).
For the spin, we recover the probability distribution function of \citet{Mo1998},
whereas for the circular velocity, our distribution is skewed towards higher circular velocities than the ones obtained from HyperLeda.
The low-mass Sd, Sm, and Irr galaxies seems to be more affected than larger spiral galaxies.
Despite the large scatter, this method yields circular velocities similar to those observed within a range of factors of one to two for most galaxies.

\acknowledgments
We acknowledge financial support to the DAGAL network from the People Programme (Marie Curie Actions) of the European Union's Seventh Framework Programme FP7/2007-2013/ under REA grant agreement number PITN-GA-2011-289313.
AYKB also thanks Professor Emeritus John E. Beckman for the valuable suggestions during private communications.
JHK thanks the Astrophysical Research Institute of Liverpool John Moores University for their hospitality, and the Spanish Ministry of Education, Culture and Sports for financial support of his visit there, through grant number PR2015-00512.
JHK acknowledges financial support from the Spanish Ministry of Economy and Competitiveness (MINECO) under grant numbers  AYA2013-41243-P and AYA2016-76219-P.
We acknowledge $\galex$view and the STScI team for the data provided.
We acknowledge the usage of the HyperLeda database (http://leda.univ-lyon1.fr).
This research made use of Astropy, a community-developed core Python package for Astronomy \citep{Astropy2013}.
We are grateful to our anonymous referee for the very constructive and excellent suggestions.

\appendix
\section{Appendix~A} \label{appA}
\subsection{Deriving $\Sigma_{\star}$ from $\mu_{[3.6]}$}
In this work, we make use of the stellar mass surface density $\Sigma_{\star}$ 
(solar mass per square parsec) distribution that is obtained from the 3.6 micron 
surface brightness (AB mag per square arcsec) radial profiles.
We start from the definition of absolute magnitude:
\begin{equation} \label{eq:absmag}
\mathrm{M}_{[3.6],\star} = \mathrm{M}_{[3.6],\odot} - 2.5 \log_{10} \left( \frac{ L_{[3.6],\star}}{ L_{[3.6],\odot}} \right)
\end{equation}
where $\mathrm{M}_{[3.6],\star}$ and $L_{[3.6],\star}$ are the 
3.6\,$\mu$m absolute magnitude (AB) and luminosity (in $ergs \cdot s^{-1} Hz^{-1}$) of the galaxy, 
$\mathrm{M}_{[3.6],\odot}$ and $L_{[3.6],\odot}$ are 
the solar 3.6\,$\mu$m absolute magnitude and luminosity.

We also need the following expression for the mass-to-light ratio:
\begin{equation} \label{eq:mass2light}
\frac{M_{\star}}{L_{[3.6],\star}} = \Upsilon_{[3.6]}
\end{equation}
where the mass-to-light ratio of the Sun ($M_{\odot}/L_{[3.6],\odot}$) is unity, and
where $M_{\star}$ is the stellar-mass of the galaxy,
$L_{[3.6],\star}$ is the luminosity at 3.6\,$\mu$m, 
$M_{\odot}$ is a solar mass, and 
$L_{\odot,3.6}$ is the solar luminosity at 3.6\,$\mu$m.
$\Upsilon_{[3.6]}$ is the mass-to-light ratio at 3.6 micron as obtained by 
\citet{Meidt2014} and is equal to 0.6
(assuming a Chabrier IMF).

Rearranging eq.\ref{eq:absmag} we have:
\begin{equation} \label{eq:absmag2}
\frac{L_{[3.6],\star}}{L_{[3.6],\odot}} = 10^{-0.4(\mathrm{M}_{[3.6],\star} - \mathrm{M}_{[3.6],\odot})}
\end{equation}

Rearranging eq.\ref{eq:mass2light}, 
and adding the conversion factor $a_{\mathrm{IMF}}$ for the transformation from a Chabrier IMF \citep{Chabrier2003} to Kroupa IMF \citep{Kroupa2001},
we get:
\begin{equation} \label{eq:mass2light2}
\frac{M_{\star}}{M_{\odot}} = \frac{L_{[3.6],\star}}{L_{[3.6],\odot}} \cdot \Upsilon_{[3.6]} \cdot a_{\mathrm{IMF}}
\end{equation}
where we use, in our case, $a_{\mathrm{IMF}}=\frac{M/L(Kroupa)}{M/L(Chabrier)} = 1.034$ (conversion factors from \citet{Madau2014}).

We then take the log of eq.\ref{eq:mass2light2}, 
combined with eq.\ref{eq:absmag2}:
\begin{equation}
\log_{10} \left( \frac{M_{\star}}{M_{\odot}} \right) = \log_{10} \left( \frac{L_{[3.6],\star}}{L_{[3.6],\odot}} \right) + \log_{10} (\Upsilon_{[3.6]} \cdot a_{\mathrm{IMF}})
\end{equation}
i.e$.$
\begin{equation} \label{eq:massfromabsmag}
\log_{10} \left( \frac{M_{\star}}{M_{\odot}} \right) = 0.4(\mathrm{M}_{[3.6],\odot} - \mathrm{M}_{[3.6],\star}) + \log_{10} (\Upsilon_{[3.6]} \cdot a_{\mathrm{IMF}})
\end{equation}

Finally, changing $\mathrm{M}_{[3.6],\star}$ to $\mu_{[3.6]} \mathrm{(AB\,mag\,arcsec^{-2})} + 5 - 5 \log_{10} d (\mathrm{pc})$
and converting arcsec$^{-2}$ to parsec$^{-2}$ gives
\begin{equation}
\Sigma_{\star} (M_{\odot}\,\mathrm{pc^{-2}}) = \Upsilon_{[3.6]} \cdot a_{\mathrm{IMF}} \cdot 10^{0.4\mathrm{M_{[3.6],\odot}}} \cdot 10^{-0.4 \cdot \left( \mu_{[3.6]} + 5 - 5 \log_{10} d \right)} \cdot \left( \frac{206265}{d} \right)^{2}
\end{equation}
where $\mathrm{M_{[3.6],\odot}}$ is the Sun's 3.6 micron absolute AB magnitude which is taken to be 6.03 mag
(converted to AB scale from the Vega magnitude value, $M_{\odot,3.6,Vega}=3.24$, given by equation(13) in \citet{Oh2008}).

This corresponds to a stellar mass surface density 
$M_{\star}$/area\,=\,1.045\,$M_{\odot}$/pc$^{2}$ at 
a 3.6 micron surface brightness $\mu_{[3.6]}$\,=\,27\,mag\,arcsec$^{-2}$ in the case of a Chabrier IMF,
and $M_{\star}$/area\,=\,1.080\,$M_{\odot}$/pc$^{2}$ in the case of a Kroupa IMF.
The equation, then, simplifies to the following:
\begin{equation} \label{eq:stelmass}
\log_{10}(\Sigma_{\star} (M_{\odot}\,\mathrm{pc^{-2}}) ) = 10.819 - 0.4 \mu_{[3.6]} + \log_{10} a_{\mathrm{IMF}} \\
\end{equation}
where the term $\log_{10} a_{\mathrm{IMF}}=0$ for a Chabrier IMF, $0.015$ for a Kroupa IMF, and $0.215$ for a Salpeter IMF.

\section{Appendix~B} \label{appB}
\subsection{Deriving the observed sSFR from (FUV\,$-$\,[3.6])}
We start with the SFR(UV), assuming a Salpeter IMF \citep{Salpeter1955} and using a calibration by \citet{Madau1998}, as provided by \citet{Kennicutt1998}, which we can convert to the expression for a Kroupa IMF by multiplying by $b_{\mathrm{IMF}}=0.67$, or by $b_{\mathrm{IMF}}=0.63$ for a Chabrier IMF, as reviewed and prescribed in \citet{Kennicutt2012,Madau2014}:
\begin{equation} \label{eq:ken98}
\mathrm{SFR} (M_{\odot}\,yr^{-1}) = 1.4 \times 10^{-28} \cdot b_{\mathrm{IMF}} \cdot L_{\nu}\,\mathrm{(ergs \cdot s^{-1} Hz^{-1})}
\end{equation}
and with the following expression of luminosity:
\begin{equation} \label{eq:lnu}
\begin{aligned}
L_{\nu} &=4 \pi (d (\mathrm{cm}))^{2} f_{\nu}\\
L_{\nu} &=4 \pi \left[ d (\mathrm{pc}) \cdot 3.086\times10^{18} \mathrm{(cm/pc)} \right] ^{2} 10^{-0.4(\textrm{FUV}+48.6)}
\end{aligned}
\end{equation}
Combining the two equations above gives:
\begin{equation} \label{eq:SFR}
\log_{10} (\mathrm{SFR}) (M_{\odot}\,yr^{-1}) = 2 \log_{10} d (pc) - 0.4\, \mathrm{FUV} - 9.216 + \log_{10} b_{\mathrm{IMF}}
\end{equation}
where FUV is in AB magnitudes, distance d is in parsec,
and $\log_{10} b_{\mathrm{IMF}}=0$, $-0.174$, and $-0.201$ for a Salpeter, Kroupa, and Chabrier IMF respectively.

Secondly, we need to introduce the distance modulus $m-M=5-5\log_{10} d(pc)$ into eq.~\ref{eq:massfromabsmag}, so this becomes:
\begin{equation} \label{eq:Mstar}
\log_{10} \left( \frac{M_{\star}}{M_{\odot}} \right) = \log_{10} (\Upsilon_{3.6} \cdot a_{\mathrm{IMF}}) - 0.4 ([3.6] + 5 - 5\log_{10} d - 6.03)
\end{equation}

where $[3.6]$ is the apparent AB magnitude at 3.6\,$\mu$m, 
$d$ is the distance in parsec, and 
$M_{\odot,3.6,AB} = 6.03$ (see Appendix~\ref{appA}).

Finally, we combine \ref{eq:SFR} and \ref{eq:Mstar}:
\begin{equation}
\log_{10}(\mathrm{sSFR}) = \log_{10} \left( \frac{\mathrm{SFR}}{M_{\star}} \right) = \log_{10}(\mathrm{SFR}) - \log_{10}(M_{\star})
\end{equation}

It thus becomes:
\begin{equation} \label{eq:ssfr}
\log_{10}(\mathrm{sSFR}) = -0.4 (\mathrm{FUV} - [3.6]) - 9.628 + \log_{10} b_{\mathrm{IMF}} - \log_{10} (\Upsilon_{3.6} \cdot a_{\mathrm{IMF}})
\end{equation}
We can thus obtain the logarithm of the 
specific star formation rate (in units of $year^{-1}$) from 
the (FUV\,$-$\,[3.6]) (ABmag) color.
We emphasize that these quantities would not be corrected for extinction.

\bibliographystyle{apj}

\end{document}